 \documentclass[final,5p,times,twocolumn,authoryear]{elsarticle}

\usepackage{lipsum}
\usepackage{amsmath}
\usepackage{amssymb}
\usepackage{cases}
\usepackage{mathtools}
\usepackage{mathrsfs}
\usepackage{graphicx}
\usepackage{url}
\usepackage{verbatim}
\usepackage[colorlinks=true, allcolors=blue]{hyperref}

\journal{Annals of Physics}

\begin{document}

\begin{frontmatter}



\title{A New Perspective on Kaluza-Klein Theories}



\author{L. Horoto and
F. G. Scholtz \\
Department of Physics,\\
University of Stellenbosch, Stellenbosch-7600, South Africa}

\begin{abstract}
By assuming that the geometry of spacetime is uniquely determined by the energy momentum tensor of matter alone, i.e. without any interactions, enables us to construct the Lagrangian from which the metric of higher dimensional spacetime follows. From the geodesic equations that follow it becomes clear that the incorrect mass of elementary particles predicted by Kaluza-Klein theories arises from the assumption that in the absence of gravity the solution to the Einstein field equations reduces to the Minkowski metric. From  construction of a consistent theory of $4\mathcal{D}$ electromagnetism, we find that this assumption does not only result in  the incorrect mass of elementary particles, but also the incorrect value of the cosmological constant. This suggests that these incorrect predictions, which are often regarded as major flaws of Kaluza-Klein theories, just reflects the inconsistency of some postulates of general relativity and gauge theories. Abandoning this assumption results in modifications of general relativity. We show that the unified description of fundamental interactions naturally incorporates the Higgs mechanism. For non-Abelian gauge fields, we find that the manifold comprising the extra dimensions has to be a group manifold and show that the standard model is realised in 16$\mathcal{D}$ spacetime. We show that charge and spin are the same concept, but what makes them different is that the former follows from symmetry of $4\mathcal{D}$ spacetime while the latter follows from symmetry of the internal space.
\end{abstract}




\end{frontmatter}




\section{Introduction}
\label{introduction}
Kaluza-klein theory is an attempt at providing a unified description of natural phenomena by treating the fundamental interactions as general relativity in higher dimensions. The pioneers of this theory (Theodore Kaluza \cite{kaluza2018unification} and Oscar Klein \cite{klein1991quantum, klein1926atomicity} ) had an idea of unifying gravity and electromagnetism by treating the two forces as nothing but attributes of 5$\mathcal{D}$ space-time. Since its inception, this idea has undergone considerable modifications \cite{thiry1948geometry} with the aim of encompassing the weak and strong nuclear force \cite{kerner1968generalization,salam1982kaluza,weinberg1983charges} . The idea is simply the extension of general relativity to $(4+d)$-dimensions. However; contrary to general relativity, in this theory one chooses a specific ansatz (supplemented by the cylinder condition) to be the solution to the Einstein field equations. The geodesic equations arising from this ansatz comprise the Lorentz force law or its generalization to non-Abelian gauge theories \cite{kerner1968generalization}. The conserved charges correspond to momenta in the extra dimensions while the higher dimensional Einstein equations reduce to 4$\mathcal{D}$ Einstein equations and the equations of motion  for gauge fields albeit with the constraint on the field strength tensor which requires introduction of a scalar field (radion/dilaton) to relax it  \cite{zee2013einstein}. This theory is mired with problems, some of which are explored in \cite{Wesson1998SpaceTimeMatterMK}. However; the one that dealt a major blow to the theory is arguably its failure to describe any of the known particles and the prediction of the cosmological constant, which is at odds with observation.
There are proposed solutions to these problems \cite{Wesson1998SpaceTimeMatterMK,freund1980dynamics,witten1981search, duff1984consistency,bailin1993introduction, collins1989particle} . However; all of these require additional gauge fields and reject the gauge fields arising from the ansatz. Thus they do not approach unification in the sense envisaged by Kaluza \cite{collins1989particle}.
In this work we propose a solution that does not sacrifice Kaluza's idea.  We assume that the geometry of spacetime is uniquely determined by the energy momentum tensor of matter alone, i.e. without any interactions. This enables us to construct the Lagrangian for
vector fields in (4+ d )- dimensional spacetime. From the Lagrangian follows the equations of motion which imply that
we are dealing with killing vector fields.
 Solving the killing equation grants us
the form of these killing vector fields
which we utilize in deriving the metric.
With the metric known, we find the geodesic equations.
From the conserved charges that follow we find that
we are dealing with particles whose mass is of the order of Planck mass $(M_{pl})$. It also becomes evident that this results from one of the postulates of general relativity, which is the assumption
that in the absence of gravity the solution to the Einstein field equations reduces to the Minkowski metric. This is contrary to the generally accepted idea that this problem is the result of compactifying the extra dimensions. We show that the
 culprit is indeed the assumption that $g_{\mu\nu}=\eta_{\mu\nu}$ by considering 4D electromagnetism. We find that the Lagrangian that respects all symmetries of electromagnetism is necessarily the Lagrangian of the theory of gravity. Due to non-minimal coupling of the scalar
field, the resulting theory is not general relativity,
but reduces to general relativity if the solution to Einstein field equations is related to the solution
arising from these equations by a conformal factor
or, equivalently, when we require that the scalar field is constant. From this we find that indeed the  vacuum energy is of the order of $M_{pl}^4$ and  that the mass  of the resulting particles is of the order of $M_{pl}$.

Having shown that these incorrect predictions of Kaluza-Klein theories are not their flaws, but the result of inconsistent assumptions of general relativity and gauge theories, we construct the unified theory of gravity and electromagnetism in 5D spacetime. The resulting theory naturally incorporates the Higgs mechanism and in the non-relativistic limit results in Schrödinger's equation. This is quite remarkable since there is no place in the formalism where we invoke any of the postulates of quantum  mechanics.
Generalization to non-Abelian gauge fields entails introducing vector fields whose number is equal to the number of extra dimensions. These killing vector fields have to be linearly independent as vector fields, as such the internal space is the group manifold and not a coset manifold in general. We demonstrate that these $ d $   copies of electromagnetism in $(4+ d)$ dimensions become Yang-Mills theory in $4$ dimensions and for spin-0 fields we get a non-Abelian Higgs model.
We provide the way of constructing the metric of the internal space, given a gauge group, in addition to constructing electroweak theory. We then apply this to the standard model gauge group.

We conclude by demonstrating that the existence of extra dimensions introduces a  possibility of a unified description  of particles in the sense that particles of any spin can be associated with 1-forms dual to non-equivalent sets of Killing vector fields that form a representation of the Lie algebra of SU(2).

 \section{Matter and curved spacetime are different ways of viewing the same thing}
In Kaluza-Klein theories one starts with a higher dimensional Einstein-Hilbert action, obtains the Einstein vacuum equations, chooses a specific ansatz (without any justification), plugs it in the field equations and acquires results which are brimming with inconsistencies. What then follows is the unsuccessful program in taming these inconsistencies. We take this as an indication that in general the ansatz is not the solution of the equations of motions that follow from the said Lagrangian. On the other hand, the geodesic equations that follow from the ansatz are equations of motion for a test particle under the influence of both electromagnetic/Yang-Mills and gravitational fields (see \cite{kaluza2018unification, kerner1968generalization}), which means that the ansatz has something to do with the geometric theory of gravity and gauge theories. This implies that the Einstein-Hilbert Lagrangian is not necessarily the Lagrangian of the unified theory of the forces of nature.

Our argument in support of the aforementioned conclusion is as follows: From Electroweak theory we know that a unified picture is apparent when none of the gauge symmetries are broken. In  our geometric description we treat all symmetries to be the result of spacetime symmetries, that is the spacetime metric is invariant under coordinate transformation or equivalently they imply the existence of Killing vector fields. In accordance with this, unification is apparent in the most symmetric of all spacetimes. One can show that such spacetimes follow from the Lagrangian \cite{blau2018lecture}
\begin{equation}\label{RicciCosmos}
    \mathfrak{L} = \Tilde{\mathfrak{R}} + 2\Lambda
\end{equation}
where $\Lambda$ is the cosmological constant. While in general the solution to the $4\mathcal{D}$ Einstein field equations posses no symmetry, the non-Abelian gauge theories are gauge invariant. This means that although $4\mathcal{D}$ spacetime may not have any Killing vector fields the Higher dimensional spacetime has Killing vector fields. Therefore not every solution of the general higher dimensional Einstein field equations is a solution to the field equations of higher dimensional gravity. The higher dimensional Lagrangian should reflect this. Accordingly, the higher dimensional Lagrangian is somewhere between the the Einstein Lagrangian for arbitrary matter fields and \eqref{RicciCosmos} and reduces to \eqref{RicciCosmos} at the unification scale. Our goal is to find such a Lagrangian.
Instead of postulating the complete Lagrangian we assemble it piece by piece from our postulate that there is a complete equivalence between spacetime geometry and the energy momentum tensor of matter alone.   The remainder of this section is devoted to constructing this postulate.

In the absence of forces the Einstein field equations are
\begin{equation}\label{EFEnoForce}
    \mathfrak{R}_{\mu\nu}-\frac{1}{2}g_{\mu\nu}\mathfrak{R} = \kappa\mathfrak{T}_{\mu\nu}^{m}.
\end{equation}
Where $\mathfrak{T}^{m}_{\mu\nu}$ is the energy momentum tensor of matter fields (our definition of matter fields excludes force fields. That is, by matter we refer to the source term in equations of gauge fields). From this equation it is clear that if the metric is known all properties of matter encoded in the energy-momentum tensor can be deduced. The known properties of matter are determined from the behaviour of matter in the presence of forces. As such; if like gravity all other fundamental forces can be treated as curvature of spacetime, all properties of matter are embedded in the energy-momentum tensor. Consequently; owing to \eqref{EFEnoForce}, if we know the metric we can determine all properties of matter. Thus, there is no clear distinction between matter and the resulting spacetime which makes matter and spacetime different ways of viewing the same thing. This conclusion contradicts the postulate of general relativity which implies that in the absence of matter spacetime is flat since it implies that in the absence of matter there is no spacetime. We will later see that a consistent theory of electromagnetism backs this conclusion.

We assume that all forces of nature are just different manifestation of curvature of spacetime. The consequence of this assumption is that all conserved quantities arise from the isometries of the metric. In accordance with this, in $n$-dimensional spacetime, they are components of either linear, radial and/or angular $n$-momentum when spacetime is viewed as a submanifold of Minkowski spacetime. In addition to this, no contradiction should arise if equations involving matter fields are reformulated in terms of the geometric quantities.

In $4\mathcal{D}$, at most we can have the theory of gravity, a scalar field and/or spinor field. This is possible due to the fact that the conserved quantities that are associated with these fields are part of $4\mathcal{D}$ total angular momentum. Since all the ten components of $4\mathcal{D}$ energy momentum tensor are associated with gravity, the theory of scalar and/or spinor fields can be achieved by introducing the metric conformal to the solution of the Einstein Field Equations.  For this reason, to accommodate all known forces of nature, our assumption necessitates the existence of extra dimensions. 

We  start by constructing the unified theory of gravity and electromagnetism. Apart from it being simple compared to its non-Abelian counterparts, the motivation for this approach is that electromagnetism can be treated as non-Abelian gauge theory with the matrix of the gauge potential having only one non-zero entry (we show this later). As such this construction will afford us clues on how to generalize to non-Abelian gauge theories.

\section{Unification of electromagnetism and gravity}
This section is aimed at finding the Lagrangian of the unified theory of gravity and electromagnetism. We go about achieving this in the following way:
We utilize our assumption regarding matter and curvature of spacetime to construct the Lagrangian of $5\mathcal{D}$ electromagnetism. This Lagrangian helps us determine the metric of $5\mathcal{D}$ spacetime in addition to the $\mathfrak{R}^{a}_{5}$ components of the Ricci curvature tensor. Simplifying the equations we find that the $\mathfrak{R}^{\mu}_{5}$ can be expressed in terms of the source term of $4\mathcal{D}$ Maxwell's equations. This confirms that we can write curvature in terms of matter fields. In order to find its expression, we find the general form of currents in the Maxwell's equations and show that it is indeed the general form by considering the case of spin-0 and spin-$\frac{1}{2}$ matter fields in \autoref{4delectrodynamics}. Having established this, we construct the simple form of the Lagrangian for a theory that respects all the symmetries of $4\mathcal{D}$ electromagnetism in \autoref{WeylImplyGravity}. The resulting theory is that of $4\mathcal{D}$ gravity. This construction is pivotal as it will be evident that while the resulting theory is consistent with classical theories from which quantum field theories are constructed, it produces results which are in contradiction with observation if general relativity is regarded as a theory of pure gravity. It is only when it is regarded as the theory of gravity and the Higgs field that we get the results consistent with experiment. This shows that the particles whose masses are of the order of planck mass predicted by higher dimensional theories are not their flaw, but the result of this inconsistency.
In \autoref{ElectroGrav5D} we demonstrate that the Lagrangian of $5\mathcal{D}$ gravity consistent with our assumption regarding  curvature and matter results in the Lagrangian proposed in \autoref{WeylImplyGravity}. We also show that the inconsistency of this Lagrangian with the aforementioned assumption of general relativity is the source of the cosmological constant problem. We show the connection of the developed theory with classical physics by considering the case where $4\mathcal{D}$ spacetime is regarded as Minkowski spacetime. In addition to this, we construct the equations of motion for $4\mathcal{D}$ gravity by using the equation of motion of the Higgs field found from the reduced Lagrangian. The purpose of this is to confirm that the Lagrangian is indeed consistent and thus provide justification of some introduced constants (which otherwise seem arbitrary) since a different choice results in inconsistencies. We note that consistency of our assumption regarding the $5\mathcal{D}$ gravity and the reduced lagragian results in a constraint equation for the electromagnetic field. We demonstrate that the source of this is our assumption that the source term of $5\mathcal{D}$ electromagnetism comprises only the generalization of the source of $4\mathcal{D}$ electromagnetism. It turns out that $4\mathcal{D}$ electromagnetism is a source of $5\mathcal{D}$ electromagnetism which qualifies it as matter in line with our definition. Taking this into consideration, we get the revised $5\mathcal{D}$ Lagrangian for which one can easily check that the equations of motion are consistent with those of the reduced Lagrangian without any constrained. We conclude by showing that the gauge transformation is indeed the result of coordinate transformation and rewriting some expressions in a way that enables us to generalize them to the non-Abelian case.

\subsection{5D Metric and Equations of motion}
Since in addition to $4$-momentum, only the electric charge has to arise from the isometries of the metric, increasing the dimensions of spacetime to $5$ should suffice. Vacuum electrodynamics can be extended  to  $5\mathcal{D}$ by introducing the Lagrangian
\begin{equation}
    \mathfrak{L} = -\frac{1}{4}\mathrm{F}_{ab}\mathrm{F}^{ab}
\end{equation}
with the Greek indices ranging from 0 to 3 while the Latin indices include $5$ and we skip $4$ in accordance with the established convention. In line with our conclusion that matter and curved spacetime are just two sides of the same coin, in the presence of matter the $5\mathcal{D}$ Lagrangian is
\begin{equation}\label{matterAsCurvedSpaceTimeL}
    \mathfrak{L} = -\sqrt{\mathrm{G}}\frac{1}{4}\mathrm{F}_{ab}\mathrm{F}^{ab} + \sqrt{\mathrm{G}}\Tilde{\mathfrak{R}}^{ab}A_{a}A_{b}
\end{equation}
where $\mathrm{G}$ is the absolute value of the determinant of the unknown $5\mathcal{D}$ metric $\mathrm{G}_{ab}$. At first sight this Lagrangian appears as if it is not gauge invariant. However, since conservation of charge, which arises from gauge invariance, is a consequence of invariance of the metric under a certain type of coordinate transformation, gauge invariance is the by-product of such a coordinate transformation. As the above Lagrangian is coordinate invariant, it must be the case that it is gauge invariant.

The equations of motion arising from \eqref{matterAsCurvedSpaceTimeL} are
\begin{equation}\label{gravityAsSource}
    \partial_{a}\left(\sqrt{\mathrm{G}}\mathrm{F}^{ab}\right) = -2\sqrt{\mathrm{G}}\mathfrak{R}^{ab}A_{a}.
\end{equation}
Evidently these equations imply that  $A^{a}$ is the Killing vector field (see \cite{weinberg1972gravitation}). That is, it obeys the Killing equation
\begin{equation}\label{abelKillingE}
    A^{a}\partial_{a}A_{b} + \partial_{b}A^{a}A_{a} = 0.
\end{equation}
$A^{a}$ is associated with the translation symmetry of the extra dimension. Accordingly, we can write the solution \eqref{abelKillingE} as
\begin{equation}\label{KillingSoln}
    A^{a} =\frac{\zeta^2}{\lambda^2} \delta_{5}^{a}A_{5}^{-1}.
\end{equation}
where $\lambda$ and $\zeta$ are constants with ${\zeta}/{\lambda}$ having the same dimensions as $A_{a}$.
This is not the only solution. However; due to the fact that the maximum number of linearly independent Killing vector fields in one dimension is one, all other solutions are related to this one by a coordinate transformation. That being so, this choice of solution amounts to choosing a coordinate system. By substituting the solution in \eqref{abelKillingE} one can verify that in this coordinate system the Killing equation can be written in the form
\begin{equation}\label{F5u0}
    \mathrm{F}_{5\mu} = 0.
\end{equation}
While \eqref{abelKillingE} is covariant, it should be noted that \eqref{F5u0} is not. However; there exists a coordinate transformation under which it is covariant, we will show that it is that coordinate transformation which results in a gauge transformation.

Construction of the metric in this coordinate system proceeds as follows: \eqref{KillingSoln} makes it clear that the coordinate axis of the extra dimension is aligned with the Killing vector field making it possible to express the metric in terms of the Killing vector field. Given $A^{a}$ we can use the metric to find the expression for $A_{a}$. This gives us
\begin{equation}
    \begin{split}
       A_{a} &= \mathrm{G}_{ab}A^{b} \\
             &= \frac{\zeta^2}{\lambda^2}\mathrm{G}_{a5}A_{5}^{-1}
    \end{split}
\end{equation}
from which we conclude that
\begin{equation}
    \mathrm{G}_{a5} =\frac{\lambda^2}{\zeta^2} A_{a}A_{5}.
\end{equation}
Since $\mathrm{G}^{\mu a}A_{a} = A^{\mu} = 0$, we have
\begin{equation}
    \mathrm{G}^{5\mu} = -A_{5}^{-1}\mathrm{G}^{\mu\nu}A_{\nu}
\end{equation}
while $\mathrm{G}^{5a}A_{a} = A_{5}^{-1}$ implies that
\begin{equation}
    \mathrm{G}^{55} = A_{5}^{-2}\left(\frac{\zeta^2}{\lambda^2} - A_{5}\mathrm{G}^{5\mu}A_{\mu}\right).
\end{equation}
We make the definition
\begin{equation}
    \mathrm{G}^{\mu\nu} = g^{\mu\nu}
\end{equation}
and introduce $g_{\mu\nu}$ such that $g^{\mu\rho}g_{\rho\nu} = \delta^{\mu}_{\nu}$. Using $\mathrm{G}^{\mu a}\mathrm{G}_{a\nu} = \delta^{\mu}_{\nu}$, we can express $\mathrm{G}_{\mu\nu}$ in terms of $g_{\mu\nu}$ in the following way:
\begin{equation}
    \begin{split}
   \delta^{\mu}_{\nu} &=  \mathrm{G}^{\mu a}\mathrm{G}_{a\nu}\\
                &= \mathrm{G}^{\mu\rho}\mathrm{G}_{\rho\nu} +\mathrm{G}^{\mu 5}\mathrm{G}_{5\nu}\\
                &=g^{\mu\rho}\left(\mathrm{G}_{\rho\nu} -\frac{\lambda^2}{\zeta^2}A_{\rho}A_{\nu}\right)\\
                &=g^{\mu\rho}g_{\rho\nu}
   \end{split}
\end{equation}
therefore
\begin{equation}
    \mathrm{G}_{\mu\nu} = g_{\mu\nu} + \frac{\lambda^2}{\zeta^2}A_{\mu}A_{\nu}.
\end{equation}
We can write the equations compactly as
\begin{equation}\label{general5Dmetric}
    \boxed{\begin{aligned}
        \mathrm{G}_{ab} &= \delta_{a}^{\mu}\delta_{b}^{\nu}g_{\mu\nu} + \frac{\lambda^2}{\zeta^2}A_{a}A_{b}\\
        \mathrm{G}^{\mu\nu} &= g^{\mu\nu}\\
        \mathrm{G}^{5a} &= \delta_{5}^{a}A_{5}^{-2}\frac{\zeta^2}{\lambda^2} - A_{5}^{-1}\mathrm{G}^{\mu a}A_{\mu}.
    \end{aligned}}
\end{equation}
$\mathrm{G}_{ab}$ has fifteen independent parameters, as such it is the most general form of a $5\mathcal{D}$ metric. Any metric can be written in this form. In constructing $\mathrm{G}_{ab}$ we made no use of the fact that $A^{a}$ is a Killing vector field as such $\mathrm{G}_{ab}$ is not necessarily consistent with  \eqref{gravityAsSource}. For $\mathrm{G}_{ab}$ to result in \eqref{gravityAsSource}, its Lie derivative along $A^{a}$ must vanish. That is
\begin{equation}\label{lieof5dMetric}
    \begin{split}
        0 &= L_{A}\mathrm{G}_{ab},\\
          &= \delta_{a}^{\mu}\delta_{b}^{\nu} L_{A}g_{\mu\nu} +\frac{\lambda^2}{\zeta^2} A_{a}L_{A}A_{b} + \frac{\lambda^2}{\zeta^2}A_{b}L_{A}A_{a}.\\
    \end{split}
\end{equation}
As \eqref{general5Dmetric} and \eqref{lieof5dMetric} are covariant, it is clear that the theory is valid under an arbitrary coordinate transformation. That is there is no preferred frame of reference.
Since $L_{A}A_{b} = 0$ and $A^{\mu} = 0$, \eqref{lieof5dMetric} reduces to
\begin{equation}
    \partial_{5}g_{\mu\nu} = 0.
\end{equation}
Therefore $A_{a}$ in $\mathrm{G}_{ab}$ is a solution of \eqref{gravityAsSource} if $g_{\mu\nu}$ is independent of the extra dimension and $\mathrm{F}_{5\mu} = 0$. $g_{\mu\nu}$ can be regarded as either a solution to the $4\mathcal{D}$ Einstein field equations or related to such a solution by a conformal factor. When it is regarded as the solution of the Einstein field equation, it follows that 
\begin{equation}
    \frac{\lambda^2}{\zeta^2} = \frac{2\kappa}{\mu_{0}} = \frac{4 e^2}{\alpha M_{pl}^2}.
\end{equation}

When $A_{5}$ is constant $\mathrm{G}_{ab}$ reduces to the Kaluza-Klein metric and $L_{A}\mathrm{G}_{ab} = 0$ reduces to the cylinder condition .  From this it is clear that the cylinder condition is a result of introducing a coordinate system that is adapted to the Killing vector field. In the non-Abelian case there is no coordinate system adapted to all Killing vectors thus we can not invoke the cylinder condition, ignoring this fact is the reason why all attempts at constructing a realistic non-Abelian gauge theory have been unsuccessful .

The expression for the Christoffel symbols is 
\begin{equation}
    \begin{split}
    \Tilde{\Gamma}^{a}_{bc} &=\frac{\mathrm{G}^{ad}}{2}\left(\partial_{b}\mathrm{G}_{dc} +\partial_{c}\mathrm{G}_{db}-\partial_{d}\mathrm{G}_{bc}\right)\\
&=\delta^{\mu}_{b}\delta^{\nu}_{c}\frac{\mathrm{G}}{2}^{a\rho}\left(\partial_{\mu}g_{\rho\nu}+ \partial_{\nu}g_{\rho\mu} -\partial_{\rho}g_{\mu\nu}\right) + \frac{\lambda^{2}}{2\zeta^{2}}\mathrm{G}^{ad}\big(\partial_{b}A_{d}A_{c}\\ & + A_{d}\partial_{b}A_{c}+ \partial_{c}A_{b}A_{d} +A_{b}\partial_{c}A_{d}-\partial_{d}A_{b}A_{c} - A_{b}\partial_{d}A_{c}\big)\\
    &= \delta^{\mu}_{b}\delta^{\nu}_{c}\mathrm{G}^{a\rho}\Gamma_{\mu\nu\rho} + \frac{\lambda^2}{\zeta^2}\frac{\mathrm{G}}{2}^{a\rho}\left(\mathrm{F}_{b\rho}A_{c} +\mathrm{F}_{c\rho}A_{b}\right) + \frac{\delta_{5}^{a}}{2}A_{5}^{-1}\mathfrak{S}_{bc}.\\
    \end{split}
\end{equation}
Using this expression and simplifying the results, we find that the geodesic equations
\begin{equation}\label{5dgeodesicE}
 \frac{d^2x^{a}}{d\tau^2} +\Gamma^{a}_{bc}\frac{dx^{b}}{d\tau}\frac{dx^{c}}{d\tau}  = 0 
\end{equation}
reduce to
\begin{equation}\label{reduced5Dgeodesics}
    \begin{split}
        \frac{d^2x^{\mu}}{d\tau^2} +\Gamma^{\mu}_{\alpha\beta}\frac{dx^{\alpha}}{d\tau}\frac{dx^{\beta}}{d\tau} +\frac{\lambda^2}{\zeta^2} A_{c}\frac{dx^{c}}{d\tau}\mathrm{F}_{\hspace{4pt}\nu}^{\mu}\frac{dx^{\nu}}{d\tau} &= 0,\\
        A_{5}^{-1}\frac{d}{d\tau}\left(A_{a}\frac{dx^{a}}{d\tau}\right) &= 0 .
    \end{split}
\end{equation}
When $g_{\mu\nu} = \eta_{\mu\nu}$ this reduces to the Lorentz force Law with charge to mass ratio of the test particle given by
\begin{equation}\label{chargeToMassR}
    \frac{e}{m} = \frac{\lambda^2}{\zeta^2} A_{c}\frac{dx^{c}}{d\tau}.
\end{equation}
Multiplying \eqref{5dgeodesicE} by $\mathrm{G}_{ab}\frac{dx^b}{d\tau}$ we get
\begin{equation}
    \begin{split}
    0 &=\mathrm{G}_{ab}\frac{d^2x^a}{d\tau^2}\frac{dx^{b}}{d\tau^2} +\frac{1}{2}\frac{d}{d\tau}\left(\mathrm{G}_{ab}\right)\frac{dx^a}{d\tau}\frac{dx^b}{d\tau},\\
    &= \frac{d}{d\tau}\left(\mathrm{G}_{ab}\frac{dx^a}{d\tau}\frac{dx^b}{d\tau}\right).
    \end{split}
\end{equation}
Thus $$\frac{m^2_{(5)}}{m^2} =-\mathrm{G}_{ab}\frac{dx^a}{d\tau}\frac{dx^b}{d\tau}$$ is a conserved quantity. Making use of \eqref{chargeToMassR} we get
\begin{equation}
   m^2 g_{\mu\nu}\frac{dx^{\mu}}{d\tau}\frac{dx^{\nu}}{d\tau} = - m_{(5)}^2 - \frac{\zeta^2}{\lambda^2}e^2.
\end{equation}
When $m_{(5)}^{2} = 0$ we get $m = {\sqrt{\alpha}M_{pl}}/{2}$. We see that this is the unrealistic mass that one gets from Kaluza-Klein Theory. It is worth noting that at this point we made no assumption regarding the nature of the $5^{th}$- dimension apart from the requirement that it is homogeneous. It can be extended or compact but as long as $m_{(5)} = 0$ we will get the same results as long as we keep the assumption that in the absence of gravity the solution to the Einstein field equations reduces  the Minkowski metric.

Making use of the fact that the absolute value of the determinant of $G_{ab}$ is
$\mathrm{G} = g \frac{\lambda^2}{\zeta^2}A_{5}^{2}$ -where $g$ is the absolute value of the determinant of $g_{\mu\nu}$-  we can simplify the $4\mathcal{D}$ part of \eqref{gravityAsSource} in the following way
\begin{equation}
    \begin{split}
      2 \frac{\zeta}{\lambda}\sqrt{g}\Tilde{\mathfrak{R}}^{\mu}_{5} &= -\partial_{a}\left(\sqrt{\mathrm{G}}\mathrm{F}^{a\mu}\right)\\
        &=-\frac{\lambda}{\zeta}\partial_{a}\left(\sqrt{g}A_{5}\mathrm{G}^{a\nu}\mathrm{F}_{\nu}^{\hspace{4pt}\mu}\right) \\
        &=-\frac{\lambda}{\zeta}A_{5}\partial_{\nu}\left(\sqrt{g}\mathrm{F}^{\nu\mu}\right) -\frac{\lambda}{\zeta}\sqrt{g}\left(\partial_{\nu}A_{5} - \partial_{5}A_{\nu}\right)\mathrm{F}^{\nu\mu}\\
        &=-\frac{\lambda}{\zeta}A_{5}\partial_{\nu}\left(\sqrt{g}\mathrm{F}^{\nu\mu}\right) .      
    \end{split}
  \end{equation}
  Therefore
  \begin{equation}\label{5DcurvSource4Dem}
    \boxed{ \frac{1}{\sqrt{g}} \partial_{\nu}\left(\sqrt{g}\mathrm{F}^{\nu\mu}\right) =-\frac{2\zeta^2}{\lambda^2}\Tilde{\mathfrak{R}}^{\mu}_{5}A_{5}^{-1}.} 
  \end{equation}
Comparing with the Maxwell's equations, the current that is conserved in the ordinary sense is
\begin{equation}
   \boxed{ \mathfrak{J}^{\mu} = 2\sqrt{g}\frac{\zeta^2}{\lambda^2A_{5}}\Tilde{\mathfrak{R}}^{\mu}_{5}.}
\end{equation}
We note that in $4\mathcal{D}$ spacetime, the product $\sqrt{g}g^{\mu\rho}g^{\nu\delta}$ is invariant under the Weyl transformation 
\begin{equation}
    g_{\mu\nu} \longmapsto e^{2\alpha(x)}g_{\mu\nu}.
\end{equation}
As such $\partial_{\nu}\left(\sqrt{g}\mathrm{F}^{\nu\mu}\right)$ is invariant under Weyl transformation and so is $\mathfrak{J}^{\mu}$. In order to find the expression of the Ricci curvature tensor, we have to find the general form of the current density for the Maxwell equations. We achieve this through the use of gauge and Weyl-invariance.
\subsection{General Form of Current Density in Maxwell's equations}\label{4delectrodynamics}
  In $4\mathcal{D}$ spacetime we can derive Maxwell's equations from the Lagrangian
  \begin{equation}
      \mathfrak{L} = -\frac{1}{4}\mathrm{F}_{\mu\nu}\hat{\mathrm{F}}^{\mu\nu} + \mathfrak{J}^{\mu}A_{\mu}.
  \end{equation}
  The notation $\hat{\mathrm{F}}^{\mu\nu}$ indicates that the indices are raised by $\gamma^{\mu\nu}$ which in this case is equal to the Minkowski metric $\eta^{\mu\nu}$.
  The above Lagrangian is not gauge invariant. However, it can be made gauge invariant if in place of $A_{\mu}$ we substitute
  \begin{equation}
      A_{\mu}-\frac{1}{e}\partial_{\mu}\mathcal{S}
  \end{equation}
  with $\mathcal{S}$ behaving in the following way
  \begin{equation}
      \mathcal{S} \longmapsto \mathcal{S} -e\chi
  \end{equation}
  under the gauge transformation
  \begin{equation}
      A_{\mu}\longmapsto A_{\mu} -\partial_{\mu}\chi.
  \end{equation}
  The gauge invariant Lagrangian is therefore
  \begin{equation}\label{GaugeCovariantEMLagrangian}
      \mathfrak{L} = -\frac{1}{4}\mathrm{F}_{\mu\nu}\hat{\mathrm{F}}^{\mu\nu} +\mathfrak{J}^{\mu}\left(A_{\mu}-\frac{1}{e}\partial_{\mu}\mathcal{S}\right).
  \end{equation}
  While this substitution does not affect Maxwell's equation, it provides a way of finding the expression of $\mathfrak{J}^{\mu}$ since the equation of motion of $\mathcal{S}$ results in
  \begin{equation}\label{JisConjugateMomentaOfS}
      \mathfrak{J}^{\mu} = -e\frac{\partial\mathfrak{L}}{\partial{\partial_{\mu}\mathcal{S}}}.
  \end{equation}
   Weyl invariance and conservation of $\mathfrak{J}^{\mu}$ together with gauge invariance imply that the most general form of the Lagrangian for $\mathcal{S}$ is 
  \begin{equation}\label{lmatter}
      \mathfrak{L}_{M} = \frac{1}{2} \phi^2\eta^{\mu\nu}\left(\partial_{\mu}\mathcal{S} - eA_{\mu}\right)\left(\partial_{\nu}\mathcal{S} - eA_{\nu}\right) +\partial_{\mu}\mathfrak{J}^{\mu\nu}\left(\partial_{\nu}\mathcal{S}-eA_{\nu}\right) + \mathrm{V}
  \end{equation}
  with $\phi \longmapsto e^{-\alpha(x)}\phi$ as the behaviour of $\phi$ under the Weyl transformation.
  In addition to $\mathfrak{J}^{\mu\nu}$ being anti-symmetric , $\mathrm{V}$ and $\mathfrak{J}^{\mu\nu}$ may depend on $\phi$ in such a way that they are Weyl invariant but they can not depend on $\mathcal{S}$ and $A_{\mu}$.
  This means that the electromagnetic current can always be written in the form
  \begin{equation}\label{CurrentDensity}
     \boxed{ \mathfrak{J}^{\mu} = -e\phi^2\eta^{\mu\nu}\left(\partial_{\nu}\mathcal{S} - eA_{\nu}\right) -e\partial_{\nu}\mathfrak{J}^{\nu\mu}}.
  \end{equation}
  In line with \eqref{JisConjugateMomentaOfS}, the conjugate momenta of $\mathcal{S}$ is the momentum density of matter particles. Thus $\partial_{\nu}\mathfrak{J}^{\nu\mu}$ does not contribute to linear momentum since it is a total derivative. However; the angular momentum density is
  \begin{equation}\label{AngularMomentumDensity}
          (\mathcal{J}^{0})^{\mu\nu} = \phi^2\left[x^{\mu}\left(\partial^{\nu}\mathcal{S} - eA^{\nu}\right) -x^{\nu}\left(\partial^{\mu}\mathcal{S} - eA^{\mu}\right)\right] + 2\mathfrak{J}^{\mu\nu}
  \end{equation}
  if we ignore the total derivative term. $\mathfrak{J}^{\mu\nu}$ is the part of angular momentum density independent of spacetime coordinates therefore it is spin angular momentum density.
  
  To demonstrate that $\mathfrak{J}^{\mu}$ is indeed the general form of current in Maxwell's equations we consider the case where the sources are spin-0 and spin-$\frac{1}{2}$ particles. 
  For spin-$0$ particles, $\mathfrak{J}^{\mu\nu} = 0$.
  In this case we have
  \begin{equation}
      \begin{aligned}
          \mathfrak{J}_{\mu} &= -e\phi^2\left(\partial_{\mu}\mathcal{S} - eA_{\mu}\right)\\
          &= \frac{\imath e}{2}\left[\phi e^{\imath\mathcal{S}}\partial_{\mu}\left(\phi e^{-\imath\mathcal{S}}\right)-\phi e^{-\imath\mathcal{S}}\partial_{\mu}\left(\phi e^{\imath\mathcal{S}}\right) +2\imath eA_{\mu}\phi^2\right].       
      \end{aligned}
  \end{equation}
  Making the substitution $\psi = \frac{1}{\sqrt{2}}\phi e^{\imath\mathcal{S}}$, we get
  \begin{equation}
      \boxed{\mathfrak{J}_{\mu} = \imath e\left(\partial_{\mu}\psi^{*}\psi - \psi^{*}\partial_{\mu}\psi + 2 \imath e A_{\mu}\psi\psi^{*}\right)}
  \end{equation}
  which  is the current density in scalar electrodynamics. With $\psi = \frac{1}{\sqrt{2}}\left(\phi_{1}+\imath\phi_{2} \right) $ and $A_{\mu}=A_{\mu}^{3} $, we can rewrite the current density as
  \begin{equation}
      \begin{aligned}
          \mathfrak{J}^{\mu}_{3} &= -e\left(\delta^{k}_{i}\partial^{\mu} + e\varepsilon_{3 i}^{\hspace{9pt}k}A_{3}^{\mu} \right)\phi_{k}\varepsilon_{3}^{\hspace{4pt}ij}\phi_{j}\\
          &=\imath e\left(\mathfrak{D}^{\mu}\Phi^{\dagger} \right)\tau_{3}\Phi.
      \end{aligned}
  \end{equation}
   For spin-$\frac{1}{2}$, we replace $\psi$ by a spinor $\frac{1}{\sqrt{m}}\Psi$ and  set $\mathfrak{J}^{\mu\nu} =\frac{1}{m} \bar{\Psi}\sigma^{\nu\mu}\Psi$.
This results in
\begin{equation}\label{FermionCUrrent}
    \mathfrak{J}^{\mu} = \frac{\imath e}{m}\eta^{\mu\nu}\left(\partial_{\nu}\bar{\Psi}\Psi - \bar{\Psi}\partial_{\nu}\Psi + 2 \imath e A_{\nu}\bar{\Psi}\Psi\right) + \frac{e}{m}\partial_{\nu}\left(\bar{\Psi}\sigma^{\nu\mu}\Psi\right).
\end{equation}
    where $\sigma^{\mu\nu} = \frac{\imath}{2}\left[\gamma^{\mu},\gamma^{\nu}\right]$, with $\gamma^{\mu}$ as Dirac matrices and $\left[\hspace{3pt},\hspace{3pt}\right]$ the commutator. Since $\gamma^\mu\gamma^\nu = \frac{1}{2}[\gamma^\mu,\gamma^\nu]_+ +\frac{1}{2}[\gamma^\mu,\gamma^\nu]$ and $\eta^{\mu\nu} =\frac{1}{2}[\gamma^\mu,\gamma^\nu]_+ $, where $[,]_+$ is an anti-commutator, assuming that the particles obey the Dirac equation we get
    \begin{equation*}
        \begin{split}
           \mathfrak{J}^{\mu} &= -\frac{2\imath e}{m}\left(\bar{\Psi}\eta^{\mu\nu}\partial_{\nu}\Psi - \imath e\bar{\Psi}\eta^{\mu\nu}A_{\nu}\Psi\right) +\frac{e}{m}\partial_{\nu}\left[\bar{\Psi}\left(\imath\eta^{\mu\nu} +\sigma^{\nu\mu}\right)\Psi\right]\\
           &= -\frac{\imath e}{m}\Big(\bar{\Psi}\gamma^{\mu}\not\!{\partial}\Psi + \bar{\Psi}\gamma^{\nu}\gamma^{\mu}\partial_{\nu}\Psi - \imath e[\gamma^{\mu},\gamma^{\nu}]_{+}A_{\nu}\Psi - \bar{\Psi}\overleftarrow{\not\!{\partial}}\gamma^{\mu}\Psi \\ &-\bar{\Psi}\gamma^{\nu}\gamma^{\mu}\partial_{\nu}\Psi \Big) \\
           &=-\frac{\imath e}{m}\bar{\Psi}\gamma^{\mu}\left(\not\!{\partial}-\imath e \not\!{A}\right)\Psi +\frac{\imath e}{m}\bar{\Psi}(\overleftarrow{\not\!{\partial}} +\imath e \not\!{A})\gamma^{\mu}\Psi\\
        \end{split}
    \end{equation*}
\begin{equation}\label{UsualFormFermionC}
    \boxed{\mathfrak{J}^{\mu}= 2e \bar{\Psi}\gamma^{\mu}\Psi}.
\end{equation}
With $\Psi = \frac{1}{\sqrt{2}}\left(\phi_{1}+\imath\phi_{2} \right) $ and $A_{\mu}=A_{\mu}^{3} $, by making use of the fact that the Dirac equation can be written as
\begin{equation}
    \mathfrak{D}^{\nu}\Phi = \sigma^{\mu\nu}\mathfrak{D}_{\mu}\Phi -\imath m\gamma^{\nu}\Phi
\end{equation}
we can write the current density as
\begin{equation}
    \mathfrak{J}_{3}^{\mu} =2e\bar{\Phi}\gamma^{\mu}\tau_{3}\Phi.
\end{equation}
This shows that electromagnetism can be thought of as Yang-Mills theory with the gauge potential having a single non zero component.
This prompts an immediate generalization to non-Abelian gauge theories which proceeds as follows:
For non Abelian gauge theories the gauge potential transform in the following way under the gauge transformation
\begin{equation}
    \begin{aligned}
    A_{\mu}^{i} \longmapsto  \Tilde{A}_{\mu}^{i} = A_{\mu}^{i}  - \partial_{\mu}\epsilon^{i}- g \mathrm{C}_{jk}^{i}A_{\mu}^{k}\epsilon^{j}.
    \end{aligned}
\end{equation}
We can re-write this equation as
\begin{equation}
   \Tilde{A}_{\mu}^{i}  = A_{\mu}^{i} - \mathfrak{D}_{\mu}\epsilon^{i}
\end{equation}
with  $\delta_{j}^{i}\mathfrak{D}_{\mu} = \delta_{j}^{i}\partial_{\mu} + g\mathrm{C}_{jk}^{i} A_{\mu}^{k}$ . Thus the expression
\begin{equation}
    A_{\mu}^{i} - \frac{1}{g}\mathfrak{D}_{\mu}\mathcal{S}^{i}
\end{equation}
with the behaviour of $\mathcal{S}^{i}$ under the gauge transformation as
\begin{equation}\label{YMmatters}
    \mathcal{S}^{i} \longmapsto \mathcal{S}^{i} - g\epsilon^{i}
\end{equation}
is gauge invariant.
From this we conclude that the gauge invariant Lagrangian for non-Abelian gauge theories is
\begin{equation}
  \mathfrak{L} = -\frac{1}{4}\mathfrak{F}_{\mu\nu}^{i}\mathfrak{F}^{\mu\nu}_{i} + \mathfrak{J}_{i}^{\mu}\left(A_{\mu}^{i} -\frac{1}{g}\mathfrak{D}_{\mu}\mathcal{S}^{i}\right).  
\end{equation}
Equation \eqref{YMmatters} makes it clear that $e^{\imath\tau\cdot\mathcal{S}}$ (where $[\tau_{i},\tau_{j}] = \imath\mathrm{C_{ij}^{k}\tau_{k}}$) is gauge covariant. Thus the general form of a gauge covariant field is
\begin{equation}
    \Psi = |\Psi| e^{\imath\tau\cdot\mathcal{S}}.
\end{equation}
Since each column of $\Psi$ is gauge covariant one can consider the columns $(\psi_{i})$ of $\Psi$ as matter fields.

  \subsection{4D Gravity as a consequence of Weyl Invariance}\label{WeylImplyGravity}
  Due to the fact that it will later be evident that in order for us to get results which do not contradict observation we have to revise some assumptions of general relativity, we believe it would be to our advantage if we could show that a consistent theory of electromagnetism leads to the same results. To achieve this, we construct the complete Lagrangian of the theory of electromagnetism which respect all symmetries of the theory.

  Since $\phi$ is dynamical, the complete Lagrangian for the theory of electromagnetism in the presence of matter must include its kinetic term.
  The kinetic term is not  Weyl invariant. In fact under the infinitesimal Weyl transformation  we get
  \begin{equation}
    \begin{split}
      \sqrt{\gamma}\gamma^{\mu\nu}\partial_{\mu}\phi\partial_{\nu}\phi &\longmapsto \left(1 +\alpha\right)^{2}\sqrt{\gamma}\gamma^{\mu\nu}\partial_{\mu}\left(\phi - \alpha\phi\right)\partial_{\nu}\left(\phi - \alpha\phi\right)\\
      &\longmapsto \sqrt{\gamma}\gamma^{\mu\nu}\left(\partial_{\mu}\phi\partial_{\nu}\phi - 2\phi\partial_{\mu}\phi\partial_{\nu}\alpha\right)\\
       &\longmapsto \sqrt{\gamma}\gamma^{\mu\nu}\partial_{\mu}\phi\partial_{\nu}\phi +\phi^2\partial_{\mu}\left(\sqrt{\gamma}\gamma^{\mu\nu}\partial_{\nu}\alpha\right) \\ &-\partial_{\mu}\left(\phi^2\sqrt{\gamma}\gamma^{\mu\nu}\partial_{\nu}\alpha\right)
      \end{split}
  \end{equation}
  Thus under Weyl transformation, the variation of the kinetic term is
  \begin{equation}
      \delta\left(\sqrt{\gamma}\gamma^{\mu\nu}\partial_{\mu}\phi\partial_{\nu}\phi\right) = \phi^2\partial_{\mu}\left(\sqrt{\gamma}\gamma^{\mu\nu}\partial_{\nu}\alpha\right)-\partial_{\mu}\left(\phi^2\sqrt{\gamma}\gamma^{\mu\nu}\partial_{\nu}\alpha\right)
  \end{equation}
  while that of the total derivative term $\partial_{\mu}\left(\phi\sqrt{\gamma}\gamma^{\mu\nu}\partial_{\nu}\phi\right)$ is
\begin{equation}
    \begin{split}
   \delta\partial_{\mu}\left(\phi\sqrt{\gamma}\gamma^{\mu\nu}\partial_{\nu}\phi\right) =&\partial_{\mu}\left[\delta\left(\phi\sqrt{\gamma}\gamma^{\mu\nu}\right)\partial_{\nu}\phi\right] +\partial_{\mu}\left(\phi \sqrt{\gamma}\gamma^{\mu\nu}\partial_{\nu}\delta\phi\right)\\
    =& - \partial_{\mu}\left(\phi^2\sqrt{\gamma}\gamma^{\mu\nu}\partial_{\nu}\alpha\right).
   \end{split}
\end{equation}
As
\begin{equation}
    \phi\sqrt{\gamma}\Box\phi = \partial_{\mu}\left(\phi\sqrt{\gamma}\gamma^{\mu\nu}\partial_{\nu}\phi\right)-\left(\sqrt{\gamma}\gamma^{\mu\nu}\partial_{\mu}\phi\partial_{\nu}\phi\right),
\end{equation}
we have 
\begin{equation}
   \boxed{ \delta\left(\sqrt{\gamma}\phi\Box\phi\right)  =-\phi^2\sqrt{\gamma}\Box\alpha}
\end{equation}
with $\Box = \nabla_{\mu}\nabla^{\mu}$.
  Under the Weyl transformation the variation of the Ricci scalar is
  \begin{equation*}
    \begin{split}
      \delta(\sqrt{\gamma}\phi^2\mathfrak{R}^{(g)}) &=\sqrt{\gamma}\phi^2\gamma^{\mu\nu}\delta\mathfrak{R}_{\mu\nu}^{(g)}\\
      &=\sqrt{\gamma}\phi^2\left(\nabla^{\mu}\nabla^{\nu}-\gamma^{\mu\nu}\nabla^{2}\right)\delta \gamma_{\mu\nu}\\
      \end{split}
  \end{equation*}
\begin{equation}
    \boxed{      \delta(\sqrt{\gamma}\phi^2\mathfrak{R}^{(g)})=-6\sqrt{\gamma}\phi^{2}\Box{\alpha}.}
\end{equation}
 Accordingly
 \begin{equation}
    \sqrt{\gamma}\left(\frac{1}{6}\mathfrak{R}^{(g)}\phi^2 -\phi\Box\phi\right)  
 \end{equation}
 is Weyl invariant. In addition to this we note that the term $\sqrt{\gamma}\phi^{4}$  is also Weyl invariant. Since
 \begin{equation}
     \phi\Box\phi = \frac{1}{2}\Box\phi^2 -\gamma^{\mu\nu}\partial_{\mu}\phi\partial_{\nu}\phi,
 \end{equation}
 the most general yet simple Lagrangian for a consistent theory of electromagnetism is:
 \begin{equation}\label{completeElectromagnetism}
    \boxed{\begin{aligned}   
     \mathfrak{L} =& -\frac{1}{4}\hat{\mathrm{F}}_{\mu\nu}\mathrm{F}^{\mu\nu} + \frac{1}{12}\phi^2\mathfrak{R}^{(g)} -\frac{1}{4}\Box\phi^2 +\frac{1}{2}\gamma^{\mu\nu}\partial_{\mu}\phi\partial_{\nu}\phi + \\ &\frac{1}{2} \phi^2 \gamma^{\mu\nu}\left(\partial_{\mu}\mathcal{S} - eA_{\mu}\right)\left(\partial_{\nu}\mathcal{S} - eA_{\nu}\right)+\frac{\zeta^2 e^2}{24}\phi^{4} +\\ &\frac{1}{\sqrt{\gamma}}\partial_{\mu}\left(\sqrt{\gamma}\mathfrak{J}^{\mu\nu}\right)\left(\partial_{\nu}\mathcal{S}-eA_{\nu}\right).
     \end{aligned}}
 \end{equation}
 Where the constant in front of the $\phi^{4}$ term is added for later convenience. In our attempt to construct the consistent theory of electromagnetism, we are naturally led to the theory of electromagnetism, gravity together with a scalar field in the presence of matter. This is not Einstein gravity due to the non-minimal coupling to gravity. However; if the solution to Einstein field equations is $g^{\mu\nu} = \frac{12}{\lambda^2\phi^2}\gamma^{\mu\nu} $,
         the Ricci scalar in the Einstein field equations is given by
         \begin{equation}
             \frac{\sqrt{g}}{\lambda^2}\mathfrak{R} = \frac{\sqrt{\gamma}\phi^2}{12}\left(\mathfrak{R}^{(g)} - 6\phi^{-1}\Box\phi \right).
         \end{equation}
   The Lagrangian becomes     
         \begin{equation}
    \begin{aligned}   
     \mathfrak{L} =&\sqrt{g}\big[ -\frac{1}{4\zeta^2}{\mathrm{F}}_{\mu\nu}{\mathrm{F}}^{\mu\nu} + \frac{1}{\lambda^2}\mathfrak{R} \\ & +  \frac{6e^2}{\lambda^2} g^{\mu\nu}\left({A_{\mu}-\frac{1}{e}\partial_{\mu}\mathcal{S}} \right)\left({A_{\nu}-\frac{1}{e}\partial_{\nu}\mathcal{S}}\right)  + 6\frac{\zeta^2 e^2}{\lambda^4} +\frac{e}{2}{\mathfrak{J}}^{\mu\nu}\mathrm{F}_{\mu\nu}\big].
     \end{aligned}
     \end{equation}
 From here it is clear that we have the vacuum energy density of the order of $M_{pl}^4$. When $g^{\mu\nu} =\eta^{\mu\nu} $, introducing the gauge invariant potential $B_{\mu}=A_{\mu}-e^{-1}\partial_{\mu}\mathcal{S}$, makes it evident that for $\mathfrak{J}^{\mu\nu}=0$ we are dealing with superconductivity. Calculating the effective mass makes it clear that we are dealing with particles whose mass are of the order of Planck mass. We can get around this problem of assigning the wrong mass to elementary particles if  the solution to the Einstein field equations is conformally flat in the absence of gravity. This establishes the scalar field $\phi$ as physical. In the following section it will be clear that $\phi$ is the Higgs field.
\subsection{Electromagnetism and Gravity from 5D Spacetime}\label{ElectroGrav5D}

 We intend to show that the Lagrangian \eqref{completeElectromagnetism} follows from the reduction of the $5\mathcal{D}$ Lagrangian. It should be noted that the covariant derivative $\nabla$ is that of $4\mathcal{D}$ spacetime in all its instances.
 In accordance with our claim that matter and curvature of spacetime are different ways of viewing the same thing, the $5\mathcal{D}$ Ricci scalar for spin-0 particles must be given by
 \begin{equation}\label{5DRScalarOMatter}
     \frac{1}{12}\Tilde{\mathfrak{R}} = -\frac{1}{2}\mathrm{G}^{ab}\partial_{a}\mathcal{S}\partial_{b}\mathcal{S}.
 \end{equation}
 for a general $5\mathcal{D}$ metric $\mathrm{G}_{ab}$. This is the Hamilton-Jacobi equation with the Ricci scalar playing the role of mass term. However; in this case we think of the 1-form $\partial_{\mu}\mathcal{S}$ as the momentum density in line with the results of  \autoref{4delectrodynamics}. 
 This means that the action $\mathfrak{A}$ of $5\mathcal{D}$ gravity is
 \begin{equation}\label{5daction}
     \mathfrak{A} = \int d^{5}x\sqrt{G}\big(\frac{1}{12}\Tilde{\mathfrak{R}} + \frac{1}{2}\mathrm{G}^{ab}\partial_{a}\mathcal{S}\partial_{b}\mathcal{S}\big).
 \end{equation}
 
 As such the $5\mathcal{D}$ Einstein equations are
 \begin{equation}
     \Tilde{\mathfrak{R}}^{ab} - \frac{1}{2}\mathrm{G}^{ab}\Tilde{\mathfrak{R}} = -{6}\left(\mathrm{G}^{ac}\partial_{c}\mathcal{S}\mathrm{G}^{bd}\partial_{d}\mathcal{S} -\frac{1}{2}\mathrm{G}^{ab}\mathrm{G}^{cd}\partial_{c}\mathcal{S}\partial_{d}\mathcal{S}\right).
 \end{equation}
 The general $5\mathcal{D}$ metric can always be brought to the form \eqref{general5Dmetric}. For a universe in which electromagnetic force exists, \eqref{general5Dmetric} has to be accompanied by the Killing equation \eqref{abelKillingE} since only $5\mathcal{D}$ manifolds with at least one translation symmetry result in effects consistent with electromagnetism. Now we consider performing a gauge transformation such that the dependence on the extra dimension lies in $\chi$
 only. That is
 \begin{equation}
     \begin{split}
         A_{\mu}(x^{\alpha},x^{5}) &= A_{\mu}(x^{\alpha}) + \partial_{\mu}\chi(x^{\alpha},x^{5})\\
         \mathcal{S}(x^{\alpha},x^{5}) &=\mathcal{S}(x^{\alpha}) + e\chi(x^{\alpha},x^{5}) 
     \end{split}
 \end{equation}
 From $\mathrm{F}_{5\mu} = 0 $ we get $A_{5} = \partial_{5}\chi$ which in turn implies that
 \begin{equation}\label{A5Eq}
     \partial_{5}\mathcal{S} = e A_{5}.
 \end{equation}
 From this we get
 \begin{equation}
     \mathrm{G}^{\mu a}\partial_{a}\mathcal{S} = g^{\mu\nu}\big(\partial_{\nu}\mathcal{S} - e A_{\nu}\big).
 \end{equation}
 Since 
 \begin{equation}
     \Tilde{\mathfrak{R}}^{\mu}_{5} = -{6}\mathrm{G}^{\mu a}\partial_{a}\mathcal{S}\partial_{5}\mathcal{S},
 \end{equation}
 and in the absence of gravity $4\mathcal{D}$ spacetime is Minkowski spacetime, for \eqref{5DcurvSource4Dem} to be consistent with $4\mathcal{D}$ Maxwell's equations $g^{\mu\nu}$ must be conformal to the solution of $4\mathcal{D}$ equations of motion for the gravitational field $\gamma^{\mu\nu}$. That is 
 \begin{equation}
     g^{\mu\nu} = \frac{12}{\lambda^{2}\phi^2} \gamma^{\mu\nu}.
 \end{equation}
 The $5\mathcal{D}$ Ricci curvature tensor is 
 \begin{equation}
     \begin{split}
         \Tilde{\mathfrak{R}}_{ab} &= \delta_{a}^{\mu}\delta_{b}^{\nu}\mathfrak{R}_{\mu\nu} +\frac{\lambda^2}{2\zeta^2}\left(A_{a}\nabla_{\rho}\mathrm{F}^{\hspace{4pt}\rho}_{b}+A_{b}\nabla_{\rho}\mathrm{F}^{\hspace{4pt}\rho}_{a}-\mathrm{F}_{a\rho}\mathrm{F}_{b}^{\hspace{4pt}\rho}\right) \\ &+\frac{\lambda^{4}}{4\zeta^4}\mathrm{F_{\mu\nu}}\mathrm{F}^{\mu\nu}A_{a}A_{b}.
     \end{split}
 \end{equation}
 and the $5\mathcal{D}$ Ricci scalar is
 \begin{equation}
     \Tilde{\mathfrak{R}} = \mathfrak{R}-\frac{\lambda^2}{4\zeta^2}\mathrm{F_{\mu\nu}}\mathrm{F}^{\mu\nu}.
 \end{equation}
 In terms of the Ricci scalar we have from the equations of motion of the gravitational field
 \begin{equation}
    \frac{1}{12} \Tilde{\mathfrak{R}} = \frac{1}{\lambda^2}\phi^{-2}\mathfrak{R}^{(g)}-\frac{6}{\lambda^2}\phi^{-3}\Box\phi -\frac{12}{4\zeta^2\lambda^2}\phi^{-4}\mathrm{F_{\mu\nu}}\hat{\mathrm{F}}^{\mu\nu} 
 \end{equation}
with the indices raised by $\gamma^{\mu\nu}$. Substituting this into \eqref{5daction}
we get
\begin{equation}
    \begin{aligned}
    \mathfrak{A} &= \int d^5x{\zeta^{-1}\lambda^3}A_{5}\sqrt{\gamma}\Big(\frac{\phi^2}{12}\mathfrak{R}^{(g)} -\frac{1}{2}\phi\Box\phi  - \frac{1}{4\zeta^2}\mathrm{F}_{\mu\nu}\hat{\mathrm{F}}^{\mu\nu}\\ &+ \frac{\lambda^2\phi^4}{24}\mathrm{G}^{ab}\partial_{a}\mathcal{S}\partial_{b}\mathcal{S}\Big).
    \end{aligned}
\end{equation}
Since the expression in the bracket and $\sqrt{\gamma}$ is independent of $x^5$ we have
\begin{equation}\label{Effective4Daction}
   \boxed{ \begin{aligned}
        \mathfrak{A} &= \int d^4x\sqrt{\gamma}\Big( -\frac{1}{4\zeta^2}\mathrm{F}_{\mu\nu}\hat{\mathrm{F}}^{\mu\nu} + \frac{1}{12}\phi^2\mathfrak{R}^{(g)} -\frac{1}{4}\Box\phi^2\\ & +\frac{1}{2}\gamma^{\mu\nu}\partial_{\mu}\phi\partial_{\nu}\phi + \frac{1}{2} \phi^2 \gamma^{\mu\nu}\left(\partial_{\mu}\mathcal{S} - eA_{\mu}\right)\left(\partial_{\nu}\mathcal{S} - eA_{\nu}\right)\\ &+\frac{\zeta^2 e^2}{24}\phi^{4}\Big)\int_{-\infty}^{\infty}dx^5\frac{\lambda^3}{\zeta}\partial_{5}\chi 
    \end{aligned}.}
\end{equation}
When
\begin{equation}\label{ConditionForReduction}
  \int_{-\infty}^{\infty}dx^5\frac{\lambda^3}{\zeta}\partial_{5}\chi  = 1  
\end{equation}
it is clear that the above action is that of a theory developed in \autoref{4delectrodynamics}. While in \eqref{completeElectromagnetism} the quartic interaction term is added by hand in the resulting effective action \eqref{Effective4Daction} the term arises naturally. It is clear that when \eqref{ConditionForReduction} is valid $A_{5}$ depends only on $x^{5}$. In this case, in addition to $A_{5}$, only $\mathcal{S}$ depends on the extra dimension. Thus unlike in Kaluza-Klein theory we do not get an infinite tower of massive tensor and vector fields when the extra dimension is compactified.

If we ignore contributions from the boundary, \eqref{Effective4Daction} is equivalent to 
\begin{equation}
    \begin{split}
    \mathfrak{A} &= \int d^4x\sqrt{\gamma}\Big( -\frac{1}{4\zeta^2}\mathrm{F}_{\mu\nu}\hat{\mathrm{F}}^{\mu\nu} + \frac{1}{12}\phi^2\mathfrak{R}^{(g)} +\frac{1}{2}\partial_{\mu}\phi\partial^{\mu}\phi\\& + \frac{1}{2} \phi^2 \gamma^{\mu\nu}\left(\partial_{\mu}\mathcal{S} - eA_{\mu}\right)\left(\partial_{\nu}\mathcal{S} - eA_{\nu}\right)+\frac{\zeta^2 e^2}{24}\phi^{4}\Big).
    \end{split}
\end{equation}
Making the substitution $\psi = \frac{1}{\sqrt{2}}\phi e^{\imath\mathcal{S}}$ we get
\begin{equation}
   \boxed{ \begin{aligned}
  \mathfrak{A} &=   \int d^4x\sqrt{\gamma}\Big( -\frac{1}{4\zeta^2}\mathrm{F}_{\mu\nu}\hat{\mathrm{F}}^{\mu\nu} +  \gamma^{\mu\nu}\left(\partial_{\mu}\mathcal{\psi^{\star}} - \imath eA_{\mu}\psi^{\star}\right)\\&\left(\partial_{\nu}\mathcal{\psi} +\imath eA_{\nu}\psi\right) +\frac{1}{6}\mathfrak{R}^{(g)}|\psi|^2  +\frac{1}{6}{\zeta^2 e^2}|\psi|^{4}\Big).
  \end{aligned}}
\end{equation}
 which is the action of an Abelian Higgs model for constant $\mathfrak{R}^{(g)}$      \cite{schwartz2014quantum}.

 From \eqref{Effective4Daction} we get the following equation of motion for $\phi$
 \begin{equation}\label{phiEOM}
    \Box\phi -  \gamma^{\mu\nu}\left(\partial_{\mu}\mathcal{S} - eA_{\mu}\right)\left(\partial_{\nu}\mathcal{S} - eA_{\nu}\right)\phi-\frac{1}{6}\mathfrak{R}^{(g)}\phi -\frac{1}{6}{\zeta^2 e^2}\phi^3 = 0.
 \end{equation}
 For constant $\phi$, Weyl invariance implies that $\mathfrak{R}^{(g)} = 0$ and we can have $\gamma^{\mu\nu} = \eta^{\mu\nu}$. In this case the relativistic Hamilton-Jacobi equation implies that
 $m^2 = \frac{1}{6}{\zeta^2e^2}\phi^{2}$. Thus the ground state of $\phi$ is
 \begin{equation}
     |\langle\phi\rangle| = {\sqrt{6}}\frac{m}{\zeta e}.
 \end{equation}
 We may write $\phi$ as
 \begin{equation}
     \phi(x) = {\sqrt{6}}\frac{m}{\zeta e} + \sigma(x).
 \end{equation}
 Substituting this in \eqref{Effective4Daction} and considering general relativity to be valid for constant $\phi(x)$, we get the action of general relativity and electromagnetism as
 \begin{equation}
     \begin{split}
    \mathfrak{A} = \int d^4x & \sqrt{\gamma}\Big( -\frac{1}{4\zeta^2}\mathrm{F}_{\mu\nu}\hat{\mathrm{F}^{\mu\nu}} + \frac{1}{2}\frac{m^2}{\zeta^2e^2}\mathfrak{R}^{(g)}  +\\&  \frac{3 m^2}{\zeta^2e^2}\gamma^{\mu\nu}\left(\partial_{\mu}\mathcal{S} - eA_{\mu}\right)\left(\partial_{\nu}\mathcal{S} - eA_{\nu}\right)+\frac{3 m^4}{2\zeta^2e^2}\Big).
    \end{split}
 \end{equation}
 Comparing with the Einstein-Hilbert action, we find that
 \begin{equation}
     m^2 = \frac{\alpha}{2} M_{pl}^2
 \end{equation}
 and the Cosmological constant is
 \begin{equation}
     \Lambda = \frac{3}{4}\alpha M_{pl}^2,
 \end{equation}
 where $\alpha$ is fine structure constant and $M_{pl}$ the Planck mass. Thus the theory describes particles whose mass are of the order of planck mass just as in Kaluza-Klein theory and the prediction of the cosmological constant is in contradiction with the observed value.
 It should be noted that this is not the problem of higher dimensional theories but an indication that some assumption of general relativity are inconsistent with  electromagnetism since in \autoref{WeylImplyGravity} we demonstrated that this Lagrangian follows from the symmetries of Electromagnetism.

 If we abandon the assumption that the Einstein field equations imply that in the absence of gravity spacetime is flat, then $\mathfrak{R}^{(g)}$ is not the Ricci scalar from the Einstein field equations, but is related to it by a conformal factor. Thus the Ricci scalar in the Einstein field equations is $\mathfrak{R} = \frac{\lambda^2m^2}{12\zeta^2e^2}\mathfrak{R}^{(g)}$ for some yet to be determined constants $\lambda$  and $\zeta$. Comparison with the Einstein-Hilbert action results in
 \begin{equation*}
     \frac{\lambda^2}{\zeta^2} = \frac{4e^2}{\alpha M_{pl}^2}.
 \end{equation*}
 The solution to the Einstein Field equations is then
 \begin{equation}
     g^{\mu\nu} = \frac{\alpha M_{pl}^2}{2 m^2}\gamma^{\mu\nu}
 \end{equation}
 with $\phi(x) = m +\sigma(x)$, $\lambda = \sqrt{\frac{2}{\alpha}}{M_{pl}^{-1}}$ and $\zeta = \frac{1}{\sqrt{2}}e^{-1}$. In this case the Cosmological constant is
 \begin{equation}
     \Lambda = \frac{1}{4\alpha}\frac{m^4}{m_{pl}^2}.
 \end{equation}
 
 We realize that in order for us to have a consistent theory of gravity and electromagnetism, which does not contradict experimental observations, general relativity has to be regarded as a theory which associate not only gravity but also a Higgs field to the curvature of spacetime. That is the Ricci scalar $\mathfrak{R}$ in the Einstein field equations can be decomposed into the sum of Ricci scalar of pure gravity $\mathfrak{R}^{(g)}$ and d'Alembertian of a Higgs field $\phi$ scaled by a conformal factor which is a function of $\phi$.
 \begin{equation}
    \frac{1}{2} \mathfrak{R} =  \frac{6}{\lambda^2}\phi^{-2}\mathfrak{R}^{(g)}-\frac{6^2}{\lambda^2}\phi^{-3}\Box\phi .
 \end{equation}
 To make contact with classical physics, we consider the case where $\gamma_{\mu\nu} = \eta_{\mu\nu}$ and we consider the theory to be in Minkowski i.e. raising and lowering of indices is performed by Minkowski metric and the connection utilized is that of flat spacetime. Since in this case $\mathfrak{R}^{(g)} = 0$, From Weyl invariance we must have $\Box\phi = 0$. From the equations of motion of $\phi$ we get
 \begin{equation}\label{EnergyMomentumConsrvation}
    \begin{aligned}
         0 &= \partial^{\mu}\phi\left[\phi\left(\partial\mathcal{S} - eA\right)^2 + \phi^{3}\right]\\
         &=\frac{1}{2}\partial^{\mu}\phi^2\left(\partial\mathcal{S} - eA\right)^2 +\partial^{\mu}(\frac{\phi^4}{4})
          \\&= - \partial^{\mu}\frac{\phi^{4}}{4} -\partial_{\nu}\left[\phi^2\left(\partial^{\nu}\mathcal{S}-eA^{\nu}\right)\left(\partial^{\mu}\mathcal{S}-eA^{\mu}\right)\right] +\hat{\mathrm{F}}^{\mu}_{\hspace{4pt}\nu}\mathfrak{J}^{\nu}\\
         &= -\partial_{\nu}\left(T^{\mu\nu}_{\phi} + T^{\mu\nu}_{M} + T^{\mu\nu}_{EM}\right).
         \end{aligned}
 \end{equation}
 Where 
 \begin{equation}
    \begin{split}
     T^{\mu\nu}_{\phi} &= \frac{\phi^4}{4}\eta^{\mu\nu},\\
     T^{\mu\nu}_{M} &= \phi^2\left(\partial^{\nu}\mathcal{S}-eA^{\nu}\right)\left(\partial^{\mu}\mathcal{S}-eA^{\mu}\right)
     \end{split}
 \end{equation}
 and  $T^{\mu\nu}_{EM}$ is the electromagnetic energy momentum tensor. Momentum of classical particles in the electromagnetic field has the form
 \begin{equation}
    P_{\mu} = \partial_{\mu}\mathcal{S}- eA_{\mu}. 
 \end{equation}
 Comparing $T^{\mu\nu}_{M}$ with the energy momentum tensor of a point particle at $x'$      \cite{weinberg1972gravitation}, we see that 
 \begin{equation}
     \phi^2(x') = P_{0}^{-1}\delta^{(3)}\left(x-x'\right).
 \end{equation}
 Thus at least in the classical limit 
 \begin{equation}
     \phi^2(x) = \frac{n(x)}{P_{0}} 
 \end{equation}
 where $n(x)$ is the number density. From this we conclude that when $\phi^2 = m^2$ we are not dealing with a single particle of mass $m$ but a uniform distribution of such  point particles that fills the entire space! Since the metric of spacetime is $g^{\mu\nu} =\frac{12}{\lambda^2\phi^{2}}\gamma^{\mu\nu}$ it is clear that in the absence of matter ($\phi = 0$) the notion of spacetime no longer makes sense since spacetime is defined as a pseudo-Riemannian manifold equipped with a metric which is none degenerate. This supports our assumption of equivalence between matter and spacetime.

 Substituting the conformally flat metric in the first line of \eqref{reduced5Dgeodesics} results in the following expression for the geodesic equations
 \begin{equation}\label{correctGeodesicE}
     \frac{d}{d\tau}\left(\phi^2\frac{dx^{\mu}}{d\tau}\right) - \partial^{\mu}\phi \phi\eta_{\alpha\beta}\frac{dx^{\alpha}}{d\tau}\frac{dx^{\beta}}{d\tau} +\frac{A_{a}}{\zeta^2}\frac{dx^{a}}{d\tau}\hat{\mathrm{F}}_{\nu}^{\hspace{4pt}\mu}\frac{dx^{\nu}}{d\tau} = 0.
 \end{equation}
 Requiring that for constant $\phi$ we get the Lorentz force law, we see that $\frac{A_{a}}{\zeta^2}\frac{dx^{a}}{d\tau} = e m $ and
\begin{equation}
 \phi^2\eta_{\alpha\beta}\frac{dx^{\alpha}}{d\tau}\frac{dx^{\beta}}{d\tau} = -m^2.
 \end{equation}
 With 
 \begin{equation}
     \xi_{a} = \frac{\lambda^2}{\zeta^2}A_{a},
 \end{equation}
 we have 
 \begin{equation}
     \xi_{a}\frac{dx^a}{d\tau} = {e m}{\lambda^2}
 \end{equation}
 and 
 \begin{equation}
     \xi_{a}\mathrm{G}^{ab}\partial_{b}\mathcal{S} = A_{5}^{-1}\partial_{5}\mathcal{S} = e.
 \end{equation}
 From this we conclude that
 \begin{equation}
     \frac{dx^{a}}{d\tau} = m \lambda^2\mathrm{G}^{ab}\partial_{b}\mathcal{S} + B^{a}
 \end{equation}
 with $\xi_{a}B^{a} = 0 $. When\[\Box\phi = \frac{1}{6}\mathfrak{R}^{(g)}\phi\] we have
 \begin{equation}
     \phi^{-2}\left(\partial\mathcal{S}-eA\right)^{2} = -1.
 \end{equation}
 Therefore in this case $B^{a} = 0$.
 When $\phi = \lambda^{-1}$, which would be the case if the Ricci curvature tensor of general relativity contains information regarding gravity only, we get $m = \lambda^{-1}$. This is twice the results one obtains from Kaluza-Klein theory. The reason for this is that in line with the equation of motion of $\phi$ in flat spacetime,  massless $5\mathcal{D}$ implies that $\phi = 0$ while for $\phi$ constant, the particles are massive in $5\mathcal{D}$ with  mass half the value in $4\mathcal{D}$.
 
 In terms of classical momentum we can write \eqref{correctGeodesicE} as
 \begin{equation}\
     \boxed{\frac{dP^{\mu}}{d\tau} +e\hat{\mathrm{F}}_{\nu}^{\hspace{4pt}\mu}\frac{dx^{\nu}}{d\tau} + \frac{m}{\phi}\partial^{\mu}\phi = 0}.
 \end{equation}
 The third term is the correction to Lorentz force Law. In the classical limit one expects this to be negligible since $\phi$ is slowly varying. However, at short distances its effects should be appreciable.

 In order to find the equations of motion of $\gamma_{\mu\nu}$ we make use of the fact that the equations of motion of $A_{\mu}$, $\phi$ and $\mathcal{S}$ are known and proceed as follows: Let $\mathfrak{E}_{\mu\nu}^{(g)}$ be the Einstein tensor for the metric $\gamma_{\mu\nu}$. We have
 \begin{equation}\label{curvedEnergyMomentumConserv}
     \begin{split}
        \nabla^{\mu}\left(\frac{\phi^2}{6}\mathfrak{E}_{\mu\nu}^{(g)}\right) =& \frac{1}{6}\left[\nabla_{\mu},\nabla_{\nu}\right]\nabla^{\mu}\phi^2 -\frac{1}{6}\mathfrak{R}^{(g)}\phi\nabla_{\nu}\phi \\
        =&\frac{1}{6}\nabla^{\mu}\left(\nabla_{\mu}\nabla_{\nu} - \gamma_{\mu\nu}\Box\right)\phi^2 -\\ &\left(\Box\phi - (\partial\mathcal{S}-eA)^2\phi -2\zeta^2e^2\phi^3\right)\nabla_{\nu}\phi\\
        =&\nabla^{\mu}\left[\frac{1}{6}\left(\nabla_{\mu}\nabla_{\nu} - \gamma_{\mu\nu}\Box\right)\phi^2-\nabla_{\mu}\phi\nabla_{\nu}\phi + \frac{1}{2}\gamma_{\mu\nu}(\nabla\phi)^2\right]\\ & - \left[(\partial\mathcal{S}-eA)^2\phi +2\zeta^2e^2\phi^3\right]\partial_{\nu}\phi\\
        =&\nabla^{\mu}\left(T_{\mu\nu}^{\phi} +T_{\mu\nu}^{EM}+T_{\mu\nu}^{M}\right).
     \end{split}
 \end{equation}
 In the second line of \eqref{curvedEnergyMomentumConserv} we made use of \eqref{phiEOM} while in the last line we used \eqref{EnergyMomentumConsrvation}.
 \begin{equation}
     \begin{split}
         T_{\mu\nu}^{\phi} &= -\nabla_{\mu}\phi\nabla_{\nu}\phi +\frac{\gamma_{\mu\nu}}{2}\left[(\nabla\phi)^2 +\frac{1}{2}\phi^{4}\right] +\frac{1}{6}\left(\nabla_{\mu}\nabla_{\nu} -\gamma_{\mu\nu}\Box\right)\phi^2,\\
         T_{\mu\nu}^{M} &=-\phi^2\left(\nabla_{\mu}\mathcal{S}-eA_{\mu}\right)\left(\nabla_{\nu}\mathcal{S}-eA_{\nu}\right) +\frac{\gamma_{\mu\nu}\phi^2}{2}\left(\nabla\mathcal{S}-eA\right)^2.
     \end{split}
 \end{equation}
 Therefore the equations of  motion of $\gamma_{\mu\nu}$ are
 \begin{equation}
     \frac{1}{6}\phi^2\mathfrak{E}^{(g)}_{\mu\nu} = T_{\mu\nu}^{\phi} + T_{\mu\nu}^{M} + T_{\mu\nu}^{EM}.
 \end{equation}
 We can show that the above equations are 
 \begin{equation}
     \frac{\delta \mathfrak{A}\left[\gamma^{\mu\nu}(x)\right]}{\delta\gamma^{\mu\nu}(y)} = 0
 \end{equation}
 by making use of
 \begin{equation*}
     \begin{split}
         \frac{\delta(\phi^2\sqrt{\gamma}\mathfrak{R}^{(g)})}{\delta\gamma^{\mu\nu}(y)} =& \phi^2\Big(\frac{\delta\sqrt{\gamma(x)}}{\delta\gamma^{\mu\nu}(y)}\mathfrak{R}^{(g)}(x) +\sqrt{\gamma}\frac{\delta\gamma^{\alpha\beta}(x)}{\delta\gamma^{\mu\nu}(y)}\mathfrak{R}^{(g)}_{\alpha\beta}(x) +\\ & \sqrt{\gamma}\gamma^{\alpha\beta}(x)\frac{\delta\mathfrak{R}^{(g)}_{\alpha\beta}(x)}{\delta\gamma^{\mu\nu}(y)}\Big)\\
         =&\sqrt{\gamma}\phi^2\left(\mathfrak{E}_{\mu\nu}^{(g)} - \nabla_{\mu}\nabla_{\nu} +\gamma_{\mu\nu}\Box\right)\delta^{(4)}(x-y)\\
     \end{split}
 \end{equation*}
 \begin{equation}
     \boxed{ \frac{\delta(\phi^2\sqrt{\gamma}\mathfrak{R}^{(g)})}{\delta\gamma^{\mu\nu}(y)}=\sqrt{\gamma}\delta^{(4)}(x-y)\left(\mathfrak{E}_{\mu\nu}^{(g)} - \nabla_{\mu}\nabla_{\nu} +\gamma_{\mu\nu}\Box\right)\phi^2.}
 \end{equation}
We derived the equations in this way to show that our knowledge of dynamics of matter fields is sufficient to determine the field equations of the theory of gravity. In fact, in terms of the Ricci scalar of general relativity, \eqref{phiEOM} is
\begin{equation}
    \boxed{\mathfrak{R} = -g^{\mu\nu}\left(\partial_{\mu}\mathcal{S}-eA_{\mu}\right)\left(\partial_{\nu}\mathcal{S}-eA_{\nu}\right) -6\alpha M_{pl}^2}.
\end{equation}
It is completely determined by the energy momentum tensor of matter fields and vacuum energy. 

In line with our postulate, we have $\mathfrak{L} = 0$ for fields obeying the equation of motion. Comparing $\mathfrak{L}$ with \eqref{phiEOM} we get
\begin{equation}\label{EMConstraint}
    -\frac{1}{4}\mathrm{F}_{\mu\nu}\hat{\mathrm{F}}^{\mu\nu} =\zeta^2\phi^{4}.
\end{equation}

To understand why we get this constraint  we express $5\mathcal{D}$ Maxwell's equations in terms of matter fields. With raising of indices done by $\mathrm{G}^{ab}$ we get
\begin{equation*}
    \begin{split}
        \partial_{a}\left(\sqrt{\mathrm{G}}\mathrm{F}^{ab}\right) &= \delta_{5}^{b}\partial_{a}\left(\sqrt{g}A_{5}\mathrm{G}^{a\mu}\mathrm{G}^{5\nu}\mathrm{F}_{\mu\nu}\right) +\delta_{\mu}^{b}\partial_{a}\left(\sqrt{g}A_{5}\mathrm{G}^{a\nu}\mathrm{F}_{\nu}^{\hspace{4pt}\mu}\right)\\
        &=-\delta_{5}^{b}\partial_{a}\left(\sqrt{g}\mathrm{G}^{a\nu}\mathrm{F}_{\nu}^{\hspace{4pt}\mu}A_{\mu}\right) +\delta_{\mu}^{b}\partial_{a}\left(\sqrt{g}A_{5}\mathrm{G}^{a\nu}\mathrm{F}_{\nu}^{\hspace{4pt}\mu}\right)\\
        &=-\delta_{5}^{b}\left[\partial_{\nu}\left(\sqrt{g}\mathrm{F}^{\nu\mu}\right)A_{\mu} +\frac{\sqrt{g}}{2}\mathrm{F}_{\mu\nu}\mathrm{F}^{\mu\nu}\right] \\&+\delta_{\mu}^{b}A_{5}\partial_{\nu}\left(\sqrt{g}\mathrm{F}^{\nu\mu}\right)\\
    \end{split}
\end{equation*}
\begin{equation}\label{Unconstrained5DEM}
    \boxed{\partial_{a}\left(\sqrt{\mathrm{G}}\mathrm{F}^{ab}\right)= \sqrt{g}\partial_{5}\mathcal{S}\mathrm{G}^{ba}\partial_{a}\mathcal{S} - \delta_{5}^{b}\frac{\sqrt{g}}{2}\left(\mathrm{F}_{\mu\nu}\mathrm{F}^{\mu\nu} + \frac{2\zeta^2e^2}{\lambda^2}\right).}
\end{equation}
$4\mathcal{D}$ electromagnetism constitute part of the $\mathfrak{J}^{5}$ component of the 5-current density. Since $4\mathcal{D}$ electromagnetism is the source of $5\mathcal{D}$ electromagnetism, it can be regarded as matter just like ordinary matter which is the source term in $4\mathcal{D}$ electrodynamics. Consequently, while in $4\mathcal{D}$ there is a clear distinction between ordinary matter and electromagnetic radiation (the former is regarded as particles while the latter is a force), in $5\mathcal{D}$ such  a distinction does not exist. They are both particles and there are no forces.   
Taking this into account, \eqref{5daction} gets modified to 
\begin{equation}
  \boxed{  \mathfrak{L} = \frac{1}{\lambda^2}\Tilde{\mathfrak{R}} -\frac{1}{4\zeta^2}\mathrm{F}_{ab}\mathrm{F}^{ab} +\frac{1}{\zeta^2}\Tilde{\mathfrak{R}}^{ab}A_{a}A_{b} + \frac{6}{\lambda^2}\mathrm{G}^{ab}\partial_{a}\mathcal{S}\partial_{b}\mathcal{S}}
    \end{equation}
One can easily show that the reduced Lagrangian does not change.

We show that the gauge transformation is a result of coordinate transformation as follows:
Since $\xi^{a}$ is a Killing vector field, the metric is invariant under the infinitesimal coordinate transformation
\begin{equation}
    x^{a}\longmapsto y^{a}= x^{a} + \epsilon\xi^{a}
\end{equation}
for constant $\epsilon$. This is a global coordinate transformation.  We can make it local if we let $\epsilon = \epsilon(x)$. It can be easily checked that the metric is no longer invariant, however the line element and all other scalars are still invariant. The transformation law of $A_{a}$ in this case is
\begin{equation*}
    \begin{split}
        A_{a}(x) \longmapsto & A_{b}(x - \epsilon\xi)\left(\delta_{a}^{b} -\epsilon\partial_{a}\xi^{b} - \xi^{b}\partial_{a}\epsilon\right)\\
        &=A_{a}(x) -\epsilon\left(\xi^{b}\partial_{b}A_{a} +\partial_{a}\xi^{b}A_{b}\right) -A_{b}\xi^{b}\partial_{a}\epsilon\\ 
    \end{split}
\end{equation*}
\begin{equation}
    \boxed{A_{a}(x) \longmapsto A_{a}(x) - \partial_{a}\epsilon}
\end{equation}
while that of $\mathcal{S}$ is
\begin{equation*}
    \begin{split}
    \mathcal{S}(x)\longmapsto &\mathcal{S}(x- \epsilon\xi)\\
    &= \mathcal{S}(x) -\epsilon\xi^{a}\partial_{a}\mathcal{S}\\
    \end{split}
\end{equation*}
\begin{equation}
    \boxed{\mathcal{S}(x)\longmapsto\mathcal{S}(x) - e\epsilon.}
\end{equation}
These are the Abelian gauge transformations proposed in \autoref{4delectrodynamics}.
Exponentiating  $\mathcal{S}$ we get
\begin{equation}
    e^{\imath\mathcal{S}} \longmapsto e^{\imath e\epsilon}e^{\imath\mathcal{S}}
\end{equation}
from which we conclude that $\hat{\Psi} =e^{\imath\mathcal{S}}$ is gauge covariant and so is 
\begin{equation}\label{gaugeCovDOMatter}
    \left( \partial_{\mu} -\imath e
    A_{\mu}\right)\hat{\Psi}.
\end{equation}
We note that 
\begin{equation}\label{ChargeEigenVE}
   \boxed{ \xi^{a}\partial_{a}\hat{\Psi} =\imath  e\hat{\Psi}}
\end{equation}
thus the Killing vector field $-\imath\xi$, which generates translations in the internal space, is the electric charge operator and $\hat{\Psi}$ is its eigenfunction. 
We can rewrite \eqref{GaugeCovariantEMLagrangian} as
\begin{equation}
    \mathfrak{L} = -\frac{1}{4}\mathrm{F}_{\mu\nu}\hat{\mathrm{F}}^{\mu\nu} +\mathfrak{J}^{\mu}\left(A_{\mu} +\frac{\imath}{e}\hat{\Psi}^{*}\partial_{\mu}\hat{\Psi}\right).
\end{equation}
As such in place of $\mathcal{S}$, we may use $\hat{\Psi} = e^{\imath\mathcal{S}}$ as matter field. We call $\hat{\Psi}$ written in this way a wavefunction. Since we can write $\hat{\Psi} = \hat{\phi}_{1} + \imath\hat{\phi}_{2}$ for some real functions $\hat{\phi}_{i}$, we can equivalently think of $\hat{\Psi}$ as a vector
\begin{equation}
  \hat{\Psi} =  \begin{bmatrix}
      \hat{\phi}_{1}\\
      \hat{\phi}_{2}.
  \end{bmatrix}
\end{equation}
In which case \eqref{gaugeCovDOMatter} becomes
\begin{equation}
    \left(\partial_{\mu}\delta^{i}_{k} + e\varepsilon^{i}_{\hspace{4pt}3k}A_{\mu}\right)\hat{\phi}^{k}
\end{equation}
where $A_{\mu}^{\hspace{4pt}i} =\delta^{i}_{3} A_{\mu}$. In which case it is evident that electromagnetism is the special case of non-Abelian gauge theory.

\section{Gravity and non-Abelian Gauge Theories}
Having established the theory of electromagnetism and gravity by considering the world to be $5\mathcal{D}$, we generalize the formalism to arbitrary dimensions in order to attain a geometric description of non-Abelian gauge theories. This entails increasing the number of internal dimensions and that of gauge fields from 1 to $n-4$. The resulting theory is that of $n-4$ neutral vector fields in  n-dimensions with curvature of n-dimensional spacetime as the source but in 4$\mathcal{D} $ spacetime it is the usual Yang-Mills theory. Just as in the case of electromagnetism, we use the equations following from the Lagrangian of these vector fields to determine the components of the metric. We proceed to show that the non-Abelian gauge transformation is the coordinate transformation. In addition to this, we show that the quantity $\mathrm{F}^{\mu k}_{\hspace{9pt}b}$ appearing in the Lagrangian is related to the Yang-Mills field strength tensor $\mathfrak{F}_{\nu b}^{k}$ by
\begin{equation}
    \mathrm{F}^{\mu k}_{b} = g^{\mu\nu}\mathfrak{F}_{\nu b}^{k}
\end{equation}
and the gauge covariant derivative $\mathfrak{D}^{\rho}$ is just the ordinary derivative in n-dimensional spacetime. That is,
\begin{equation}
    \mathfrak{D}^{\rho} = \mathrm{G}^{\rho a}\partial_{a}.
\end{equation}
It becomes clear that the self interaction terms arises from the non-vanishing Lie bracket (commutator) of the Killing vector fields and the dependence of the gauge potentials (the duals of the Killing vector fields) on the extra dimensions. Since the Killing vectors considered here are linearly independent (this is clear from the fact that the matrix formed from their components is invertible) and hence span the tangent space of the internal space, their non-commutativity is the clear indication that the internal space is curved. In short, we get the non-Abelian gauge theory if the killing fields do not commute. From this, it is clear that with the cylinder condition imposed it is impossible to have the non-Abelian gauge theory. The cylinder condition can be imposed if the Killing fields have a vanishing Lie bracket. This is because this condition is equivalent to formulating the theory in the coordinate system adapted to all the Killing vector fields (that is, the coordinate system in which the integral curves of the Killing vector fields are the coordinate lines) and such a coordinate system does not exist for non-commuting Killing vector fields. 
We end the \autoref{dimNTheory} by calculating the expressions for the components of the Ricci curvature tensor and showing that they are consistent with the equations that follow from the Lagrangian of the vector fields from which  we got the expression for the metric. These then lead us to the Lagrangian of the theory of gravity and non-Abelian gauge theories.

In \autoref{standardM} we construct the unified theory of gravity and the gauge Bosons of the standard model of particle physics. From \autoref{dimNTheory} it will be clear that once the linearly independent Killing vector fields with the same commutation relations as the Lie algebra of the group in question are known, the geometric description of the gauge theory is achieved. This is because the matrix of Killing vector fields whose components  $\xi_{i}^{\hspace{4pt}j}$ are defined by $\xi_{i}=\xi_{i}^{\hspace{4pt}j}\partial_{j}$ can be constructed. The inverse of this matrix is the matrix of 1-forms whose components are
$A_{ij} = \frac{\zeta^2}{\lambda^2}g_{ik}\xi_{j}^{\hspace{4pt}k}$. This matrix is then used to construct the metric of the internal space since $g_{ik}=\frac{\lambda^2}{\zeta^2}A_{i}^{\hspace{4pt}j}A_{kj}$.
In constructing the electroweak theory, we start with the theory of two Abelian gauge Bosons. This helps us guess the metric of the internal space. We motivate why such  a choice is reasonable and find the Killing vector fields from which  that metric is constructed. We show that the Lie bracket of these Killing field is that of $\mathfrak{u}(1)\times\mathfrak{su}(2)$. We use the equations of conservation of charge to construct the gauge covariant derivative. With all of the aforementioned quantities known the construction of electroweak theory is complete.

In order to facilitate construction of the complete theory, we provide a general way of finding the set of linearly independent Killing vector fields that span the tangent space of any group manifold. This makes it evident that the semi-simple groups should be realized as sub-manifolds of non-semi-simple groups obtained by imposing constraints which break some of the symmetries of non-semi-simple groups. We end \autoref{standardM} by constructing the set of linearly independent Killing vector fields for the standard model gauge group
$U(1)\times SU(2)\times SU(3)$.

\subsection{n-dimensional metric and equations of motion}\label{dimNTheory}
The Lagrangian \eqref{matterAsCurvedSpaceTimeL} generalizes to 
\begin{equation}\label{nonAbelMatterAsCurvedL}
    \mathfrak{L} = -\sqrt{\mathrm{G}}\frac{1}{4}\mathrm{F}_{ab}\cdot\mathrm{F}^{ab} +\sqrt{\mathrm{G}}\Tilde{\mathfrak{R}}^{ab}A_{a}\cdot A_{b}
\end{equation}
where $\mathrm{F}_{ab}^{k} = \partial_{a}A_{b}^{\hspace{4pt}k}-\partial_{b}A_{a}^{\hspace{4pt}k}$ and $A_{a}\cdot A_{b} =  \delta_{ij} A_{a}^{\hspace{4pt}i}A_{b}^{\hspace{4pt}j}$. While $a,b,\cdots, h$ designate n-dimensional quantities,  $i,j,\cdots,z$ run from 5 to $n$ in $n$-dimensional spacetime. It is clear that $\cdot$ means that the summation runs over the internal indices only. It can be easily confirmed that $A_{a}\cdot A_{b} = tr(A_{a}^{i}\tau_{i}\tau_{j}A_{b}^{j})$ with the matrices $\tau_{i}$ normalised so that $tr(\tau_{i}\tau_{j}) = \delta_{ij}$, thus our notation is equivalent to the notation in the usual construction of non-Abelian gauge theories. It should also be noted that $\lbrace A_{b}^{i}\rbrace$ can be treated as components of an $(n-4)\times n$- matrix. 

We see that in higher dimensional spacetime this is just the theory of $n$-uncoupled vector fields with curvature as their source.  Just as it was the case in the Abelian theory, the resulting equations of motion imply that the set of vector fields $\lbrace A^{i a} = \mathrm{G}^{ab}A_{b}^{i}\rbrace$ where $i$ is a label, are Killing vector fields.
Following the same procedure as in the Abelian case to construct the metric, we end up with
\begin{equation}\label{nonAbelMetricN}
    \boxed{\begin{aligned}
    \mathrm{G}_{ab} &= \delta_{a}^{\mu}\delta_{b}^{\nu}g_{\mu\nu} + \frac{\lambda^2}{\zeta^2}A_{a}\cdot A_{b}\\
    \mathrm{G}^{\mu\nu} &= g^{\mu\nu}\\
    \mathrm{G}^{ai} &= \delta^{a}_{k}\xi^{k}\cdot\xi^{i}\frac{\zeta^2}{\lambda^2} - \mathrm{G}^{a\mu}A_{\mu}\cdot\xi^{i}
    \end{aligned}}
\end{equation}
with $\xi^{i}\cdot A_{k} = \delta_{k}^{i}$. We note that 
\[\xi^{i}_{\hspace{4pt}b} = \frac{\lambda^2}{\zeta^2}A_{b}^{i}. \] Thus $\lbrace\xi^{ia}= \frac{\lambda^2}{\zeta^2}\mathrm{G}^{ab}A_{b}^{i}\rbrace$ are Killing vectors in $n$-dimensional spacetime while $\lbrace\xi^{ik}\rbrace$ can be regarded as Killing vectors in the internal space. As the internal indices are raised and lowered by $\delta_{ij}$, it does not matter if we make them subscripts or superscripts.

The Lie derivative of the  Killing vector $\xi_{i}$ along the Killing vector $\xi_{j}$ is
\begin{equation}
    L_{\xi_{i}}\xi_{j}^{\hspace{4pt}k} = \xi_{i}^{\hspace{4pt}l}\partial_{l}\xi^{\hspace{4pt}k}_{j} -\xi_{j}^{\hspace{4pt}l}\partial_{l}\xi^{\hspace{4pt}k}_{i}.
\end{equation}
Evidently this is a Lie bracket. Since the Killing vectors are closed under Lie bracket
\begin{equation}\label{LieKVector}
 L_{\xi_{i}}\xi_{j}^{\hspace{4pt}k} =  \sigma\mathrm{C}^{l}_{\hspace{4pt}ij}\xi_{l}^{\hspace{4pt}k} 
\end{equation}
with $\mathrm{C}^{l}_{\hspace{4pt}ij}$ as the structure constants and $\sigma$ is the coupling constant. Using the relation between $\xi^{i}_{\hspace{4pt}k}$ and $A^{k}_{\hspace{4pt}l}$ we get
\begin{equation}
    \begin{split}
    g\mathrm{C}^{k}_{\hspace{4pt}mn}A_{j}^{\hspace{4pt}n}A_{i}^{\hspace{4pt}m} &=\left(\xi_{m}^{\hspace{4pt}l}\partial_{l}\xi^{\hspace{4pt}p}_{n} -\xi_{n}^{\hspace{4pt}l}\partial_{l}\xi^{\hspace{4pt}p}_{m}\right)A_{j}^{\hspace{4pt}n}A_{p}^{\hspace{4pt}k}A_{i}^{\hspace{4pt}m}\\
    &= - \left(\xi_{m}^{\hspace{4pt}l}\xi_{n}^{\hspace{4pt}p}\partial_{l}A_{p}^{\hspace{4pt}k} -\xi_{n}^{\hspace{4pt}l}\xi_{m}^{\hspace{4pt}p}\partial_{l}A_{p}^{\hspace{4pt}k}\right)A_{j}^{\hspace{4pt}n}A_{i}^{\hspace{4pt}m} \\
     &= - \left(\xi_{m}^{\hspace{4pt}l}\partial_{l}A_{j}^{\hspace{4pt}k} -\xi_{m}^{\hspace{4pt}p}\partial_{j}A_{p}^{\hspace{4pt}k}\right) A_{i}^{\hspace{4pt}m}\\
     &= \partial_{j}A_{i}^{\hspace{4pt}k}-\partial_{i}A_{j}^{\hspace{4pt}k}.
    \end{split}
\end{equation}
With  
\begin{equation}
   \boxed{ \mathfrak{F}_{ab}^{k} = \partial_{a}A_{b}^{\hspace{4pt}k}-\partial_{b}A_{a}^{\hspace{4pt}k} +\sigma\mathrm{C}^{k}_{\hspace{4pt}mn}A_{a}^{\hspace{4pt}m}A_{b}^{\hspace{4pt}n}}
\end{equation}
we realise that the Killing equation demands that
\begin{equation}\label{IntegrabilityCs}
    \mathfrak{F}_{aj}^{k} = 0,
\end{equation}
which is the generalization of $\mathfrak{F}_{a5} = 0$ of the Abelian theory. Clearly, \eqref{IntegrabilityCs} are the integrability conditions of $\xi^{i}\cdot A_{k} = \delta_{k}^{i}$ found in      \cite{eisenhart2003continuous,landau2013classical}. It should be noted that $\mathfrak{F}_{ab}^{k}$ is not  $\mathrm{F}_{ab}^{k}$ which appears in the higher dimensional lagrangian \eqref{nonAbelMatterAsCurvedL}  but is related to it by
\begin{equation}
    \mathfrak{F}_{ab}^{k} = \mathrm{F}_{ab}^{k}+\sigma\mathrm{C}^{k}_{\hspace{4pt}mn}A_{a}^{\hspace{4pt}m}A_{b}^{\hspace{4pt}n}.
\end{equation}
We show that the  non-Abelian gauge transformation is also the result  of coordinate transformation as follows;
$\mathrm{G}_{ab}$ is invariant under the global coordinate transformation
\begin{equation}
    x^{a} \longmapsto y^{a} = x^{a} + \epsilon^{i}\xi_{i}^{\hspace{4pt}a}
\end{equation}
for constant $\epsilon$. When $\epsilon^{i} = \epsilon^{i}(x)$, one can show that the line element is still invariant. The gauge potential undergoes the following transformation
\begin{equation*}
    \begin{split}
    A_{a}^{\hspace{4pt}k}(x)\longmapsto &A_{b}^{k}(x-\epsilon\cdot\xi)\left(\delta_{a}^{b} - \epsilon\cdot\partial_{a}\xi^{b} - \partial_{a}\epsilon\cdot\xi^{b}\right)\\
    &=A_{a}^{k}(x) -\epsilon\cdot\left(\xi^{b}\partial_{b}A_{a}^{\hspace{4pt}k} +\partial_{a}\xi^{b}A_{b}^{\hspace{4pt}k}\right) -A_{b}^{\hspace{4pt}k}\xi^{\hspace{4pt
    }b}_{i}\partial_{a}\epsilon^{i}\\
    &=A_{a}^{k}(x) -\epsilon\cdot\xi^{b}\left(\partial_{b}A_{a}^{\hspace{4pt}k} -\partial_{a}A_{b}^{\hspace{4pt}k}\right) -\partial_{a}\epsilon^{k}\\
    &=A_{a}^{k}(x) + \sigma\epsilon\cdot\xi^{b}\mathrm{C}^{k}_{\hspace{4pt}mn}A_{a}^{\hspace{4pt}m}A_{b}^{\hspace{4pt}n} -\partial_{a}\epsilon^{k}.
    \end{split}
\end{equation*}
\begin{equation}
    \boxed{ A_{a}^{\hspace{4pt}k}(x)\longmapsto  A_{a}^{k}(x) + \sigma\epsilon^{n}\mathrm{C}^{k}_{\hspace{4pt}mn}A_{a}^{\hspace{4pt}m}-\partial_{a}\epsilon^{k}}
\end{equation}
which is the non-Abelian gauge transformation. 
In terms of the matrices $\tau_{i}$ with the Lie commutator $[\tau_{i},\tau_{j}] = \imath\mathrm{C}_{\hspace{4pt}ij}^{k}\tau_{k} $ we can write the above transformation as
\begin{equation}
 \tau\cdot A_{a}\longmapsto (\mathbb{I} +\sigma\epsilon\cdot\tau) \tau\cdot A_{a}(\mathbb{I}-\sigma\epsilon\cdot\tau) -\frac{i}{\sigma}\partial_{a}(\mathbb{I}+\sigma\epsilon\cdot\tau) (\mathbb{I}-\sigma\epsilon\cdot\tau). 
\end{equation}
As such for a finite coordinate transformation we have
\begin{equation}\label{UsualATrans}
   \boxed{ \tau\cdot A_{a}\longmapsto \mathrm{U}\tau\cdot A_{a}\mathrm{U}^{-1} -\frac{\imath}{\sigma}\partial_{a}\mathrm{U}\mathrm{U}^{-1}}
\end{equation}
where $\mathrm{U} = e^{\imath\epsilon\cdot\tau}$.
$\xi_{k}^{\hspace{4pt}a}\mathfrak{F}_{a\nu}^{j} = 0 $ and 
\begin{equation}
    \begin{aligned}
        L_{\xi_{l}}\mathfrak{F}^{k}_{\mu\nu} &= \xi_{l}^{\hspace{4pt}a}\partial_{a}\mathfrak{F}^{k}_{\mu\nu} +\partial_{\mu}\xi_{l}^{\hspace{4pt}a}\mathfrak{F}_{a\nu}^{k}+\partial_{\nu}\xi_{l}^{\hspace{4pt}a}\mathfrak{F}_{a\mu}^{k}\\
        &=\xi_{l}^{\hspace{4pt}a}\partial_{a}\mathfrak{F}^{k}_{\mu\nu}\\
        &=\xi_{l}^{\hspace{4pt}m}\left(\partial_{\mu}\partial_{m}A_{\nu}^{k}-\partial_{\nu}\partial_{m}A_{\mu}^{k} + \sigma\mathrm{C}_{ji}^{k}\left(\partial_{m}A_{\mu}^{\hspace{4pt}j}A_{\nu}^{\hspace{4pt}i}+\partial_{m}A_{\nu}^{\hspace{4pt}j}A_{\mu}^{\hspace{4pt}i} \right) \right)\\
        &= -\sigma\mathrm{C}_{lp}^{k}\left(\partial_{\mu}A_{\nu}^{\hspace{4pt}p}-\partial_{\nu}A_{\mu}^{\hspace{4pt}p} \right) +\sigma\mathrm{C}_{\hspace{4pt}np}^{k}\xi_{l}^{\hspace{4pt}m}\Big[\left(\partial_{\mu}A_{m}^{\hspace{4pt}n}-\partial_{m}A_{\mu}^{\hspace{4pt}n} \right)A_{\nu}^{\hspace{4pt}p}\\ & -\left(\partial_{\nu}A_{m}^{\hspace{4pt}n}-\partial_{m}A_{\nu}^{\hspace{4pt}n} \right)A_{\mu}^{\hspace{4pt}p}  \Big] \\
        &= -\sigma\mathrm{C}_{\hspace{4pt}lp}^{k}\mathfrak{F}_{\mu\nu}^{p},
    \end{aligned}
\end{equation}
where in the fifth line we made use of \eqref{IntegrabilityCs} and the Jacobi identity. Therefore under the above coordinate transformations, $\mathfrak{F}_{\mu\nu}^{k} $ transforms as
\begin{equation}
    \mathfrak{F}_{\mu\nu}^{k}\longmapsto \left(\delta_{j}^{k} -\sigma \epsilon^{i}\mathrm{C}_{\hspace{4pt}ij}^{k} \right)\mathfrak{F}_{\mu\nu}^{j}.
\end{equation}
In terms of the matrices $\tau_{i}$ we can write the above transformation as
\begin{equation}
    \tau\cdot\mathfrak{F}_{\mu\nu} \longmapsto (\mathbb{I} + \imath\sigma\epsilon\cdot\tau)\tau\cdot\mathfrak{F}_{\mu\nu}(\mathbb{I} - \imath\sigma\epsilon\cdot\tau). 
\end{equation}
Thus, under the finite coordinate transformation we have
\begin{equation}\label{UsualFTrans}
   \boxed{ \tau\cdot\mathfrak{F}_{\mu\nu}\longmapsto \mathrm{U}\tau\cdot\mathfrak{F}_{\mu\nu}\mathrm{U}^{-1}.}
\end{equation}
The equations \eqref{UsualATrans} and \eqref{UsualFTrans} are the usual transformation laws of the gauge potential and the field strength tensor respectively     \cite{ryder1996quantum}. This proofs that the gauge transformation  is a consequence of coordinate transformations which are isometries of spacetime.

With 

   \[ \mathfrak{S}_{\hspace{4pt}ab}^{k} =  \partial_{a}A_{b}^{\hspace{4pt}k}+\partial_{b}A_{a}^{\hspace{4pt}k}\]
the Christoffel symbols are 
\begin{equation}
    {\Tilde{\Gamma}^{a}_{bc} = \delta^{\mu}_{b}\delta^{\nu}_{c}\mathrm{G}^{a\rho}\Gamma_{\mu\nu\rho} + \frac{\lambda^2}{\zeta^2}\frac{\mathrm{G}}{2}^{a\rho}\left(\mathfrak{F}_{b\rho}\cdot A_{c} +\mathfrak{F}_{c\rho}\cdot A_{b}\right) + \frac{\delta_{k}^{a}}{2}\xi^{k}\cdot\mathfrak{S}_{bc}.}
\end{equation}
Using this to calculate the Ricci curvature tensor we get (see \ref{AppRicci})

 \begin{equation}
    \boxed{\begin{aligned}
     \Tilde{\mathfrak{R}}_{bc} =& \delta_{b}^{\mu}\delta_{c}^{\nu}\mathfrak{R}_{\mu\nu} + \frac{\lambda^{2}}{2\zeta^2}\left(A_{b}\cdot\bar{\mathfrak{D}}_{\rho}\mathfrak{F}_{c}^{\hspace{4pt}\rho} + A_{c}\cdot\bar{\mathfrak{D}}_{\rho}\mathfrak{F}_{b}^{\hspace{4pt}\rho} -\mathfrak{F}_{b\rho}\cdot\mathfrak{F}_{c}^{\hspace{4pt}\rho}\right)\\
     &+\frac{\lambda^4}{4\zeta^4}\mathfrak{F}_{\rho\delta}\cdot A_{b} \mathfrak{F}^{\rho\delta}\cdot A_{c} +\frac{1}{4}\sigma^2\mathrm{C}^{q}_{\hspace{4pt}mn}\mathrm{C}^{m}_{\hspace{4pt}pq}A_{b}^{\hspace{4pt}n}A_{c}^{\hspace{4pt}p}
     \end{aligned}}
 \end{equation}
 where \[\bar{\mathfrak{D}}_{\rho}\mathfrak{F}_{b}^{k\hspace{4pt}\rho} = \nabla_{\rho}\mathfrak{F}_{b}^{k\hspace{4pt}\rho} +\sigma A_{\rho}^{\hspace{4pt}m}\mathrm{C}^{k}_{mn}\mathfrak{F}_{b}^{n\hspace{4pt}\rho}.\]
 From this we find the following expression for the Ricci Scalar
 \begin{equation}\label{ReducednRicciScalar}
    \boxed{ \Tilde{\mathfrak{R}} = \mathfrak{R} -\frac{\lambda^2}{4\zeta^2}\mathfrak{F}_{\rho\delta}\cdot\mathfrak{F}^{\rho\delta} +\frac{\sigma^2\zeta^2}{4\lambda^2}\mathrm{C}^{q}_{\hspace{4pt}mn}\mathrm{C}^{\hspace{4pt}mn}_{q}.}
 \end{equation}
 The contravariant components of the Ricci scalar split into 
 \begin{equation}\label{ContravariantRicci}
     \begin{split}
         \Tilde{\mathfrak{R}}^{\mu\nu}&=\mathfrak{R}^{\mu\nu} -\frac{\lambda^2}{2\zeta^2}\mathfrak{F}^{\mu}_{\hspace{4pt}\rho}\cdot\mathfrak{F}^{\nu\rho},\\
         \Tilde{\mathfrak{R}}^{\mu k} &= -\xi^{k}\cdot A_{\nu}\Tilde{\mathfrak{R}}^{\mu\nu} +\frac{1}{2}\xi^{k}\cdot\bar{\mathfrak{D}}_{\rho}\mathfrak{F}^{\mu\rho},\\
         \Tilde{\mathfrak{R}}^{ik} &= \xi^{i}\cdot A_{\mu}\xi^{k}\cdot A_{\nu}\Tilde{\mathfrak{R}}^{\mu\nu} +\frac{1}{4}\mathfrak{F}_{\rho\delta}\cdot\xi^{i}\mathfrak{F}^{\rho\delta}\cdot\xi^{k}, \\ &+\frac{\zeta^4\sigma^2}{4\lambda^4}\mathrm{C}_{\hspace{4pt}m}^{q}\cdot\xi^{i}\mathrm{C}_{\hspace{4pt}q}^{m}\cdot\xi^{k}.
     \end{split}
 \end{equation}
 Since 
 \begin{equation}
     \begin{split}
         \partial_{a}\big(\sqrt{\mathrm{G}}\mathrm{G}^{ae}\mathrm{G}^{bd}\mathrm{F}_{de}^{k}\big) &=\sqrt{\mathrm{G}}\big(\delta_{\delta}^{b}-\delta_{j}^{b}\xi^{j}\cdot A_{\delta}\big)\bar{\mathfrak{D}}_{\rho}\mathfrak{F}^{k\delta\rho} \\ &+\frac{\sqrt{\mathrm{G}}\delta_{j}^{b}}{2}\big(\mathfrak{F}_{\mu\nu}^{k}\mathfrak{F}^{\mu\nu}\cdot\xi^{j} +\frac{\sigma^2\zeta^2}{\lambda^2}\mathrm{C}_{mn}^{\hspace{8pt}k}\mathrm{C}^{mnl}\xi_{l}^{\hspace{4pt}j}\big),
     \end{split}
 \end{equation}
 the equations on the second and third lines of \eqref{ContravariantRicci} are equations of motion that follow from \eqref{nonAbelMatterAsCurvedL}, just as in the Abelian case, proving that the theory is consistent.
 
This leads us to the conclusion that the complete Lagrangian for the $n$-dimensional theory of gravity is
\begin{equation}
    \boxed{\mathfrak{L} = \frac{1}{\lambda^2}\Tilde{\mathfrak{R}} -\frac{1}{4\zeta^2}\mathrm{F}_{ab}\cdot\mathrm{F}^{ab} +\frac{1}{\zeta^2}\Tilde{\mathfrak{R}}^{ab}A_{a}\cdot A_{b} + \frac{6}{\lambda^2}\mathrm{G}^{ab}\partial_{a}\mathcal{S}\cdot\partial_{b}\mathcal{S}}.
    \end{equation}
We saw in \autoref{4delectrodynamics} that $\mathcal{S}$ is not gauge covariant but $\hat{\Psi} = e^{\imath\tau\cdot\mathcal{S}}$ is gauge covariant and so are all its columns $\hat{\phi}_{i}$. In terms of $\hat{\phi}_{i}$ the above Lagrangian is
\begin{equation}
    \boxed{\mathfrak{L} = \frac{1}{\lambda^2}\Tilde{\mathfrak{R}} -\frac{1}{4\zeta^2}\mathrm{F}_{ab}\cdot\mathrm{F}^{ab} +\frac{1}{\zeta^2}\Tilde{\mathfrak{R}}^{ab}A_{a}\cdot A_{b} + \frac{6}{\lambda^2}\mathrm{G}^{ab}\partial_{a}\hat{\phi}\cdot\partial_{b}\hat{\phi}}.
    \end{equation}
 Making use of the fact that $\hat{\phi}$ is gauge covariant, we get
 \begin{equation}
   \mathrm{G}^{a\mu}\partial_{a}\hat{\phi}_{i}  =g^{\mu\nu}\mathfrak{D}_{\nu}\hat{\phi}_{i}
 \end{equation}
 and 
 \begin{equation}
   \mathrm{G}^{ab}\partial_{a}\hat{\phi}\cdot\partial_{b}\hat{\phi} = g^{\mu\nu}\mathfrak{D}_{\mu}\hat{\phi}\cdot\mathfrak{D}_{\nu}\hat{\phi}  + \frac{\zeta^{2}\sigma^2}{\lambda^2}
 \end{equation}
 where in the last term we made use of the fact that we can choose $\sigma$ so that $\mathrm{C}_{jk}^{i}\mathrm{C}_{i\hspace{4pt}l}^{\hspace{4pt}j}=\delta_{kl} $ and $\hat{\phi}\cdot\hat{\phi}=1 $. Since $g^{\mu\nu}=\frac{12}{\lambda^2|\Psi|^2}\gamma^{\mu\nu}$, we get from \eqref{ReducednRicciScalar}
 \begin{equation}
    \begin{aligned}
    \frac{1}{12} \Tilde{\mathfrak{R}} &= \frac{1}{\lambda^2}|\Psi|^{-2}\mathfrak{R}^{(g)}-\frac{6}{\lambda^2}|\Psi|^{-3}\Box|\Psi| -\frac{12}{4\zeta^2\lambda^2}|\Psi|^{-4}\mathfrak{F_{\mu\nu}}\cdot\hat{\mathfrak{F}}^{\mu\nu}\\ & + \frac{n\sigma^2\zeta^2}{48\lambda^2}.
    \end{aligned}
 \end{equation}
 For fields obeying the field equations 
 \begin{equation}
     \frac{1}{4\zeta^2}\mathrm{F}_{ab}\cdot\mathrm{F}^{ab} =\frac{1}{\zeta^2}\Tilde{\mathfrak{R}}^{ab}A_{a}\cdot A_{b}.
 \end{equation}
 As such the $4\mathcal{D}$ Lagrangian becomes
\begin{equation}
  \begin{split}
    \mathfrak{L} &=  -\frac{1}{4\zeta^2}\mathfrak{F}_{\mu\nu}\cdot\hat{\mathfrak{F}}^{\mu\nu} + \frac{1}{12}|\Psi|^2\mathfrak{R}^{(g)} +\frac{1}{2}\partial_{\mu}|\Psi|\partial^{\mu}|\Psi|\\& + \frac{1}{2} |\Psi|^2 \gamma^{\mu\nu}\mathfrak{D}_{\mu}\hat{\phi}\cdot\mathfrak{D}_{\nu}\hat{\phi}+\frac{\zeta^2 \sigma^2}{24}\left(1+\frac{k}{2}\right) |\Psi|^{4}
    \end{split}  
\end{equation}
 and reduces to
 \begin{equation}
    \boxed{\begin{aligned}
         \mathfrak{L} &=  -\frac{1}{4\zeta^2}\mathfrak{F}_{\mu\nu}\cdot\hat{\mathfrak{F}}^{\mu\nu} + \frac{1}{12}|\Psi|^2\mathfrak{R}^{(g)} + \frac{1}{2}  \gamma^{\mu\nu}\mathfrak{D}_{\mu}\Psi\cdot\mathfrak{D}_{\nu}\Psi \\ &+\frac{\zeta^2 \sigma^2}{24}\left(1+\frac{k}{2}\right) |\Psi|^{4}
     \end{aligned}}
 \end{equation}
 where $k=\mathrm{C}^{q}_{\hspace{4pt}mn}\mathrm{C}^{\hspace{4pt}mn}_{q}$. For  $\mathfrak{R}^{(g)} = constant$, this is a non-Abelian Higgs Model.
 
Instead of undergoing the tedious process of finding the expression for the Ricci curvature tensor, one could simply find the $n$-dimensional Einstein equations from the variation principle and use the equations of motion of $A_{a}$ together with the expression of the metric that follows from them  to reduce the equations to the above form.

\subsection{Electroweak and Strong interactions}\label{standardM}
Having developed a consistent theory of gravity and gauge theories, we are faced with a task of showing that the resulting equations have a solution and they are consistent with physical reality. 

Since the number of Killing fields is the same as the dimensions of the internal space and the Lie derivative of a Killing field along another Killing vector field is a Killing vector field, the non-Abelian gauge theory can be realized in $3$ or more dimensional internal space. This is clear from the fact that the Killing vector fields have to be linearly independent since their matrix has to be invertible(see \autoref{dimNTheory}). 

\subsubsection{U(1)$\times$ U(1) Theory}
We want to eventually explore the non-Abelian gauge theory that arises in this geometric description. We start by constructing the precursor of this theory which is the Abelian gauge theory for which the internal space is $2$ dimensional. Since we saw that in the absence of gravity and gauge fields $4\mathcal{D}$ spacetime is conformally flat, we assume that the internal space is also conformally flat. In line with \eqref{nonAbelMetricN}  we have
\begin{equation}\label{n-ballmetric}
    A_{i}\cdot A_{j} = \frac{\delta_{ij}}{\sigma^2|x|^2}.
\end{equation}
The Killing vectors that we require are
\begin{equation}\label{WeakKillingEx}
        \xi_{j} = \sigma\big(-x_{5}\partial_{j} +x_{j}\partial_{5}+\delta_{5j}x^{i}\partial_{i} \big)
\end{equation}
which are radial and angular momentum operators in  Euclidean space although in the internal space they are just translations. One can easily show that they commute thus have vanishing Lie bracket which confirms that we are dealing with an Abelian gauge theory. One can verify that 
\begin{equation}
    \partial_{i}A_{j}^{\hspace{4pt}k}-\partial_{j}A_{i}^{\hspace{4pt}k} =0
\end{equation}
by making use of the fact that it is the inverse of $\xi_{i}^{\hspace{4pt}k}$ in line with the previous subsection or by substituting
\begin{equation}
    A_{j}^{\hspace{4pt}k} =\frac{1}{\sigma|x|^2}\big(\delta_{j}^{k}x_{5} -\delta_{5}^{k}x_{j} + \delta_{5j}x^{k}\big).
\end{equation}

For the Killing vector fields for which there exists a Killing field whose Lie derivative along their direction is non-zero, they can be written as
\begin{equation}\label{KVFExpression}
    \xi_{i}^{\hspace{4pt}k} = \sigma \mathrm{C}_{ij}^{\hspace{10pt}k}x^{j} = -\sigma \imath\mathrm{\tau}_{i|j}^{\hspace{8pt}k}x^{j}
\end{equation}
where $\lbrace\tau_{i}\rbrace$ are anti-symmetric matrices. We show this in \eqref{LieDerivOfAntisym}. The matrix form can be adapted even when the structural coefficients are vanishing. In this case the matrices are
\begin{equation}
    \begin{aligned}
        \tau_{5}= \begin{pmatrix}
            1 && 0\\
            0 && 1
        \end{pmatrix} \hspace{15pt} &\tau_{6} = \begin{pmatrix}
            0 && \imath\\
            -\imath &&  0
        \end{pmatrix}
    \end{aligned}
\end{equation}
In line with \eqref{ChargeEigenVE}, the conserved charges are
\begin{equation}\label{AbelianCharge conservation}
    \xi_{j}^{\hspace{4pt}k}\partial_{k}\mathcal{S} = \sigma\tau_{j}
\end{equation}
where $\mathcal{S}=\tau\cdot\mathcal{S}$ and from \eqref{extraDGaugeC} we see that the gauge covariant derivative can be written as
\begin{equation}
    \mathfrak{D}_{a} = \partial_{a} -\imath\sigma\tau\cdot A_{a}.
\end{equation}
Thus we have successfully constructed a theory of two neutral gauge fields since all that is left is plugging these results in the equations of motion determined in \autoref{dimNTheory}. In this case the Higgs field is a vector field in internal space and has the form
\begin{equation}
    \Psi = \phi e^{\imath \tau\cdot\mathcal{S}},
\end{equation}
which can be verified from the reduced Lagrangian.
\subsubsection{U(1)$\times$ SU(2) Theory}
In constructing the non-Abelian gauge theory, we maintain our assumption regarding the internal space. As such, the metric of the internal space is still \eqref{n-ballmetric} but with the indices running from 5 to 8. We start by showing that such an  assumption is reasonable.

We start with $n$-dimensional Euclidean space. The isometries of the Euclidean metric are $n$ translations and  $\frac{1}{2}n(n-1)$ rotations. As such it has $\frac{1}{2}n(n+1)$ Killing vector fields which makes Euclidean space a maximal symmetric space. The translation Killing vector fields are 
\begin{equation}\label{transkill}
    \xi_{i}^{\hspace{4pt}k} = m\delta_{i}^{\hspace{4pt}k}
\end{equation}
and the conserved charges associated with them are components of linear momentum
\begin{equation}
    P^{k} = \xi_{i}^{\hspace{4pt}k}\frac{dx^{i}}{d\tau}.
\end{equation}
From the geodesic equation we see that on a curved spacetime linear momentum is not conserved in the ordinary sense which implies that in general \eqref{transkill} are not Killing vector fields on a curved spacetime. Thus we can construct the curved manifolds as submanifolds of Euclidean space by breaking the symmetries of the Euclidean metric. Since our interest is in non-Abelian gauge theories and translations represent the Abelian part of spacetime symmetries, the most logical thing to do is to break all translation symmetries while retaining the rotational symmetries. The most obvious way of achieving this is to introduce the conformal factor $f(x^2)$ so that the metric of the internal space is
\begin{equation}\label{ConformallyFlatMetric}
    g_{ij} =f(x^2)\delta_{ij}.
\end{equation}
That rotational invariance is retained is evident from the fact that both the Euclidean metric and the conformal factor are invariant under rotations. If in addition, we restrict the coordinate so that $f(x^2) = \text{const}$, the dimensions of the resulting manifold are $d = n-1$ and the manifold is maximally symmetric since it is endowed with $\frac{1}{2}d(d+1)$ Killing vector fields. This manifold is a coset space $\mathbf{SO}(d+1)/\mathbf{SO}(d)$. For this manifold to be the internal space, it must be a Lie group, which means that $\mathbf{SO}(d)$ must be the invariant subgroup of $\mathbf{SO}(d+1)$. This is the case for $d= 1,3$. $d= 1$ is the Abelian gauge theory.
As such, apart from $\mathrm{S}^{3}$, all other maximally symmetric spaces fail to provide us with a gauge theory.
Therefore to attain a realistic non-Abelian  gauge theory we have to consider less symmetric manifolds.

The manifold we have been dealing with is the $n$-sphere.
We can construct a less symmetric n-dimensional manifold which is the foliation of $n-1$-spheres. The isometries of the metric of this space are $\frac{1}{2}n(n-1)$ rotations and one radial translation. Thus in addition to the generators of angular momentum, it has
\begin{equation}
    \xi_{5} = x^{k}\partial_{k}
\end{equation}
as the Killing vector field. By solving the Killing equation with \eqref{ConformallyFlatMetric} as the metric of this manifold, we get
\begin{equation}
    f(x^2) = \frac{1}{\sigma^2x^2}
\end{equation}
and we end up with \eqref{n-ballmetric}.

The Killing vector fields that we are looking for are
\begin{equation}\label{WeakKillingE}
        \xi_{j} =\sigma\big(x_{j}\partial_{5} -x_{5}\partial_{j}  +\delta_{5j}x^{i}\partial_{i} \pm \delta_{j}^{N}\varepsilon_{\hspace{4pt}NP}^{M}x^{P}\partial_{M}\big)
\end{equation}
where the range of capital letters excludes 5.
This is clear from the fact that \eqref{n-ballmetric} is constructed from the 1-forms dual to them which implies that \eqref{n-ballmetric} is constant along their integral curves.
 Their Lie bracket is 
\begin{equation}
    \begin{split}
    L_{\xi_{M}}\xi_{5}^{\hspace{4pt}i } =[\xi_{M},\xi_{5}]^{i} &= 0,\\
     L_{\xi_{M}}\xi_{N}^{\hspace{4pt}i} =[\xi_{M},\xi_{N}]^{i} &= -2\sigma \varepsilon_{MN}^{\hspace{14pt}P}\xi_{P}^{\hspace{4pt}i},
    \end{split}
\end{equation}
which are the commutation relations of $\mathfrak{su}(2)\times \mathfrak{u}(1)$.
The matrices that follow from \eqref{WeakKillingE} are
\begin{equation}
    \begin{split}
    \tau_{5|i}^{\hspace{12pt}k} &= \delta_{i}^{k},\\
    -\imath\tau_{M|i}^{\hspace{12pt}k} &= \delta_{Mi}\delta_{5}^{k}-\delta_{M}^{k}\delta_{5i}\pm\delta_{N}^{k}\delta_{i}^{P}\epsilon^{N}_{\hspace{4pt}MP},
    \end{split}
\end{equation}
or more explicitly
\begin{equation}\label{4by4Electroweak}
    \begin{aligned}
        \tau_{5} &= \begin{pmatrix}
            1 && 0 && 0 && 0\\
            0 && 1 && 0 && 0\\
            0 && 0 && 1 && 0\\
            0 && 0 && 0 && 1
        \end{pmatrix}
         &\tau_{6} = \begin{pmatrix}
            0 && -\imath && 0 && 0\\
            \imath && 0 && 0 && 0\\
            0 && 0 && 0 && \mp\imath\\
            0 && 0 && \pm\imath && 0
        \end{pmatrix}\\
        \tau_{7} &= \begin{pmatrix}
            0 && 0 && -\imath && 0\\
            0 && 0 && 0 && \mp\imath\\
            \imath && 0 && 0 && 0\\
            0 && \pm\imath && 0 && 0
        \end{pmatrix}
        &\tau_{8} = \begin{pmatrix}
            0 && 0 && 0 && -\imath\\
            0 && 0 && \mp\imath && 0\\
            0 && \pm\imath && 0 && 0\\
            \imath && 0 && 0 && 0
        \end{pmatrix}
    \end{aligned}
\end{equation}
This is the representation of $\mathfrak{su}(2)\times \mathfrak{u}(1)$. The Pauli matrices and the identity matrix have the same commutation relations thus they form a complex representation of the said algebra. In this case the $\tau_{j}$s do not commute as such the equation of conservation of charge 
\begin{equation}\label{spin}
    \boxed{\xi_{j}^{\hspace{4pt}k}\partial_{k}\hat{\Psi} = \imath\sigma\tau_{j}\hat{\Psi}}
\end{equation}
can not be reduced to \eqref{AbelianCharge conservation}. Thus it is the fundamental equation from which \eqref{AbelianCharge conservation} follow.
The solution of this equation is
\begin{equation}
    \hat{\Psi} =\mathcal{P} exp\big({\imath\sigma\int\tau\cdot A_{k}dx^{k}}\big)
\end{equation}
where $\mathcal{P}$ denotes path ordering.
From \eqref{spin} we conclude that the gauge covariant derivative is
\begin{equation}
    \mathfrak{D}_{a} = \partial_{a} - \imath\tau\cdot A_{a}. 
\end{equation}
With the gauge covariant derivative known, we proceed in the same manner as in the Abelian case. 

Given the Lie group, the problem of finding the geometric description of a gauge theory boils down to finding the 1-forms $A_{i}=A_{ij}dx^{j}$ such that
\begin{equation}
    \partial_{i}\hat{\Psi} = \imath\sigma\tau\cdot A_{i}\hat{\Psi}
\end{equation}
where $\hat{\Psi}$ is the general group element. Since we can always normalize the matrices $\lbrace\tau_{i}\rbrace$ such that $tr(\tau_{i}\tau_{j}) =\alpha\delta_{ij}$ for some constant $\alpha$, we have
\begin{equation}
    \begin{aligned}
        A_{ji} &=\frac{1}{\alpha}tr(\tau_{j}\tau\cdot A_{i})\\
        &=-\imath\frac{\sigma}{\alpha}tr(\tau_{j}\partial_{i}\hat{\Psi}\hat{\Psi}^{-1}).
    \end{aligned}
\end{equation}
This means to formulate the gauge theory all we need is the expression of the general element of the gauge group. One way of constructing the general element of the group is the Euler angle decomposition      \cite{bertini2006euler}. For the case of $\mathbf{SU}(2)$ this is done explicitly in      \cite{biedenharn1984angular} while for the case of $\mathbf{SU}(3)$ it is carried out in      \cite{byrd1997geometry} in addition to computing the 1-forms $A_{ij}$. 
We will not follow this approach since in our case, after constructing the 1-forms in this way we will have to find the relation between Euler angles and Cartesian coordinates. This is due to the fact that in the usual formalism, the gauge theories are constructed in Minkowski spacetime with Cartesian coordinates thus to facilitate comparison we should work in Cartesian coordinates. We develop a different approach and in \autoref{spinAsSTSym} it will be evident that in the case of $\mathbf{SU}(2)$ this approach is equivalent to the aforementioned approach.

Given an $n$-dimensional real representation $\lbrace\tau_{i}\rbrace$ of the Lie algebra of an $n$-dimensional group, we can use the matrices $\lbrace\tau_{i}\rbrace$ to compute the Killing vector fields which form the representation of the Lie algebra. These Killing vector fields are
\begin{equation}
    \xi_{i} = -\imath\tau_{i|j}^{\hspace{9pt}k}\mathbf{x}^{j}\partial_{k}
\end{equation} 
 where $\tau_{i|j}^{\hspace{9pt}k}$ are the components of the matrix $\tau_{i}$.  The lie derivative of these Killing vector fields is
 \begin{equation}\label{LieDerivOfAntisym}
    \begin{aligned}
     [\xi_{i},\xi_{j}] &= -\tau_{i|k}^{\hspace{9pt}j}\tau_{l|m}^{\hspace{9pt}n}[x^{k}\partial_{j},x^{m}\partial_{n}] \\
     &= -\tau_{i|k}^{\hspace{9pt}j}\tau_{l|j}^{\hspace{9pt}n}(\delta_{j}^{m}x^{k}\partial_{n}-\delta_{n}^{k}x^{m}\partial_{j})\\
     &=(\tau_{i|j}^{\hspace{9pt}n}\tau_{l|k}^{\hspace{9pt}j}-\tau_{i|k}^{\hspace{9pt}j}\tau_{l|j}^{\hspace{9pt}n})x^{k}\partial_{n}\\
     &=\imath \mathrm{C}_{il}^{\hspace{9pt}j}\tau_{j|k}^{\hspace{9pt}n}x^{k}\partial_{n}\\
     &=-\mathrm{C}_{il}^{\hspace{9pt}j}\xi_{j}.
    \end{aligned}
 \end{equation}
 Clearly, these Killing vector fields form the representation of the  Lie algebra of the group in question. Accordingly they belong to the tangent space of the group manifold.
 For semi-simple groups
 \begin{equation}
     \tau_{i|j}^{\hspace{9pt}k} = \imath \mathrm{C}_{ij}^{\hspace{9pt}k}.
 \end{equation}
Since the structure constants are anti-symmetric in the first two indices, we have
\begin{equation}
    x^{i}\xi_{i} = 0
\end{equation}
in the case of semi-simple groups. This means that these Killing vector fields are not linearly independent therefore the matrix formed out of them is not invertible. In other words they do not constitute the basis for the tangent space of the group manifold.
We can get around this by realizing the semi-simple group $\mathbf{G}$ as the subgroup of the $d$-dimensional group $\mathbf{F}$ with invariant Abelian subgroup.  To construct this group, we consider the $d$-dimensional matrix representation of the Lie algebra of the semi-simple group in question with $d> n$ and take the representation of the Lie algebra of the Abelian subgroup $\mathfrak{u}(1)$ to be the $d$-dimensional identity matrix. Since $d\ge n+1$; for $d\neq n+1$, we find the $d-(n+1)$-dimensional group $\mathrm{H}$ whose Lie algebra has $d$-dimensional representation so that $\mathbf{F}$ is
\begin{equation}
    \mathrm{F} = \mathbf{U}(1)\times\mathbf{H}\times\mathbf{G}.
\end{equation}
The presence of the Abelian ideal ensures that the general element of the Lie algebra of $\mathbf{F}$ is invertible which in turn implies that the Killing vector fields constructed from the $d$-dimensional matrix representation of the Lie algebra of $\mathrm{F}$ are linearly independent. With the Linearly independent Killing vector fields known,  we impose the constraints on the coordinates $\lbrace x^{i}\rbrace$ which reduce the dimensions from $d$ to $n$ and break the $ \mathbf{U}(1)\times\mathbf{H}$ symmetry only. The remaining Linearly independent Killing vector fields will be the ones whose commutation relations correspond to those of the Lie Algebra of $\mathbf{G}$. The matrix of these Killing vector fields is in general invertible. Even if this matrix is not invertible for some values of the coordinates, this should not be a cause for concern.
Since the Ricci scalar of the internal space
\begin{equation}
    \mathfrak{\hat{R}} = \frac{\sigma^2\zeta^2}{4\lambda^2}\mathrm{C}^{q}_{\hspace{4pt}mn}\mathrm{C}^{\hspace{4pt}mn}_{q}
\end{equation} is a constant, any singularities encountered for some values of the coordinates are just coordinate singularities. They just reflect the fact that the manifold cannot be covered by a single chart thus we can avoid them by introducing a different chart. In order to illustrate this we construct $\mathbf{U}(1)$ as a subgroup of $\mathbf{U}(1)\times\mathbf{U}(1)$ and $\mathbf{SU}(2)$ as subgroup of $\mathbf{U}(1)\times\mathbf{SU}(2)$.
Following the aforementioned procedure to construct $\mathbf{U}(1)\times\mathbf{U}(1)$ and $\mathbf{U}(1)\times\mathbf{SU}(2)$, we end up with the same results as in the preceding paragraphs where we start from the metric. In both cases we have to impose one constraint for the number of independent parameters to correspond to those of the subgroup. Since we know that the subgroups in both cases are spheres, the constraint becomes
\begin{equation}
   k x_{i}x^{i} + y^2 = 1 
\end{equation}
where $\frac{1}{\sqrt{k}}$ is the radius of the sphere, $i$ ranges from 1 to n and $y=x^{n+1}$ .
Applying this constraint, we find the Killing vector field of the $\mathrm{U}(1)$ manifold to be
\begin{equation}
    \xi_{6} = \sqrt{1-k x_{5}^2}\partial_{5}
\end{equation}
and we have the 1-form
\begin{equation}
    A_{65} = \frac{1}{\sqrt{1 - k x_{5}^{2}}}.
\end{equation}
This 1-form is singular at $\sqrt{k}x_{5} = 1$. This is expected since we need at least two charts to form an atlas for a circle. Substituting the expression for the 1-form $A_{65}$ in \eqref{ConditionForReduction}, enables us to express the radius of curvature in terms of the Planck length. With $x_{5}= \frac{1}{\sqrt{k}}\sin\theta$, when $A_{6\mu} = 0$, we get the following expression for the electric charge
\begin{equation}
    e = \frac{\lambda^2}{\sqrt{k}\zeta^2}m\Dot{\theta}.
\end{equation}
In the case of $\mathrm{U}(1)\times\mathrm{SU}(2)$ manifold applying the constraint results in the Killing vector fields
\begin{equation}
    \xi_{i} = \sqrt{1-k|\mathbf{x}|^2}\partial_{i}\pm\sqrt{k}\varepsilon_{ij}^{\hspace{9pt}k}x^{j}\partial_{k}
\end{equation}
which are the linearly independent Killing vector fields of the metric of $\mathrm{SU}(2)$ for a specific choice of the sign of the angular momentum operator part of the Killing vector fields.  From this we get the inverse of the matrix of these Killing vector fields to be the matrix 1-forms whose components are  
\begin{equation}
    A_{ij}=\sqrt{1-k|\mathbf{x}|^2}\left(\delta_{ij} +\frac{k x_{i}x_{j}}{1-k |\mathbf{x}|^2}\right)\pm\sqrt{k}\varepsilon_{ijk}x^{k}.
\end{equation}
From this matrix 1-forms we get the metric of $\mathrm{SU}(2)$ to be
\begin{equation}\label{s3metric}
    g_{ik} = A_{ij}A_{k}^{\hspace{4pt}j} = \delta_{ik} +\frac{k x_{i}x_{k}}{1-k |\mathbf{x}|^2}.
\end{equation}
The conserved charges associated with these Killing fields when $A_{\mu}^{i} = 0$ are
\begin{equation}
    \mathrm{Q}_{i}= \frac{\lambda^2}{\zeta^2}\left(\sqrt{1-k|\mathbf{x}|^2}g_{ij}\frac{dx^{j}}{d\tau} \pm \sqrt{k}\varepsilon_{ijk}x^{k}\frac{dx^{j}}{d\tau}\right). 
\end{equation}
Accordingly, the Weak Hypercharge is also momentum in the internal  space.
\subsubsection{U(1)$\times$SU(2)$\times$ SU(3)}

Let $\lbrace\sigma_{a}\rbrace_{a=0}^{3}$ be the Pauli matrices with $\sigma_{0}$ as the $2\times 2$ identity matrix and $\lbrace\lambda_{i}\rbrace_{i=0}^{8}$ be the Gell-Mann matrices with  $\lambda_{0}$ as the $3\times 3$ identity matrix. The 12-dimensional matrix representation of the algebra of $\mathbf{U}(1)\times \mathbf{SU}(2)\times \mathbf{SU}(3)$ is
\begin{equation}
    \begin{aligned}
      \tau_{5} = \sigma_{0}\otimes\sigma_{0}\otimes\lambda_{0} && \tau_{6} = \sigma_{1}\otimes\sigma_{2}\otimes\lambda_{0} && &\tau_{9} =\sigma_{0}\otimes\sigma_{2}\otimes\lambda_{1}\\
      &&\tau_{7}= \sigma_{2}\otimes\sigma_{0}\otimes\lambda_{0}&& &\tau_{10}=\sigma_{0}\otimes\sigma_{0}\otimes\lambda_{2}\\
      &&\tau_{8} = \sigma_{3}\otimes\sigma_{2}\otimes\lambda_{0} && &\tau_{11}=\sigma_{0}\otimes\sigma_{2}\otimes\lambda_{3}\\
      && && &\tau_{12}=\sigma_{0}\otimes\sigma_{2}\otimes\lambda_{4}\\
      && && &\tau_{13}=\sigma_{0}\otimes\sigma_{0}\otimes\lambda_{5}\\
      && && &\tau_{14}=\sigma_{0}\otimes\sigma_{2}\otimes\lambda_{6}\\
      && && &\tau_{15}=\sigma_{0}\otimes\sigma_{0}\otimes\lambda_{7}\\
      && && &\tau_{16}=\sigma_{0}\otimes\sigma_{2}\otimes\lambda_{8}.
    \end{aligned}
\end{equation}
Making use of this representation of the Lie algebra
and making the definition
\begin{equation}
    \mathrm{L}_{i|j} =x_{i}\partial_{j}-x_{j}\partial_{i},
\end{equation}
 we get the following linearly independent Killing vector fields
 
\begin{equation}
    \begin{aligned}
        &\xi_{5} = x^{i}\partial_{i}\\
        &\xi_{6} =\mathrm{L}_{5| 14}+\mathrm{L}_{6| 15}+\mathrm{L}_{7| 16} +\mathrm{L}_{11| 8} +\mathrm{L}_{12| 9} +\mathrm{L}_{13| 10}\\
        &\xi_{7} = \mathrm{L}_{5| 11}+\mathrm{L}_{6| 12}+\mathrm{L}_{7| 13} +\mathrm{L}_{8| 14} +\mathrm{L}_{9| 15} +\mathrm{L}_{10| 16}\\
         &\xi_{8} = \mathrm{L}_{5| 8}+\mathrm{L}_{6| 9}+\mathrm{L}_{7| 10} +\mathrm{L}_{14| 11} +\mathrm{L}_{15| 12} +\mathrm{L}_{16| 13}\\
          &\xi_{9} = \mathrm{L}_{5| 9}+\mathrm{L}_{12| 14}+\mathrm{L}_{11| 15} +\mathrm{L}_{6| 8} \\
          &\xi_{10} = \mathrm{L}_{5| 6}+\mathrm{L}_{14| 15}+\mathrm{L}_{8| 9} +\mathrm{L}_{11| 12} \\
          &\xi_{11} = \mathrm{L}_{5| 8}+\mathrm{L}_{9| 6}+\mathrm{L}_{12| 15} +\mathrm{L}_{14| 11} \\
          &\xi_{12} = \mathrm{L}_{5| 10}+\mathrm{L}_{7| 8}+\mathrm{L}_{12| 16} +\mathrm{L}_{13| 14} \\
          &\xi_{13} = \mathrm{L}_{5| 7}+\mathrm{L}_{8| 10}+\mathrm{L}_{14| 16} +\mathrm{L}_{11| 13} \\
          &\xi_{14} = \mathrm{L}_{6| 10}+\mathrm{L}_{13| 15}+\mathrm{L}_{7| 9} +\mathrm{L}_{12| 16} \\
          &\xi_{15} = \mathrm{L}_{9| 10}+\mathrm{L}_{12| 13}+\mathrm{L}_{15| 16} +\mathrm{L}_{6| 7} \\
          &\xi_{16} = \frac{1}{\sqrt{3}}\left(\mathrm{L}_{5| 8}+\mathrm{L}_{6| 9}+\mathrm{L}_{11| 14} +\mathrm{L}_{12| 15} +2\mathrm{L}_{10| 7} +2\mathrm{L}_{16| 13} \right).
    \end{aligned}
\end{equation}
It should be noted that similar to the $\mathbf{U}(1)\times \mathbf{SU}(2)$ theory, substituting $\sigma_{2}\otimes\sigma_{1},\sigma_{2}\otimes\sigma_{3}\hspace{4pt} \text{and}\hspace{4pt} \sigma_{0}\otimes\sigma_{2}$ in place of $\sigma_{1}\otimes\sigma_{2},\sigma_{2}\otimes\sigma_{0}\hspace{4pt}\text{and}\hspace{4pt} \sigma_{3}\otimes\sigma_{2}$ respectively, results in a different set of Killing vector fields. Since the killing vector fields are linearly independent, we can use them to construct the metric of the $\mathbf{U}(1)\times \mathbf{SU}(2)\times \mathbf{SU}(3)$ manifold by following the same procedure outlined in the case of $\mathbf{U}(1)\times \mathbf{U}(1)$  and $\mathbf{U}(1)\times \mathbf{SU}(2)$. Thus the existence of these Killing vector fields guarantees that the standard model can be realized in $12\mathcal{D}$ internal space. 

What we have done amounts to introducing non-coordinate basis and demanding that the structure coefficients become structure constants. The requirement for this is that the space in question is at least simple transitive (the number of Killing vector fields is the same as the dimension of the manifold).

\section{Spin as spacetime symmetry}\label{spinAsSTSym}
In \autoref{4delectrodynamics} we saw that the expression for linear momentum density contains a divergence free part. While this divergence free part does not contribute to linear momentum, we saw it has a non-trivial contribution to angular momentum. Through the calculation of current density in the case of spin-$\frac{1}{2}$ particles, it was evident that this contribution is the spin part of angular momentum. This section is aimed at showing that this is not a coincidence. In addition to this, we show that spin is a consequence of spacetime symmetries just like orbital angular momentum. We show that this approach to spin results in the unified description of particles. That is both Bosons and Fermions  are vector fields, associated with non-equivalent sets of Killing vector fields forming the representation of the Lie algebra of $\mathbf{SO}(3)$.

In the previous section we saw that if the infinitesimal transformation
\begin{equation}\label{Isommetry}
    x^{\nu}\longmapsto \Tilde{x}^{\mu} = x^{\mu} + \epsilon^{i}\xi_{i}^{\hspace{4pt}\mu}
\end{equation}
is the isometry of the metric, the 1-form field $A_{i\mu}$ which is the eigen-vector of the $j^{th}$ Killing vector field $\xi_{i}$ obeys the equation
\begin{equation}\label{Kill2}
    \xi_{j}^{\hspace{4pt}\nu}\partial_{\nu}A_{i\mu} + \partial_{\mu}\xi_{j}^{\hspace{4pt}\nu}A_{i\nu} = 0.
\end{equation}
This equation just expresses the fact that $A_{i\mu}$ is invariant under this coordinate transformation. We show this as follows:
Let $\Lambda$ be the transformation matrix, we have
\begin{equation}
    A_{i\mu}(x)\longmapsto \Tilde{A}_{i\mu}(\Tilde{x}) = \Lambda_{\mu}^{\hspace{4pt}\nu}A_{i\nu}(\Lambda x).
\end{equation}
The infinitesimal transformations are
\begin{equation}
    \Lambda_{\mu}^{\hspace{4pt}\nu}=\delta_{\mu}^{\nu} +\epsilon^{i}\partial_{\mu}\xi_{i}^{\hspace{4pt}\nu}.
\end{equation}
Accordingly,
\begin{equation}\label{LocalKillT}
    \Tilde{A}_{i\mu}(\Tilde{x})= A_{i\mu}(x) +\epsilon^{j}\left(\partial_{\mu}\xi_{j}^{\hspace{4pt}\nu}A_{i\nu} +\xi_{j}^{\hspace{4pt}\nu}\partial_{\nu}A_{i\mu}\right).
\end{equation}
From this it is clear that invariance of $A_{i\mu}(x)$ under \eqref{LocalKillT} implies \eqref{Kill2}. The transformation of $A_{i\nu}$ consists of two parts:
\begin{itemize}
    \item the part that transforms the arguments and
    \item the part that transforms the components.
\end{itemize}
The generators of the transformations that act on components ($\partial_{\mu}\xi_{j}^{\hspace{4pt}\nu}$) are independent of coordinates in flat spacetime, as such they constitute spin part of the angular momentum operator in the case of rotations. The generators of transformations acting on the arguments may depend on coordinates. In the case of rotations these generators are orbital angular momentum operators.
We demonstrate this for the case of Lorentz transformations. Under the infinitesimal Lorentz transformations the coordinates transform in the following way
\begin{equation}
x^{\mu}\longmapsto \Tilde{x}^{\mu}= x^{\mu} +\alpha_{\nu}^{\hspace{4pt}\mu}x^{\nu}   
\end{equation}
where $\alpha_{\mu}^{\hspace{4pt}\nu}$ are constant parameters and $\alpha_{\mu\nu}$ are anti symmetric. In order for us to be able to identify the Killing vector fields, we express the transformations in such a way that the parameters do not carry the coordinate labels. This results in
\begin{equation}\label{InfLorentz}
    \begin{aligned}
    \Tilde{x}^{\mu}&= x^{\mu} + \frac{1}{2}\alpha^{\alpha\beta}\left(\delta_{\alpha}^{\mu}\eta_{\beta\nu}-\delta_{\beta}^{\mu}\eta_{\alpha\nu} \right)x^{\nu}\\
    &=x^{\mu} + \frac{1}{2}\alpha^{\alpha\beta}\mathrm{M}_{\alpha\beta|\nu}^{\hspace{15pt}\mu}x^{\nu}
    \end{aligned}
\end{equation}
where $\mathrm{M}_{\alpha\beta|\nu}^{\hspace{15pt}\mu} = \delta_{\alpha}^{\mu}\eta_{\beta\nu}-\delta_{\beta}^{\mu}\eta_{\alpha\nu}$.
Comparing \eqref{InfLorentz} with \eqref{Isommetry} we get the mapping
\begin{equation}
    \begin{aligned}
        i\mapsto \alpha\beta && \epsilon^{i} \mapsto \frac{1}{2}\alpha^{\alpha\beta} && \xi_{i}^{\hspace{4pt}\mu}\mapsto \mathrm{M}_{\alpha\beta|\nu}^{\hspace{15pt}\mu}x^{\nu}.
    \end{aligned}
\end{equation}
We see that the Killing vector fields are the generators of isometries of the Minkowski metric.  That is, generators of rotations and boosts in Minkowski space. The $1$-forms $A_{\alpha\beta}$ dual to these Killing vector fields have components
\begin{equation}
    A_{\alpha\beta\mu} =\mathrm{M}_{\alpha\beta|\nu\mu}x^{\nu}
\end{equation}
and the conserved charges are
\begin{equation}
    Q_{\alpha\beta} = \mathrm{M}_{\alpha\beta|\nu\mu}x^{\nu} \frac{dx^{\mu}}{d\tau}.
\end{equation}
These are the generalization of usual momenta to Minkowski space. The reader should be aware that these conserved charges are scalars since $\alpha\beta$ is just a label not coordinate index. This is contrary to classical mechanics where one views them as vectors. We know that this classical view is incorrect since  for it to hold one has to introduce the concept of a position vector and such an  object does not exist since in general coordinates do not transform like vectors. It is only in the case of rotations in which the parameters (angles of rotation) are independent of coordinates that they transform like vectors since even in the familiar case of translations the coordinate transformation is
\begin{equation}
    x^{\mu}\longmapsto \Tilde{x}^{\mu} = x^{\mu} + a^{\mu}
\end{equation}
for some constants $a^{\mu}$. Clearly this is not the transformation law of vectors. 

Substituting the expression of the Killing vector fields in \eqref{Kill2} we get
\begin{equation}
    \mathrm{M}_{\alpha\beta|\rho}^{\hspace{15pt}\mu}x^{\rho}\partial_{\mu}A_{\nu} +\mathrm{M}_{\alpha\beta|\nu}^{\hspace{15pt}\mu}A_{\mu} = 0
\end{equation}
with the labels ($\alpha\beta)$ of the $1$-forms $A_{\alpha\beta}$ suppressed since the equations are valid for all of them. We can rewrite the above equation as
\begin{equation}
    \begin{aligned}
       & \left(\delta_{\nu}^{\delta}\mathrm{M}_{\alpha\beta|\rho}^{\hspace{15pt}\mu}x^{\rho}\partial_{\mu} + \mathrm{M}_{\alpha\beta|\nu}^{\hspace{15pt}\delta}\right)A_{\delta} = 0,\\
       &\left(\mathbf{1}_{4}\otimes \mathrm{M}_{\alpha\beta|\rho}^{\hspace{15pt}\mu}x^{\rho}\partial_{\mu} + M_{\alpha\beta} \right)A = 0.\\
    \end{aligned}
\end{equation}
We realise that this is just the equation for conservation of charge \eqref{spin} we found in the previous section. It is also clear that the charge conservation equation simply means that the Lie derivative of the 1-form describing charged matter fields along the Killing vector fields associated with such charges vanishes. That is, $L_{\xi_{i}}A = 0$. Considering only spatial rotations we get
\begin{equation}\label{spin1}
    \left(\mathbf{1}_{3}\otimes\tau_{i|j}^{\hspace{8pt}k}x^{j}\partial_{k} + \tau_{i}\right)A = 0
\end{equation}
where $\tau_{i|jk} = \imath\varepsilon_{ijk}$. That is the matrices $\lbrace\tau_{i}\rbrace$ form the spin-$1$ representation of $\mathfrak{su}(2)$. 

If the matrix $A$ is the solution of \eqref{spin1}, it is clear that $\phi(|\mathbf{x}|)^{-1}A$ is also a solution. Since $A$ is invertible (because it is made up of linearly independent solutions of \eqref{spin1}) we see from
\begin{equation}
    \tau_{i} = \tau_{i|j}^{\hspace{9pt}k}x^{j}\partial_{k}A A^{-1}
\end{equation}
that for some $\hat{\Psi}\in SO(3)$ and $\phi = \phi(|\mathbf{x}|)$, we have 
\begin{equation}
    A = \phi \hat{\Psi}.
\end{equation} 
In \autoref{4delectrodynamics} we saw that a spin-$0$ charged particle is described by the complex field $\Psi =\phi e^{\imath S}$ where $\phi$ came from the conformal factor in the expression of the metric. In \autoref{ElectroGrav5D} it was shown that $\phi^{2}$ corresponds to particle number density while the phase $\hat{\psi}= e^{\imath S}$ is the eigenfunction of the charge operator (the Killing vector in the internal space). We see the same thing here except that the phase is the unitary matrix and we are considering the isometries of the usual $4\mathcal{D}$ spacetime. Thus charge and spin are similar concepts in that they arise from the isometries of space. What differentiates them is that one reflects the symmetry of the internal space while the other reflects the isometries of ordinary spacetime. 
This leads us to defining fields that describe elementary particles as $1$-forms invariant under a group of isometries of spacetime which are generated by to the set of Killing vector fields which form a representation of some Lie algebra of this group of isometries. Due to the fact that we can have sets of Killing vector fields which are not equivalent but have the same Lie algebra for a given manifold, it is clear that different types of particles correspond to different sets of Killing vector fields. In order to determine whether or not two sets of Killing vector fields are equivalent we have to find the quantity  associated with each of them which is a scalar under coordinate transformations. One such a quantity is circulation 
\begin{equation}
    C = \oint A_{i\mu}dx^{\mu} =Q_{i}\oint d\tau.
\end{equation}

For further analysis we consider the form of Killing (charge/spin conservation) equation given by \eqref{spin}.

We can rewrite \eqref{spin} as
\begin{equation}
   \xi^{\hspace{4pt}j}_{i}\partial_{j}\hat{\Psi}_{kl} = \imath\sigma\tau_{j|k}^{\hspace{9pt}m}\hat{\Psi}_{ml}.
\end{equation}
 We see that every column of $\hat{\Psi}$ is a solution to \eqref{spin}. As such, we can write \eqref{spin} in terms of one of the column vector $\hat{\varphi}$ of $\hat{\Psi}$ as
\begin{equation}\label{spinEigenVproblem}
    \mathbf{1}_{n}\otimes\xi_{i}^{\hspace{4pt}j}\partial_{j}\hat{\varphi} = \imath\sigma\tau_{i}\hat{\varphi}.
\end{equation}
We note that the condition $\hat{\Psi}^{\dagger}\hat{\Psi} = 1$ translates to  $\hat{\varphi}^{\dagger}\hat{\varphi} = 1$.
Evidently \eqref{spinEigenVproblem} is an eigenvalue equation. We go about solving it by performing the similarity transformation which diagonalizes $\tau_{i}$. In addition to this we find the coordinate system adapted to the Killing vector fields. These are coordinates in which the Killing vector fields can be written as
\begin{equation}
    \xi_{i} = \delta_{i}^{\hspace{4pt}j}\partial_{j}.
\end{equation}
That such a coordinate system exists follows from the fact that the metric is constant along the integral curves of the Killing vector fields,  thus if we choose the integral curve of $\xi_{i}$ as the coordinate line we must have
\begin{equation}
    \partial_{i}g_{kl} = 0.
\end{equation}
For non-commuting Killing vector fields, we can have coordinate system adapted to only one of them at a time. This follows from the fact that in the coordinate system adapted to Killing vector fields, the Killing vector fields are just partial derivatives and the partial derivatives commute. That partial derivatives commute is clear from the fact that the Lie derivative is defined for functions which are at least twice differentiable. Since the Lie bracket of vector fields is a vector field, if it vanishes in one coordinate system it vanishes in all coordinate systems. Therefore we can have coordinates adapted to $n$-Killing vector fields if they commute.
In these coordinates \eqref{spinEigenVproblem} becomes
\begin{equation}\label{adaptedKillingVectors}
     \mathbf{1}_{n}\otimes\partial_{i}\hat{\varphi}_{\text{\tiny{D}}} =\imath\tau^{\text{\tiny{D}}}_{i}\hat{\varphi}_{\text{\tiny{D}}}.
\end{equation}
where $\tau_{i}^{\text{\tiny{D}}} = \mathbf{S}^{-1}\tau_{i}\mathbf{S}$ and $\hat{\varphi}_{\text{\tiny{D}}} = \mathbf{S}^{-1}\hat{\varphi}$ with $\mathbf{S}$ as the similarity transformation. 
 We commence exploring the implications of these equations by considering the isometries of $3\mathcal{D}$ Euclidean metric. The Euclidean metric can be constructed from the translation Killing vector fields. Therefore we have
 \begin{equation}
     \begin{aligned}
         \xi_{i}^{\hspace{4pt}j} = \sigma\delta_{i}^{j} && \hspace{8pt}\text{and} && \tau_{i|k}^{\hspace{9pt}j} = p_{i}\delta_{k}^{j}.
     \end{aligned}
 \end{equation}
 The solution of \eqref{spin} is
 \begin{equation}\label{linear momentum}
     \hat{\Psi} = \mathbf{1}_{n}e^{\imath p\cdot x}
 \end{equation}
 Thus the free particles are described by plain waves. Making the identification
 \begin{equation}
     x^{i} = x^{i} + 2\pi l^{i}
 \end{equation}
 for some constant $l^{i}\in \mathbb{Z}$, we see that $p_{i}$ component of momentum gets quantized. Since there exists a coordinate system in which at least one of the Killing vector fields has \eqref{linear momentum} as the solution, it is clear that quantization arises from compactification of some dimensions thus if our universe is compact then all conserved charges are quantized.  
 
The other isometries of Euclidean metric are rotations for  which the Killing vector fields are $\xi_{i} = \sigma\varepsilon_{ik}^{\hspace{9pt}j} x^{k}\partial_{j}$. One can verify that their Lie derivative is non-zero as such we can find the coordinate system adapted to only one of the Killing vector fields at a time. One such coordinate system is the spherical coordinate system and the Killing vector field the coordinate system is adapted to is $\xi_{\phi} = \sigma\partial_{\phi}$.  Also the $\tau_{i}$ whose components are $\tau_{i|kj} =\imath\varepsilon_{ikj} $ do not commute thus $\hat{\varphi}$ is the eigenvector of only one of them. Following convention we choose this to be
\begin{equation}
    \tau_{3} = \begin{pmatrix}
        0 && -\imath && 0\\
        \imath && 0 && 0\\
        0 && 0 && 0
    \end{pmatrix}
\end{equation}
 and use 
 \begin{equation}
     S =\frac{1}{\sqrt{2}} \begin{pmatrix}
         -1 && 0 && 1\\
          \imath && 0 && \imath\\
         0 && \sqrt{2} && 0
     \end{pmatrix}
 \end{equation}
 to diagonalize it.
 Equation \eqref{spinEigenVproblem} becomes
 \begin{equation}
     \begin{pmatrix}
         \partial_{\phi} && 0 && 0\\
         0 && \partial_{\phi} && 0\\
         0 &&  0 &&  \partial_{\phi}
     \end{pmatrix}\begin{pmatrix}
         \hat{\varphi}_{\text{\tiny{D}}}^{-}\\
         \hat{\varphi}_{\text{\tiny{D}}}^{0}\\
         \hat{\varphi}_{\text{\tiny{D}}}^{+}
     \end{pmatrix} =\imath\begin{pmatrix}
         -\hat{\varphi}_{\text{\tiny{D}}}^{-}\\
         0\\
         \hat{\varphi}_{\text{\tiny{D}}}^{+}
     \end{pmatrix}
 \end{equation}
 the solution of which is
 \begin{equation}
    \hat{\varphi}_{\text{\tiny{D}}} = \frac{1}{\sqrt{2}}\begin{pmatrix}
        f(\theta)e^{\imath\phi}\\
        \sqrt{2}g(\theta)\\
        h(\theta)e^{-\imath\phi}
    \end{pmatrix}
 \end{equation}
 with \[|f(\theta)|^2 + 2|g(\theta)|^2 + |h(\theta)|^2 = 2.\] We recognize the solutions are the eigenvectors of $\hat{n}\cdot\tau^{\text{\tiny{D}}}$ for an arbitrary normal $\hat{n}$. As such $\hat{\varphi}_{\text{\tiny{D}}}$ describes spin-1 particles. From
 \begin{equation}
     \hat{\varphi} = \frac{h(\theta)}{{2}}\begin{pmatrix}
         1\\
         0\\
         \imath
     \end{pmatrix}e^{-\imath\phi} -\frac{f(\theta)}{{2}}\begin{pmatrix}
         1\\ 0\\ -\imath
     \end{pmatrix}e^{\imath\phi} +g(\theta)\begin{pmatrix}
         0\\ 1\\ 0
     \end{pmatrix}
 \end{equation}
we see that $\hat{\varphi}$ is the superposition of linear and circular polarization. If the spin-1 particle under consideration is a photon, then $\hat{\varphi}$ is the electromagnetic potential. The linearly independent solutions are the columns of 
\begin{equation}
    \begin{aligned}
    \hat{\Psi} =& \frac{1}{\sqrt{2}}\begin{pmatrix}
        e^{\imath\phi} && 0 && 0\\
        0 && 1 && 0 \\
        0 && 0 && e^{-\imath\phi}
    \end{pmatrix}\\ &\times\begin{pmatrix}
        1-\cos{\theta} && -\sin{\theta} && 1+\cos\theta\\
        \sqrt{2}\sin{\theta}&& \sqrt{2}\cos\theta && -\sqrt{2}\sin{\theta}\\
        1+\cos\theta && \sin\theta && 1-\cos\theta
    \end{pmatrix}.
    \end{aligned}
\end{equation}

 Having dealt with spin-1 particles, we show that the spin-$\frac{1}{2}$ particles can be described in the same manner. All we need is three Killing vector fields
 \begin{equation}
     \xi_{i} = \mathrm{C}_{i\nu}^{\hspace{9pt}\mu}x^{\nu}\partial_{\mu}
 \end{equation}
 whose Lie bracket is
 \begin{equation}
     [\xi_{i}, \xi_{j}] = \varepsilon_{ij}^{\hspace{10pt}k}\xi_{k}
 \end{equation}
 and the matrices $\tau_{i}$ whose components are
 \begin{equation}
     \tau_{i|\mu}^{\hspace{9pt}\nu} = \imath\mathrm{C}_{i\mu}^{\hspace{9pt}\nu}
 \end{equation}
 and have only two distinct eigenvalues. From the previous section we see that the Killing vector fields
 \begin{equation}\label{halfAngularM}
     \xi_{i} = \frac{1}{2}(x_{5}\partial_{i}-x_{i}\partial_{5} +\varepsilon_{ij}^{\hspace{10pt}k}x^{j}\partial_{k})
 \end{equation}
 have these properties
 and the matrices $\tau_{i}$ are $\tau_{6}, \tau_{7} \hspace{4pt}\text{and}\hspace{4pt} \tau_{8}$ from \eqref{4by4Electroweak} scaled by $\frac{1}{2}$.  To avoid confusion we use the matrices with no scale factor and refer to them by the numbers $1,2 \text{\hspace{4pt}and\hspace{4pt}} 3$ respectively. It should be noted that these Killing vector fields exist in $4\mathcal{D}$ Euclidean space not in $4\mathcal{D}$ Minkowski space. While one can verify this by computing the Lie derivative of Minkowski metric along the Killing vector fields, this is obvious from the fact that contrary to the case of Minkowski space, the first two terms form generators of rotations not boosts. In accordance with this, if we insist that space is flat we need an extra spacelike dimension to describe spin-$\frac{1}{2}$ particles in this way. While \eqref{spinEigenVproblem} assumes a simple form in coordinates adapted to the Killing vector field, finding such a coordinate system can be quite challenging. However we can get around this by finding the solution in  Cartesian coordinate system and  re-parameterize the solution so that in terms of the new parameters the Killing vector field of interest is just the partial derivative so that for that Killing vector field $\hat{\varphi}$ solves \eqref{adaptedKillingVectors}.
  Diagonalizing $\tau_{3}$ with  
 \begin{equation}
     S = \frac{1}{\sqrt{2}}\begin{pmatrix}
         1 && 1 && 0 &&  0\\
         0 && 0 && 1 &&  1\\
         0 && 0 && -\imath && \imath\\
         \imath && -\imath && 0 && 0
     \end{pmatrix}
 \end{equation}
 and solving the resulting equation, we find that the coordinate adapted to $\xi_{3}$ are the Hopf coordinates
 \begin{equation}
     \begin{aligned}
         x^{5} &= r\cos{(\beta/2)}\cos{[(\alpha +\gamma)/2]}\\
         x^{1} &= r\sin{\beta/2}\cos{([\gamma - \alpha)/2]}\\
         x^{2} &= r\sin{(\beta/2)}\sin{[(\gamma-\alpha)/2]}\\
         x^{3} &= r\cos{(\beta/2)}\sin{([\alpha +\gamma)/2]}
     \end{aligned}
 \end{equation}
with $\beta$,$\alpha$ and $\gamma$ ranging over $[0,\pi]$,  $[0,2\pi]$ and $[0,2\pi]$ respectively. We set $r =1$ restricting motion on the $3$-sphere. In this coordinate system one of the solutions is
 \begin{equation}
     \hat{\varphi}_{\text{\tiny{D}}} = \frac{1}{\sqrt{2}}\begin{pmatrix}
         e^{-\imath(\alpha +\gamma)/2}\cos{(\beta/2)}\\
         e^{\imath(\alpha-\gamma)/2}\sin{(\beta/2)}\\
        - e^{-\imath(\alpha -\gamma)/2}\sin{(\beta/2)}\\
         e^{\imath(\alpha +\gamma)/2}\cos{(\beta/2)}
     \end{pmatrix}
 \end{equation}
 which we recognize is the eigenstate of spin operator for a system consisting of two spin-$\frac{1}{2}$ particles in the direction of an arbitrary unit normal. What is remarkable about this is that when we associate particles with spacetime symmetry the spin-$\frac{1}{2}$ particles come in pairs since we can not furnish a real 2-dimensional matrix representation of $\mathfrak{so}(3)$.  The Killing vector fields \eqref{halfAngularM} in this coordinate system are
 \begin{equation}\label{LIndependentKillVFs}
   \begin{aligned}
       \xi_{3} &= -\partial_{\alpha},\\
       \xi_{2} &=  \frac{1}{\sin{\beta}}\left(\cos\alpha\cos{\beta}\partial_{\alpha} +\sin{\alpha}\sin{\beta}\partial_{\beta} - {\cos{\alpha}}\partial_{\gamma}\right),\\
       \xi_{1} &=\frac{1}{\sin{\beta}}\left(\sin{\alpha}\cos{\beta}\partial_{\alpha} - \cos{\alpha}\sin{\beta}\partial_{\beta}-{\sin{\alpha}}\partial_{\gamma}\right).
   \end{aligned}  
 \end{equation}
 In addition to this it is clear that we have another Killing vector field $\xi_{4} = \partial_{\gamma}$ which commutes with all of the above Killing vector fields.
 From this it is clear that to describe spin-$\frac{3}{2}$ all we need are the Killing vector fields with angular momentum commutation relations from which we can construct $\tau_{i}$s with four distinct eigenvalues and the same procedure is applicable to higher angular momenta. We also recognise that in $4\mathcal{D}$ space fundamental particles can have spin $0,\frac{1}{2},1$. From the fact that $\mathfrak{su}(2)$ does not have a $4\mathcal{D}$ irreducible real representation, we can not have spin-$\frac{3}{2}$ elementary particles in $4\mathcal{D}$ space. In line with these, particles with spin higher than $1$ can be realized as composite particles in $5\mathcal{D}$ spacetime. To clarify what we mean, we note that if spacetime is a $6\mathcal{D}$ flat manifold, in addition to the Killing vector fields for spin-$1$ and spin-$\frac{1}{2}$, we can have 
 \begin{equation}
    \begin{aligned}
     \xi_{1} &= x_{1}\partial_{2}-x_{2}\partial_{1} + x_{5}\partial_{3}-x_{3}\partial_{5} + \sqrt{3}\left(x_{3}\partial_{6}-x_{6}\partial_{3}\right),\\
     \xi_{2} &= x_{3}\partial_{1}-x_{1}\partial_{3} +x_{5}\partial_{2}-x_{2}\partial_{5}+\sqrt{3}(x_{6} \partial_{2} -x_{2}\partial_{6}),\\
     \xi_{3} &= x_{3}\partial_{2}-x_{2}\partial_{3} + 2(x_{5}\partial_{1}-x_{1}\partial_{5}),
     \end{aligned}
 \end{equation}
 which form the representation of $\mathfrak{so}(3)$
 together with the matrices 
 \begin{equation}
     \begin{aligned}
         \tau_{1}& = \begin{pmatrix}
             0 && 1 && 0 && 0 && 0\\
             -1 && 0 && 0 && 0 && 0\\
             0 && 0 && 0 &&-1 &&\sqrt{3}\\
             0 && 0 && 1 && 0 && 0\\
             0 && 0 &&-\sqrt{3}&& 0 && 0
         \end{pmatrix}\\
     \tau_{2} &=\begin{pmatrix}
         0 && 0 && -1 && 0 && 0\\
         0 && 0 && 0  && -1 && -\sqrt{3}\\
         1 && 0 && 0  && 0 && 0\\
         0 && 1 && 0  && 0 && 0\\
         0 && \sqrt{3}&& 0 && 0 &&0
     \end{pmatrix}\\
     \tau_{3}&=\begin{pmatrix}
         0 && 0 && 0 && 2 && 0\\
         0 && 0 && 1 && 0 && 0\\
         0 && -1 && 0 && 0 && 0\\
         2 && 0 && 0 && 0 && 0\\
         0 && 0 && 0 && 0 && 0
     \end{pmatrix}
      \end{aligned}
 \end{equation}
 which form the $j = 2$ representation of $\mathfrak{so}(3)$ and \eqref{spinEigenVproblem} in this case describes a spin-$2$ field. Thus we can have a spin-$2$ particle as an elementary particle in $6\mathcal{D}$ spacetime. In $4\mathcal{D}$ spacetime it can be realized as a composite particle in the following way;
 From the $1$-form $A$ describing a spin-$1$ particle, we can form the second rank tensor
 \begin{equation}
     T = A_{i}A_{j}dx^{i}\otimes dx^{j}.
 \end{equation}
 By construction $T_{ij}$ is invariant under a group of isometries generated by the Killing vector fields associated with spin-$1$ fields. Due to this it forms a 9-dimensional reducible representation of $\mathfrak{so}(3)$ since
 \begin{equation}
     \begin{aligned}
         L_{\xi_{j}}T_{ik} &= \xi_{j}^{\hspace{4pt}l}\partial_{l}T_{ik} +\partial_{i}\xi_{j}^{\hspace{4pt}l}T_{lk} +\partial_{k}\xi_{j}^{\hspace{4pt}l}T_{il} \\
         &=\left(\delta_{i}^{n}\delta_{k}^{p}\epsilon_{jm}^{\hspace{8pt}l}x^{m}\partial_{l} + \delta_{k}^{p}\epsilon_{ji}^{\hspace{8pt}n} +\delta_{i}^{n}\epsilon_{jk}^{\hspace{8pt}p} \right)T_{np}\\
         &=\left(\mathbf{1}\otimes\mathbf{1}\otimes \epsilon_{jm}^{\hspace{8pt}l}x^{m}\partial_{l} -\imath \tau_{j}\otimes\mathbf{1}-\imath\mathbf{1}\otimes\tau_{j} \right)_{ik}^{\hspace{8pt}np}T_{np}.
     \end{aligned}
 \end{equation}
 
 We can split $T_{ij}$ into the anti-symmetric part, the symmetric trace free part and the trace part as follows
 \begin{equation}
     T_{ik} = \frac{1}{2}\left(T_{ik}-T_{ki}\right) + \frac{1}{2}\left(T_{ik} + T_{ki} - \frac{\delta_{ik}}{3}T^{j}_{j}\right) + \frac{\delta_{ik}}{6}T^{j}_{j}.
 \end{equation}
 These three parts do not mix under the group of isometries in question. The interesting thing is that although the collective behaviour is that of a tensor product, the behaviour of individual parts is different. For instance the trace part behaves like a spin-$0$ field, the anti-symmetric part behaves like the original spin-$1$ field from which  it was constructed, since it is dual to the 1-form. The symmetric trace free part has $5$-components. Since it cannot be reduced to something with fewer components it must be the case that it is invariant under the group of isometries generated by the Killing vector fields of the spin-$2$ field. Thus in this case we realize the spin-$2$ field as arising from the interactions of two spin-1 fields thus we call the corresponding particle a composite particle.

 Since to accommodate spin-$\frac{1}{2}$ particles we need an extra dimension, it is clear that all spin-$\frac{1}{2}$ particles have charge. This is the charge associated with the gauge Boson corresponding to the extra dimension.

In the case of spin-$\frac{1}{2}$ we notice that the field is
\[\Psi = \phi_{5} e^{\imath\frac{\tau\cdot\mathbf{\phi}}{\phi_{5}}},\]
 and there exists a vector $\mathbf{u}$ and a scalar ${u}_{5}$ such that     \cite{cheng2000gauge}
\begin{equation}
    \Psi = {u}_{5} +\imath\tau\cdot\mathbf{u}.
\end{equation}
With $\phi = 1$, the similarity transformation of a vector $\mathbf{v}$ is
\begin{equation}\label{vectorSpinorRelation}
    \begin{aligned}
      \tau_{i}\otimes\mathbf{v}_{i} =  \tau\cdot\mathbf{v}\longmapsto& \left({u}_{5} +\imath\tau\cdot\mathbf{u}\right)\tau\cdot\mathbf{v}\left({u}_{5} -\imath\tau\cdot\mathbf{u}\right)\\
        &= \tau_{i}\otimes\Big[\left(1-2{\mathbf{u}}^2\right)\mathbf{1}_{3}- 2\sqrt{\left(1-\mathbf{u}^2\right)}\mathbf{u}\cdot\mathbf{L}\\ &+(\mathbf{u}\cdot\mathbf{L})^2\Big]_{i}^{\hspace{4pt}k}\mathbf{v}_{k}
    \end{aligned}
\end{equation}
where $\mathbf{L}_{i|jk} = \varepsilon_{ijk}$. One can verify that the expression in the square brackets is the general form of matrix representation of $\mathbf{SU}(2)$ for spin-1. Since the spinor is a column of $\Psi$ the above transformation provides us a means of constructing the spinor from a vector and vice versa. We will exploit this to find the equations of motion for spin-$\frac{1}{2}$ fields. Also from the above transformation we can surmise that the total angular momentum $j$ in this case ranges over both integer and half integer values. We show that this is indeed the case by finding the eigenvalues of the Casimir operator $\mathcal{J}^2 = -\xi_{i}\xi^{i}$ constructed from \eqref{halfAngularM} and for ease of comparison we use spherical coordinates.

In spherical coordinates we have 
\begin{equation}
    \begin{aligned}
        x^{5} &= r\cos{(\beta/2)},\\
        x^{1} &= r\sin\theta\cos\alpha\sin{(\beta/2)},\\
        x^{2} &= r\sin{\theta}\sin{\alpha}\sin{(\beta/2)},\\
        x^{3} &= r\cos\theta\sin(\beta/2)
    \end{aligned}
\end{equation}
 where $\alpha,\beta$ range over $[0,2\pi]$ and $\theta$ ranges over $[0,\pi]$. The Killing vector fields \eqref{halfAngularM} become
 \begin{equation}
     \begin{aligned}
         \xi_{1} &= \sin\theta\sin\alpha\partial_{\beta} + \frac{1}{2}\big(\sin\alpha + \cos\theta\cos\alpha\cot(\beta/2)\big)\partial_{\theta}\\ & +\frac{1}{2}\big(\cot\theta\cos\alpha - \cot(\beta/2)\frac{\sin\alpha}{\sin\theta}\big)\partial_{\alpha},\\
         \xi_{2} &= \sin\theta\sin\alpha\partial_{\beta} + \frac{1}{2}\big(-\cos\alpha + \cos\theta\sin\alpha\cot(\beta/2)\big)\partial_{\theta} \\ &+\frac{1}{2}\big(\cot\theta\sin\alpha +\cot(\beta/2)\frac{\cos\alpha}{\sin\theta}\big)\partial_{\alpha},\\
         \xi_{3} &= \cos\theta\partial_{\beta} - \frac{1}{2}\sin{\theta}\cot(\beta/2)\partial_{\theta} -\frac{1}{2}\partial_{\alpha}.
     \end{aligned}
 \end{equation}
Using the Killing vector fields to write the Casimir operator, we end up with 
 \begin{equation}
     \mathcal{J}^2 = -\frac{1}{4\sin^2(\beta/2)}\Big[\partial_{\beta}\big(4\sin^2(\beta/2)\partial_{\beta}\big) +\frac{1}{\sin\theta}\partial_{\theta}\big(\sin\theta\partial_{\theta}\big) +\frac{1}{\sin^2\theta}\partial_{\alpha}^2\Big].
 \end{equation}
 One can easily solve the equation
 \begin{equation}
     \mathcal{J}^2{\Psi} = j(j+1){\Psi}
 \end{equation}
by separation of variables. In order to illustrate that angular momentum in this case is $j = 0,\frac{1}{2},1,\cdots$, we consider  ${\Psi} =\Psi(\beta)$ since for constant $\beta$, $\mathcal{J}^2$ reduces to the Casimir operator of $\mathfrak{so}(3)$ in which case it is common knowledge that $j$ takes only integer values. We have
\begin{equation}\label{PsiBetaEOM}
    \partial_{\beta}\left(\sin^2{(\beta/2)}\partial_{\beta}\right)\Psi(\beta) + j(j+1)\sin^2(\beta/2)\Psi(\beta) = 0.
\end{equation}
Making the substitution \[\Psi(\beta) = \frac{B(\beta)}{\sin(\beta/2)}\]
we get 
\begin{equation}
    \partial_{\beta}^2B +(j +\frac{1}{2})^2B = 0
\end{equation}
whose solution is
\begin{equation}
    B =  a_{1}\cos{\Big[\left(j+\frac{1}{2}\right)\beta\Big]} +a_{2}\sin{\Big[\left(j+\frac{1}{2}\right)\beta\Big]}.
\end{equation}
Therefore
\begin{equation}
    \Psi^{(j)} = a_2\frac{\sin{\Big[\left(j+\frac{1}{2}\right)\beta\Big]}}{\sin(\beta/2)}.
\end{equation}
We set $a_{1} = 0$ to get rid of the divergent term. With $\Psi^{(\Tilde{j})}(\beta)$ as  another solution of \eqref{PsiBetaEOM} we get
\begin{equation}
 \begin{aligned}
    \sin^2(\beta/2)\Big(\Psi^{(\Tilde{j})}\partial_{\beta}\Psi^{(j)} &- \Psi^{(j)}\partial_{\beta}\Psi^{(\Tilde{j})}\Big)\Big|_{0}^{2\pi}\\ &=a_{2}\Tilde{a}_{2}\left(\Tilde{j}(\Tilde{j}+1) - j(j+1)\right)\\
    &\times\int_{0}^{2\pi}\sin{\Big[\left(j+\frac{1}{2}\right)\beta\Big]}\sin{\Big[\left(\Tilde{j}+\frac{1}{2}\right)\beta\Big]}d\beta\\
    &= 0
    \end{aligned}
\end{equation}
which results in 
\begin{equation}
    \sin\big[(\Tilde{j}+\frac{1}{2})2\pi\big]\cos\big[(j+\frac{1}{2})2\pi\big]-\sin\big[(j+\frac{1}{2})2\pi\big]\cos\big[(\Tilde{j}+\frac{1}{2})2\pi\big] = 0
\end{equation}
from which follows
\begin{equation}
    \begin{aligned}
    j = \frac{1}{2}(n - 1)
    \end{aligned}
\end{equation}
with $n\in\mathbb{Z}^{+}$.
Therefore in this case $j=0,\frac{1}{2},1,\cdots$ unlike in the case of $\xi_{i} = \varepsilon_{ij}^{\hspace{9pt}k}x^{j}\partial_{k}$ where it takes only integer values. If we set $\beta = constant$, the Killing vector fields reduce to $\xi_{i} = \varepsilon_{ij}^{\hspace{9pt}k}x^{j}\partial_{k}$. Thus we see that half-integer angular momentum is due to motion in the extra dimension and $\Psi^{(j)} = \Psi^{(j)}(\alpha,\beta,\theta)$ are the generalization of spherical Harmonics. Thus massless spin-1 particles have two degrees of freedom while the two spin-$\frac{1}{2}$ particles have 3 degrees of freedom. This is more apparent in the coordinate system in which the metric of the $3$-sphere is \eqref{s3metric} since in such a coordinate system the Killing vectors have the form
\begin{equation}
    \xi_{i} = \frac{1}{2}\sqrt{1-k|\mathbf{x}|^2}\partial_{i} +\frac{\sqrt{k}}{2}\varepsilon_{ij}^{\hspace{9pt}k}x^{j}\partial_{k}
\end{equation}
and we get the usual generators of rotation  for motion restricted to the 2-sphere ($k|\mathbf{x}|^2 = 1$). 

 By writing \eqref{spinEigenVproblem} as
 \begin{equation}\label{componentsSpin}
     \xi_{3}^{j}\partial_{j}\hat{\varphi}_{l} = \imath\sigma\lambda_{l}\hat{\varphi}_{l}
 \end{equation}
where $\tau_{3|ij}^{\text{\tiny{D}}} = \lambda_{i}\delta_{ij}$, it becomes clear that the components $\lbrace\hat{\varphi}_{l}\rbrace$ of $\hat{\varphi}$ are eigenfunctions $\lbrace{\Psi^{(j)}_{l}}\rbrace$ of $\xi_{3}$ for fixed angular momentum $j$. Thus with $\Psi^{(j)}$ known we can construct the spin eigenstates for spin-$j$ particle without any knowledge of the matrices $\lbrace\tau_{i}^{\text{\tiny{D}}}\rbrace$. With the complete set of orthogonal eigenstates known  we can construct the matrices $\lbrace\tau_{i}^{\text{\tiny{D}}}\rbrace$ by using them as the basis vectors. From the form of the Killing vector fields\eqref{LIndependentKillVFs} in the coordinate system adapted to one of them, it is clear that the complete set of commuting operators consists of the Casimir operator $\mathcal{J}^2$ and the Killing vector fields $\xi_{3}$ and $\xi_{4}$. Let the quantum numbers associated with them be $j,m$ and $m'$ respectively. The components of the matrix $\Psi$ for a particle with spin-$j$ are $\Psi_{mm'}^{(j)}$. In a chapter on quantum mechanics of the rotator, Eisenberg and Greiner     \cite{eisenberg1987nuclear} found the angular momentum operators equivalent to the Killing fields\eqref{LIndependentKillVFs}. They went on to show that their eigenfunctions are the Wigner D-functions. Therefore 
\begin{equation}
    \Psi_{mm'}^{(j)} = e^{\imath m\alpha} d_{mm'}^{j}(\beta) e^{\imath m'\gamma}
\end{equation}
where $d_{mm'}^{j}(\beta)$ are the Wigner d-functions. When $m'= 0$, the components become the spherical Harmonics. This is expected since in the case of vanishing $\gamma$, the Killing vector fields reduce to the usual angular momentum operators.

We know that spin-$\frac{1}{2}$ particles obey the Dirac equation. As such, to further affirm that for $j=\frac{1}{2}$, $\hat{\varphi}$ is the spinor, we show that in this case \eqref{spinEigenVproblem} implies Dirac equation. For this we make use of the relation between spinors and vectors found in \eqref{vectorSpinorRelation} and Maxwell's equations. We note that if we let $x_{5}$ be purely imaginary, the Killing vector fields in \eqref{halfAngularM} become complex and are the generators of isometries of Minkowski metric. This means that the 1-forms dual to these Killing vector fields are also complex.
We can write Maxwell's equations as 
\begin{equation}
    \left(\varepsilon_{i}^{\hspace{4pt}jk}\partial_{j} + \imath\delta_{i}^{k}\partial_{0}\right)G_{k} = 0.
\end{equation}
with $G_{k} = B_{k} + \imath E_{k}$. Contracting with $\tau_{i}$s of spin-$\frac{1}{2}$ particles given in \eqref{4by4Electroweak} we get
\begin{equation}\label{WeylEquation}
    \begin{aligned}
        (\tau^{i}\varepsilon_{i}^{\hspace{4pt}jk}\partial_{j} + \imath\tau^{k}\partial_{0}) G_{k}  = 0\\
        (\tau\cdot\partial - \partial_{0})\tau\cdot G = 0.
    \end{aligned}
\end{equation}
In the last line of the above equation we made use of the fact that $\tau_{i}\tau_{j} = \imath\varepsilon_{ijk}\tau_{k}$.
In line with the \eqref{vectorSpinorRelation}, the spinors associated with the complex vector $G_{k}$ are the columns $\hat{\varphi}$ of $\tau\cdot G$. Thus reverting  to $x_{5}$, the  equation of motion for the spinor field is
\begin{equation}
    (-\imath\tau\cdot\partial + \partial_{5})\hat{\varphi} = 0.
\end{equation}

We have been considering the case of time independent $\hat{\varphi}$. For time dependent $\hat{\varphi}$ we have to introduce another matrix $\tau_{0}$ in such a way that the matrices form a Lie algebra since we are essentially introducing another Killing vector field $\xi_{0}$.We can write the matrices $\lbrace\tau_{i}\rbrace$ in terms of the Pauli matrices $\lbrace\lambda_{i}\rbrace_{i=0}^{3}$ with $\lambda_{0}$ as the identity matrix, in the following way
\begin{equation}
    \begin{aligned}
        \tau_{5} &= \lambda_{0}\otimes\lambda_{0}\hspace{30pt} &\tau_{1} = \lambda_{0}\otimes\lambda_{2}\\
        \tau_{3} &= \lambda_{2}\otimes\lambda_{1}\hspace{30pt}&\tau_{2} = \lambda_{2}\otimes\lambda_{3}.
    \end{aligned}
\end{equation}
From this, it is apparent that for the matrices to be closed under the Lie bracket, we can have 
\begin{equation}
    \begin{aligned}
        \tau_{0} &= \lambda_{2}\otimes\lambda_{0}.
    \end{aligned}
\end{equation}
Since the Lie derivative of $\xi_{5}$ along all other Killing vectors is zero, we can always find the coordinate system adapted to it. In line with this we can make the substitution
\begin{equation}
    \partial_{5}\hat{\varphi} = \imath m\hat{\varphi}
\end{equation}
in \eqref{WeylEquation}. This substitution results in massive Dirac equation in 4$\mathcal{D}$ spacetime.

We conclude by showing that all continuous symmetries, both external and internal are spacetime symmetries and all the properties of matter are contained in the energy momentum tensor. The energy momentum tensor is not unique in the sense that given the canonical energy momentum tensor $\Tilde{\mathcal{T}}^{ab} $ we can always add the identically conserved term $\nabla_{c}\mathcal{S}^{cab}$ with $\mathcal{S}^{cab}=-\mathcal{S}^{acb} $ to get the new symmetric energy momentum tensor $\mathcal{T}^{ab} =\Tilde{\mathcal{T}}^{ab} + \nabla_{c}\mathcal{S}^{cab}$. Addition of this identically conserved term does not change the conserved charges (components of 4-momentum) since it is a boundary term. If in addition to the symmetric energy momentum tensor $\mathcal{T}^{ab}$ there exists a Killing vector field $\xi_{i}^{\hspace{4pt}a} $, we have a conserved current density 
\begin{equation}
    \mathcal{J}_{i}^{\hspace{4pt}a} = \mathcal{T}^{a}_{b}\xi_{i}^{\hspace{4pt}b}
\end{equation}
since
\begin{equation}
    \begin{aligned}
        \nabla_{a}\mathcal{J}_{i}^{\hspace{4pt}a} &= \left(\nabla_{a}\mathcal{T}^{a}_{b} \right)\xi_{i}^{\hspace{4pt}b} + \frac{1}{2}T_{ab}\left(\nabla^{a}\xi_{i}^{\hspace{4pt}b}+\nabla^{b}\xi_{i}^{\hspace{4pt}a} \right)\\
        &= 0.
    \end{aligned}
\end{equation}
In terms of $\Tilde{\mathcal{T}}^{ab}$ and the identically conserved term, the current density is
\begin{equation}
    \mathcal{J}_{i}^{\hspace{4pt}a} = \Tilde{\mathcal{T}}^{a}_{b}\xi_{i}^{\hspace{4pt}b}-\frac{1}{2}\left(\partial_{c}\xi_{ib}-\partial_{b}\xi_{ic}\right)\mathcal{S}^{cab}+ \nabla_{c}\left( \xi_{ib}\mathcal{S}^{cab}\right).
\end{equation}
Accordingly, if $\partial_{c}\xi_{ib}-\partial_{b}\xi_{ic}\neq 0 $, the identically conserved term contributes to the conserved charge
\begin{equation}
    \mathrm{Q}_{i} =\int\sqrt{\mathrm{G}}dx^{1}\wedge\cdots\wedge dx^{n-1}\mathcal{J}_{i}^{\hspace{4pt}0}.
\end{equation}
This shows that the identically conserved term has observable effects and the energy momentum tensors $\mathcal{T}^{ab}$ and $\Tilde{\mathcal{T}}^{ab}$ are distinguishable. 

In order to understand what the contribution of this identically conserved term is, we consider Minkowski spacetime. The isometries of Minkowski metric are translations, rotations and boosts. The generators of translations are Killing vector fields $\xi_{\nu}^{\hspace{4pt}\mu} = \delta_{\nu}^{\mu}$ and conserved charges associated with them are components of $4-$momentum. As expected, in this case there is no contribution from the identically conserved term. For the  generators of rotations and boosts we have 
\begin{equation}
   (\mathcal{J}^{\mu})_{\alpha\beta} = \Tilde{\mathcal{T}}^{\mu}_{\hspace{4pt}\beta} x_{\alpha}-\Tilde{\mathcal{T}}^{\mu}_{\hspace{4pt}\alpha}x_{\beta} - 2\mathcal{S}_{\alpha\hspace{4pt}\beta}^{\hspace{4pt}\mu} 
\end{equation}
if we neglect the boundary term. It is clear that $ (\mathcal{J}^{0})_{\alpha\beta}$ corresponds to \eqref{AngularMomentumDensity}. Thus $\mathcal{S}_{\alpha\hspace{4pt}\beta}^{\hspace{4pt}0} $ is the spin part of angular momentum density. 

To further affirm the aforementioned results, we consider the conserved charges arising from Noether's theorem (see \cite{noether1971invariant}). Since all symmetries are spacetime symmetries as per our assumption, the variation of any matter field is just  the Lie derivative of such a field along a vector field. That is, for a vector field $\phi_{a}$ we have
\begin{equation}
\begin{aligned}
    \delta\phi_{a} &= \epsilon^{i}L_{\xi_{i}}\phi_{a}\\
            &= \epsilon^{i}\xi_{i}^{\hspace{4pt}b}\partial_{b}\phi_{a} + \epsilon^{i}\partial_{a}\xi_{i}^{\hspace{4pt}b}\phi_{b}.
    \end{aligned}
\end{equation}
Therefore $\delta x^{b} = \epsilon^{i}\xi_{i}^{\hspace{4pt}b}$ and 
$$\Lambda_{ia} = \partial_{a}\xi_{i}^{\hspace{4pt}b}\phi_{b} = -\imath\tau_{i|a}^{\hspace{9pt}b}\phi_{b}.$$
The expression for the conserved currents \cite{ryder1996quantum} becomes
\begin{equation}\label{ConservedCurrents1}
    \mathcal{J}_{i}^{d} = -\imath\frac{\partial\mathfrak{L}}{\partial(\partial_{d}\phi_{a})} \tau_{i|a}^{\hspace{9pt}b}\phi_{b} - \left(\frac{\partial\mathfrak{L}}{\partial(\partial_{d}\phi_{a})}\partial_{b}\phi_{a} - \delta_{b}^{d}\mathfrak{L}\right)\xi_{i}^{\hspace{4pt}b}.
\end{equation}
When $\Tilde{\mathcal{T}}^{d}_{\hspace{4pt}b} = \frac{\partial\mathfrak{L}}{\partial(\partial_{d}\phi_{a})}\partial_{b}\phi_{a} - \delta_{b}^{d}\mathfrak{L}$ is symmetric, the two terms in \eqref{ConservedCurrents1} are conserved. Considering the case where $\lbrace\tau_{i}\rbrace$ form the representation of Lorentz algebra, we see that spin part of the current density is 
\begin{equation}
  \mathcal{S}_{\alpha\hspace{4pt}\beta}^{\hspace{4pt}\mu} = -\imath  \frac{\partial\mathfrak{L}}{\partial(\partial_{\mu}\phi_{a})} \tau_{\alpha\beta|a}^{\hspace{14pt}b}\phi_{b}= -\imath  \frac{\partial\mathfrak{L}}{\partial(\partial_{\mu}\mathbf{\Phi})} \tau_{\alpha\beta}\mathbf{\Phi}.
\end{equation}
We saw that the Killing vector fields in the internal space have no spacetime components. That is $\xi_{i}^{\hspace{4pt}\mu}=0 $. Due to this the conserved current densities associated with them in the absence of the gauge potentials are
\begin{equation}
     \mathcal{J}_{i}^{\mu} = -2\imath\frac{\partial\mathfrak{L}}{\partial(\partial_{\mu}\phi_{a})} \tau_{i|a}^{\hspace{9pt}b}\phi_{b}
\end{equation}
where we made use of \eqref{spinEigenVproblem}. This reiterates the already stated idea that spin and charge are one and the same thing. 

\section{Summary and conclusions}
We derived the metric of higher dimensional spacetime. This made it clear that the Kaluza-Klein ansatz is generally not the solution to vacuum Einstein equations in higher dimensions. We found that consistency of gravity and gauge theories demands doing away with one of the postulates of general relativity and this solves the problems of particles whose mass is of the order of Planck mass and the vacuum energy density of the order of $M_{pl}^{4}$ in addition to resulting in the modified Einstein field equations. It was apparent that imposing the cylinder condition is equivalent to introducing the coordinate system adapted to all Killing vectors. For non-commuting Killing vector fields such a coordinate system does not exist. Accordingly, neglecting massive modes in the original theory cannot  result in non-Abelian gauge theory. We showed that compactification does not result in infinite tower of massive modes for all fields and the fields acquire mass through the Higgs mechanism. The condition on the Killing vector fields shows that the internal space is the group manifold and that the standard model is incorporated in the generalization to non-Abelian gauge theories. The theory suggests the unified description of particles in the sense that all fundamental particles with spin can be described by 1-forms.


\appendix

\section{The Expression of the Ricci curvature Tensor of the Unified Theory}\label{AppRicci}
In order to facilitate calculating the Christoffel symbols and Ricci curvature tensor we utilise the following results
\begin{equation*}
    \begin{split}
    \mathrm{G}^{\mu a}\left(\partial_{a}A_{b}^{\hspace{4pt}k}-\partial_{b}A_{a}^{\hspace{4pt}k}\right) &=g^{\mu\nu}\Big(\partial_{\nu}A_{b}^{\hspace{4pt}k}-\partial_{b}A_{\nu}^{\hspace{4pt}k} \\&- A_{\nu}^{\hspace{4pt}i}\xi_{i}^{\hspace{4pt}j}\left(\partial_{j}A_{b}^{\hspace{4pt}k}-\partial_{b}A_{j}^{\hspace{4pt}k}
    \right)\Big)\\
    &=g^{\mu\nu}\Big(\partial_{\nu}A_{b}^{\hspace{4pt}k}-\partial_{b}A_{\nu}^{\hspace{4pt}k} \\&+ \sigma A_{\nu}^{\hspace{4pt}i}\xi_{i}^{\hspace{4pt}j}\mathrm{C}^{k}_{\hspace{4pt}mn}A_{j}^{\hspace{4pt}m}A_{b}^{\hspace{4pt}n}\Big)\\
    &=g^{\mu\nu}\left(\partial_{\nu}A_{b}^{\hspace{4pt}k}-\partial_{b}A_{\nu}^{\hspace{4pt}k} + \sigma\mathrm{C}^{k}_{\hspace{4pt}mn}A_{\nu}^{\hspace{4pt}m}A_{b}^{\hspace{4pt}n}\right)
    \end{split}
\end{equation*}
\begin{equation}
    \boxed{\mathrm{G}^{\mu a}\left(\partial_{a}A_{b}^{\hspace{4pt}k}-\partial_{b}A_{a}^{\hspace{4pt}k}\right)= g^{\mu\nu}\mathfrak{F}_{\nu b}^{k}}
\end{equation}
\begin{equation*}
    \begin{split}
        \partial_i\mathfrak{F}_{ab}^{k} &= \partial_{a}\partial_{i}A_{b}^{\hspace{4pt}k}-\partial_{b}\partial_{i}A_{a}^{\hspace{4pt}k} + \sigma\mathrm{C}^{k}_{\hspace{4pt}mn}\left(\partial_{i}A_{a}^{\hspace{4pt}m}A_{b}^{\hspace{4pt}n}+\partial_{i}A_{b}^{\hspace{4pt}m}A_{a}^{\hspace{4pt}n}\right)\\
        &=\sigma\mathrm{C}^{k}_{\hspace{4pt}mn}\big(\partial_{a}A_{i}^{\hspace{4pt}n}A_{b}^{\hspace{4pt}m} -\partial_{a}A_{b}^{\hspace{4pt}n}A_{i}^{\hspace{4pt}m} \\ &-\partial_{b}A_{i}^{\hspace{4pt}n}A_{a}^{\hspace{4pt}m} +\partial_{b}A_{a}^{\hspace{4pt}n}A_{i}^{\hspace{4pt}m}  
 +\partial_{i}A_{a}^{\hspace{4pt}m}A_{b}^{\hspace{4pt}n}+\partial_{i}A_{b}^{\hspace{4pt}m}A_{a}^{\hspace{4pt}n}\big)\\
        &=-\sigma A_{i}^{\hspace{4pt}m}\mathrm{C}^{k}_{\hspace{4pt}mn}\big(\partial_{a}A_{b}^{\hspace{4pt}n}-\partial_{b}A_{a}^{\hspace{4pt}n}\big) -\sigma^{2}\mathrm{C}^{k}_{\hspace{4pt}mn}\mathrm{C}^{n}_{\hspace{4pt}pq}A_{a}^{\hspace{4pt}p}A_{i}^{\hspace{4pt}q}A_{b}^{\hspace{4pt}m}\\
        &+\sigma^{2}\mathrm{C}^{k}_{\hspace{4pt}mn}\mathrm{C}^{n}_{\hspace{4pt}pq}A_{b}^{\hspace{4pt}p}A_{i}^{\hspace{4pt}q}A_{a}^{\hspace{4pt}m}\\
         &=-\sigma A_{i}^{\hspace{4pt}m}\mathrm{C}^{k}_{\hspace{4pt}mn}\big(\partial_{a}A_{b}^{\hspace{4pt}n}-\partial_{b}A_{a}^{\hspace{4pt}n}\big)\\& -\sigma^{2}\left(\mathrm{C}^{k}_{\hspace{4pt}qn}\mathrm{C}^{n}_{\hspace{4pt}pm}-\mathrm{C}^{k}_{\hspace{4pt}pn}\mathrm{C}^{n}_{\hspace{4pt}qm}\right)A_{b}^{\hspace{4pt}q}A_{i}^{\hspace{4pt}m}A_{a}^{\hspace{4pt}p}\\
          &=-\sigma A_{i}^{\hspace{4pt}m}\mathrm{C}^{k}_{\hspace{4pt}mn}\big(\partial_{a}A_{b}^{\hspace{4pt}n}-\partial_{b}A_{a}^{\hspace{4pt}n} +\sigma\mathrm{C}_{\hspace{4pt}pq}^{n}A_{a}^{\hspace{4pt}p}A_{b}^{\hspace{4pt}q}\big)
    \end{split}
\end{equation*}
\begin{equation}
    \boxed{\partial_i\mathfrak{F}_{ab}^{k}=-\sigma A_{i}^{\hspace{4pt}m}\mathrm{C}^{k}_{\hspace{4pt}mn}\mathfrak{F}_{ab}^{n}}
\end{equation}
or equivalently
\begin{equation}
    \mathfrak{D}_{i}\mathfrak{F}_{ab}^{k} = 0.
\end{equation}
Using this we find that
\begin{equation*}
    \begin{split}
        \mathrm{G}^{ab}\partial_{a}\mathfrak{F}_{b\nu}^{k} & = \mathrm{G}^{a\mu}\partial_{a}\mathfrak{F}_{\mu\nu}^{k}\\
        &= g^{\rho\mu}\left(\partial_{\rho} - A_{\rho}^{i}\xi_{i}^{\hspace{4pt}j}\partial_{j}\right)\mathfrak{F}_{\mu\nu}^{k}
    \end{split}
\end{equation*}
\begin{equation}
    \boxed{
        \mathrm{G}^{ab}\partial_{a}\mathfrak{F}_{b\nu}^{k}= g^{\rho\mu}\mathfrak{D}_{\rho}\mathfrak{F}_{\mu\nu}^{k}.}
\end{equation}
That being so, we see that the gauge covariant derivative is
\begin{equation}
    \mathfrak{D}^{\rho} = G^{\rho a}\partial_{a}
\end{equation}
in addition to this, it is evident that while in higher dimension the non-Abelian gauge theory is simply $n$-dimensional electromagnetism and the non-linear effects are due to the curvature of the internal space.
Furthermore, for a gauge covariant quantity $\Psi^{k}$ we have
\begin{equation}\label{extraDGaugeC}
    \boxed{\mathfrak{D}_{i}\Psi^{k} =  0.}
\end{equation}

With 

   \[ \mathfrak{S}_{\hspace{4pt}ab}^{k} =  \partial_{a}A_{b}^{\hspace{4pt}k}+\partial_{b}A_{a}^{\hspace{4pt}k}\]
the Christoffel symbols are 
\begin{equation}
    {\Tilde{\Gamma}^{a}_{bc} = \delta^{\mu}_{b}\delta^{\nu}_{c}\mathrm{G}^{a\rho}\Gamma_{\mu\nu\rho} + \frac{\lambda^2}{\zeta^2}\frac{\mathrm{G}}{2}^{a\rho}\left(\mathfrak{F}_{b\rho}\cdot A_{c} +\mathfrak{F}_{c\rho}\cdot A_{b}\right) + \frac{\delta_{k}^{a}}{2}\xi^{k}\cdot\mathfrak{S}_{bc}.}
\end{equation}

Given the Christoffel symbols we proceed to calculate the Ricci curvature tensor in  the following way:
\begin{equation}
    \begin{split}
        \Tilde{\Gamma}^{a}_{bd}\Tilde{\Gamma}^{d}_{ac} &= \Big[\delta^{\mu}_{b}\delta^{\alpha}_{d}\mathrm{G}^{a\rho}\Gamma_{\mu\alpha\rho} + \frac{\lambda^2}{\zeta^2}\frac{\mathrm{G}}{2}^{a\rho}\left(\mathfrak{F}_{b\rho}\cdot A_{d} +\mathfrak{F}_{d\rho}\cdot A_{b}\right) \\ &+ \frac{\delta_{k}^{a}}{2}\xi^{k}\cdot\mathfrak{S}_{bd}\Big]\\
        &\times\Big[\delta^{\beta}_{a}\delta^{\nu}_{c}\mathrm{G}^{d\delta}\Gamma_{\beta\nu\delta} + \frac{\lambda^2}{\zeta^2}\frac{\mathrm{G}}{2}^{d\delta}\left(\mathfrak{F}_{a\delta}\cdot A_{c} +\mathfrak{F}_{c\delta}\cdot A_{a}\right) \\ &+ \frac{\delta_{l}^{d}}{2}\xi^{l}\cdot\mathfrak{S}_{ac}\Big]\\
        &=\delta_{b}^{\mu}\delta_{c}^{\nu}\Gamma^{\beta}_{\mu\alpha}\Gamma^{\alpha}_{\nu\beta} +\frac{\lambda^2}{2\zeta^2}\Gamma^{\rho}_{\mu\delta}\left(\delta_{b}^{\mu}\mathfrak{F}_{\rho}^{\hspace{4pt}\delta}\cdot A_{c} +\delta_{c}^{\mu}\mathfrak{F}_{\rho}^{\hspace{4pt}\delta}\cdot A_{b}\right) \\
        &+\frac{\lambda^{4}}{4\zeta^4}\mathfrak{F}_{\delta}^{\hspace{4pt}\rho}\cdot A_{c}\mathfrak{F}_{\rho}^{\hspace{4pt}\delta}\cdot A_{b} + \frac{\lambda^2}{4\zeta^2}\mathrm{G}^{a\rho}\left(\mathfrak{F}_{b\rho}\cdot\mathfrak{S}_{ac} +\mathfrak{F}_{c\rho}\cdot \mathfrak{S}_{ab}\right)\\
        & + \frac{1}{4}\xi^{k}\cdot\mathfrak{S}_{bl}\xi^{l}\cdot\mathfrak{S}_{ck},
    \end{split}
\end{equation}

\begin{equation}
    \begin{split}
     \Tilde{\Gamma}^{a}_{bc}\Tilde{\Gamma}^{d}_{ad} &=\Big[\delta^{\mu}_{b}\delta^{\nu}_{c}\mathrm{G}^{a\rho}\Gamma_{\mu\nu\rho} + \frac{\lambda^2}{\zeta^2}\frac{\mathrm{G}}{2}^{a\rho}\left(\mathfrak{F}_{b\rho}\cdot A_{c} +\mathfrak{F}_{c\rho}\cdot A_{b}\right) \\
         &+ \frac{\delta_{k}^{a}}{2}\xi^{k}\cdot\mathfrak{S}_{bc}\Big] \left[\delta_{a}^{\alpha}\Gamma^{\delta}_{\alpha\delta} + \frac{\xi^{k}}{2}\cdot\mathfrak{S}_{ak}\right]\\
         &= \delta_{b}^{\mu}\delta_{c}^{\nu}\Gamma^{\alpha}_{\mu\nu}\Gamma_{\alpha\rho}^{\rho} + \frac{\lambda^2}{2\zeta^2}\Gamma_{\rho\delta}^{\delta}\left(\mathfrak{F}_{b}^{\hspace{4pt}\rho}\cdot A_{c} + \mathfrak{F}_{c}^{\hspace{4pt}\rho}\cdot A_{b}\right) +\frac{1}{2}\Big[ \\
        &
         \mathrm{G}^{a\rho}\Gamma_{\rho\mu\nu}+\frac{\mathrm{G}^{a\rho}}{2}\big(\mathfrak{F}_{b\rho}\cdot A_{c} + \mathfrak{F}_{c\rho}\cdot A_{b}\big)\frac{1}{4}\xi^{k}\cdot\mathfrak{S}_{bc}\xi^{l}\cdot\mathfrak{S}_{kl}
         \Big],
    \end{split}
\end{equation}

\begin{equation}
    \begin{split}
        \partial_{a}\Tilde{\Gamma}^{a}_{bc} &= \delta_{b}^{\mu}\delta_{c}^{\nu}\partial_{\rho}\Gamma^{\rho}_{\mu\nu} +\frac{1}{2}\partial_{b}\left(\xi^{k}\cdot\mathfrak{S}_{bc}\right) -\delta_{b}^{\mu}\delta_{c}^{\nu}\frac{\mathrm{G}^{a\rho}}{2}\Gamma_{\rho\mu\nu}\xi^{k}\cdot\mathfrak{S}_{ak} \\ & + \frac{\lambda^2}{2\zeta^2}\partial_{a}\big[\mathrm{G}^{a\rho}\big(
        \mathfrak{F}_{b\rho}\cdot A_{c} + \mathfrak{F}_{c\rho}\cdot A_{b}
        \big)
        \big]\\
        &= \delta_{b}^{\mu}\delta_{c}^{\nu}\partial_{\rho}\Gamma^{\rho}_{\mu\nu} +\frac{1}{2}\partial_{b}\left(\xi^{k}\cdot\mathfrak{S}_{bc}\right) -\delta_{b}^{\mu}\delta_{c}^{\nu}\frac{\mathrm{G}^{a\rho}}{2}\Gamma_{\rho\mu\nu}\xi^{k}\cdot\mathfrak{S}_{ak} \\ &-\frac{\lambda^2}{2\zeta^2}\big(g^{\delta\rho}\Gamma^{\mu}_{\mu\rho} + g^{\mu\delta}\Gamma^{\rho}_{\mu\delta} +\frac{\mathrm{G}^{a\rho}}{2}\xi^{k}\cdot\mathfrak{S}_{ka}\big)\left(\mathfrak{F}_{b\rho}\cdot A_{c} + \mathfrak{F}_{c\rho}\cdot A_{b}\right)\\
        &+\frac{\lambda^2}{2\zeta^2}g^{\delta\rho}\big(\mathfrak{D}_{\delta}\mathfrak{F}_{b\rho}\cdot A_{c} + \mathfrak{D}_{\delta}\mathfrak{F}_{c\rho}\cdot A_{b} + \mathfrak{F}_{b\rho}\cdot \mathfrak{F}_{\delta c}\big)\\
        &+\frac{\lambda^2}{2\zeta^2}\mathrm{G}^{a\rho}\left(\mathfrak{F}_{b\rho}\cdot\mathfrak{S}_{ac} +\mathfrak{F}_{c\rho}\cdot\mathfrak{S}_{ab}\right),
    \end{split}
\end{equation}

\begin{equation}
    \begin{aligned}
    \partial_{b}\Tilde{\Gamma^{a}_{ac}} &= \delta_{b}^{\mu}\delta_{c}^{\nu}\partial_{\mu}\Gamma^{\rho}_{\rho\nu} + \frac{1}{2}\partial_{b}\left(\xi^{k}\cdot\mathfrak{S}_{kc}\right)
    \end{aligned}
\end{equation}

\begin{equation}
    \begin{aligned}
    &\partial_{k}\big(\xi^{k}\cdot\mathfrak{S}_{bc}\big) -\partial_{b}\big(\xi^{k}\cdot\mathfrak{S}_{kc}\big)\\ &= \partial_{k}\xi^{k}\cdot\mathfrak{S}_{bc} -\partial_{b}\xi^{k}\cdot\mathfrak{S}_{kc} +\xi^{k}\cdot\left(\partial_{k}\mathfrak{S}_{bc}-\partial_{b}\mathfrak{S}_{kc}\right)\\
    &= -\xi^{k}\cdot\big(\partial_{k}A_{j} \xi^{j}\cdot\mathfrak{S}_{bc} - \partial_{b}A_{j}\xi^{j}\cdot\mathfrak{S}_{kc}\big) \\ &+\sigma\xi^{k}\cdot\mathrm{C}_{mn}\big(\partial_{k}A_{c}^{\hspace{4pt}m}A_{b}^{\hspace{4pt}n} +\partial_{k}A_{b}^{\hspace{4pt}n}A_{c}^{\hspace{4pt}m}\big)\\
    &=-\xi^{k}\cdot\big(\sigma\mathrm{C}_{mn}A_{k}^{\hspace{4pt}m}A_{l}^{\hspace{4pt}n}\xi^{l}\cdot\mathfrak{S}_{bc} + \mathfrak{S}_{kl}\xi^{l}\cdot\mathfrak{S}_{bc}\\ & - \sigma\mathrm{C}_{mn}A_{b}^{\hspace{4pt}m}A_{l}^{\hspace{4pt}n}\xi^{l}\cdot\mathfrak{S}_{kc} - \mathfrak{S}_{bl}\xi^{l}\cdot\mathfrak{S}_{kc}\\ & -\sigma^2\mathrm{C}_{mn}\mathrm{C}^{m}_{\hspace{4pt}pq}A_{c}^{\hspace{4pt}p}A_{k}^{\hspace{4pt}q}A_{b}^{\hspace{4pt}n} - \sigma\mathrm{C}_{mn}\mathfrak{S}_{ck}^{m}A_{b}^{\hspace{4pt}n}\big)\\
    &= -\xi^{k}\cdot\mathfrak{S}_{kl}\xi^{l}\cdot\mathfrak{S}_{bc} +\xi^{k}\cdot\mathfrak{S}_{bl}\xi^{l}\cdot\mathfrak{S}_{kc} \\ &+\frac{\sigma^2}{2}\mathrm{C}^{q}_{\hspace{4pt}mn}\mathrm{C}^{m}_{\hspace{4pt}pq}A_{c}^{\hspace{4pt}p}A_{b}^{\hspace{4pt}n}.
    \end{aligned}
\end{equation}
 Putting everything together and simplifying we end up with
 \begin{equation}
    \boxed{\begin{aligned}
     \Tilde{\mathfrak{R}}_{bc} =& \delta_{b}^{\mu}\delta_{c}^{\nu}\mathfrak{R}_{\mu\nu} + \frac{\lambda^{2}}{2\zeta^2}\left(A_{b}\cdot\bar{\mathfrak{D}}_{\rho}\mathfrak{F}_{c}^{\hspace{4pt}\rho} + A_{c}\cdot\bar{\mathfrak{D}}_{\rho}\mathfrak{F}_{b}^{\hspace{4pt}\rho} -\mathfrak{F}_{b\rho}\cdot\mathfrak{F}_{c}^{\hspace{4pt}\rho}\right)\\
     &+\frac{\lambda^4}{4\zeta^4}\mathfrak{F}_{\rho\delta}\cdot A_{b} \mathfrak{F}^{\rho\delta}\cdot A_{c} +\frac{1}{4}\sigma^2\mathrm{C}^{q}_{\hspace{4pt}mn}\mathrm{C}^{m}_{\hspace{4pt}pq}A_{b}^{\hspace{4pt}n}A_{c}^{\hspace{4pt}p}
     \end{aligned}}
 \end{equation}
 where \[\bar{\mathfrak{D}}_{\rho}\mathfrak{F}_{b}^{k\hspace{4pt}\rho} = \nabla_{\rho}\mathfrak{F}_{b}^{k\hspace{4pt}\rho} +\sigma A_{\rho}^{\hspace{4pt}m}\mathrm{C}^{k}_{mn}\mathfrak{F}_{b}^{n\hspace{4pt}\rho}.\]

\bibliographystyle{elsarticle-harv} 
\bibliography{sample}

\end{document}